\documentclass[a4paper,12pt]{article}
\usepackage{comment}
\usepackage{subfigure}
\usepackage{amsmath,graphicx,amssymb}
\usepackage{captcont,rotating,mathrsfs,cancel}
\usepackage{multirow,color}
\usepackage[a4paper]{geometry}
\usepackage{tabularx}
\usepackage{float}
\begin{document}
\renewcommand{\vec}[1]{\boldsymbol{\mathbf{#1}}}
\newcommand{\etal}{\textit{et al.}}
\newcommand{\Cinf}{C_{\varepsilon,\infty}}
\newcommand{\kmax}{k_\text{max}}
\newcommand{\teP}{t_{\varepsilon\vert\Pi}}
\newcommand{\tE}{t_\varepsilon}
\renewcommand{\leq}{\leqslant}
\renewcommand{\geq}{\geqslant}

\newcommand{\smeasure}[1]{\mathscr{D}#1}
\newcommand{\measure}[1]{\smeasure{\vec{#1}}}
\newcommand{\pdf}[1]{P[\vec{#1}]}
\newcommand{\gen}[1]{Z[\vec{#1}]}
\newcommand{\ord}[1]{O\left(#1\right)}
\newcommand{\bal}{\begin{align}}
\newcommand{\eal}{\end{align}}
\newcommand{\set}[1]{\{#1\}}
\newcommand{\im}{\imath}

\newcommand{\half}{\frac{1}{2}}
\newcommand{\nono}{\nonumber}
\newcommand{\nuz}{\nu_{0}\,}
\newcommand{\vep}{\varepsilon}
\newcommand{\vepu}{\varepsilon_{U,L}}

\newcommand{\RL}{R_L}
\newcommand{\Rl}{R_{\lambda}}
\newcommand{\cdim}{C_{\vep}}
\newcommand{\Ceps}{C_{\vep}}
\newcommand{\cdiminf}{C_{\vep,\infty}}
\newcommand{\cpi}{C_{\Pi}}
\newcommand{\cpinf}{C_{\Pi,\infty}}

\newcommand{\beq}{\begin{equation}}
\newcommand{\eeq}{\end{equation}}
\newcommand{\bea}{\begin{eqnarray}}
\newcommand{\eea}{\end{eqnarray}}

\newcommand{\dd}{\partial}
\newcommand{\ddt}{\frac{\partial}{\partial t}}
\newcommand{\dr}{\frac{\partial}{\partial r}}

\newcommand{\xt}{(\mathbf{x},t)}
\newcommand{\kt}{(\mathbf{k},t)}
\newcommand{\pkt}{(\mathbf{k'},t')}
\newcommand{\kjt}{(\mathbf{k-j},t)}
\newcommand{\jpt}{(\mathbf{j-p},t)}
\newcommand{\jt}{(\mathbf{j},t)}
\newcommand{\lt}{(\mathbf{l},t)}
\newcommand{\kptp}{(\mathbf{k'},t')}
\newcommand{\kp}{k^{+}}
\newcommand{\ks}{(\mathbf{k},s)}
\newcommand{\kstar}{k_{\ast}}
\newcommand{\kb}{\ov{k}}
\newcommand{\kjs}{(\mathbf{k-j},s)}
\newcommand{\jps}{(\mathbf{j-p},s)}
\newcommand{\js}{(\mathbf{j},s)}
\newcommand{\ls}{(\mathbf{l},s)}
\newcommand{\ps}{(\mathbf{p},s)}

\newcommand{\up}{u^{+}}
\newcommand{\um}{u^{-}}
\newcommand{\ua}{u_\alpha}
\newcommand{\ub}{u_\beta}
\newcommand{\ug}{u_\gamma}
\newcommand{\ud}{u_\delta}
\newcommand{\us}{u_\sigma}
\newcommand{\ur}{u^{(r)}}
\newcommand{\tilur}{\tilde{u}^{(r)}}

\newcommand{\fa}{f_\alpha}
\newcommand{\fd}{f_\delta}
\newcommand{\Mp}{M_{\alpha\beta\gamma}^{+}(\mathbf{k})}
\newcommand{\Mm}{M_{\alpha\beta\gamma}^{-}(\mathbf{k})}
\newcommand{\Mk}{M_{\alpha\beta\gamma}(\mathbf{k})}
\newcommand{\Mmk}{M_{\alpha\beta\gamma}(-\mathbf{k})}

\newcommand{\Nk}{N_{\alpha\beta\gamma}(\mathbf{k})}
\newcommand{\Mpm}{M_{\alpha\beta\gamma}^{\pm}(\mathbf{k})}
\newcommand{\Npm}{N_{\alpha\beta\gamma}^{\pm}(\mathbf{k})}

\newcommand{\frun}[1]{\texttt{#1}} 
\newcommand{\drun}[1]{\texttt{#1}} 

\newcommand{\vmod}[1]{\lvert \vec{#1} \rvert} 

\newcommand{\kmin}{0}
\newcommand{\kcut}{\Lambda} 
\newcommand{\pimax}{\varepsilon_T} 
\newcommand{\ktop}{k_{\textrm{top}}} 
\newcommand{\re}[1]{\textrm{Re} \left[ #1 \right]} 
\newcommand{\dns}{\textup{\textbf{DNS}$2012$}} 
\newcommand{\addref}{[\textbf{reference}]} 
\newcommand{\addmore}{\textbf{+more+}} 
\newcommand{\epsw}{\varepsilon_W} 
\newcommand{\eddie}{\texttt{eddie}} 
\newcommand{\tl}[1]{\texttt{#1}} 
\newcommand{\unitM}{\mathbb{I}} 
\newcommand{\tr}{\textrm{tr}} 
\newcommand{\realpart}{\textrm{Re}} 

\newcommand{\fvect}[1]{\hat{#1}} 
\newcommand{\high}{+} 
\newcommand{\low}{-} 
\newcommand{\order}[1]{O(#1)}


\newcommand{\av}[1]{\left\langle #1 \right\rangle}
\newcommand{\cond}[1]{\left\langle #1 \right\rangle_{0}}
\newcommand{\cav}[1]{\left\langle #1 \right\rangle_{c}}
\newcommand{\rav}[1]{\left\langle #1 \right\rangle^{(r)}}
\newcommand{\ov}{\overline}


\title{Does intermittency affect the inertial transfer rate in stationary isotropic turbulence?}
\author{S. R. Yoffe\footnote{SUPA Department of
Physics, University of Strathclyde, John Anderson Building, 107
Rottenrow East. Glasgow G4 0NG.}\, and W. D. McComb,\\
SUPA School of Physics and Astronomy,\\
Peter Guthrie Tait Road, \\
University of Edinburgh,\\
EDINBURGH EH9 3JZ.\\
Email: wdm@ph.ed.ac.uk}

\maketitle 
\thispagestyle{empty} 
\begin{abstract} 
Direct numerical simulations of the forced Navier-Stokes equations were performed, in which each shell-averaged quantity evolved  from a value appropriate to an initial Gaussian state, to fluctuate about a mean value. Once the transient had passed, mean values (and their associated statistics) were obtained by sampling the evolved time-series at intervals of  the order of an eddy-turnover time. This was repeated  for a range of Taylor-Reynolds numbers  from 10.6 to 335.2. With increasing Reynolds number, our results for energy spectra, transfer spectra and inertial flux supported the Kolmogorov-Obukhov picture of turbulent energy transfer. In particular, we observed the onset of scale-invariance of the inertial flux, accompanied by the onset of the -5/3 power law in the energy spectrum for the corresponding inertial range of wavenumbers. Detailed comparisons showed that our results were in agreement with those found in many other investigations. Flow visualization methods were used to study the internal intermittency. This phenomenon is seen in single realisations but was found to average out with increasing number of realisations under ensemble-averaging. Following a critical review of the literature relating to the controversy about intermittency effects versus finite-Reynolds number corrections, it was concluded that, for the case of stationary isotropic turbulence, internal intermittency cannot affect the Kolmogorov-Obukhov picture, as this is constructed entirely in terms of ensemble-averaged mean quantities.
\end{abstract}

\newpage
\tableofcontents
\newpage

\section{Introduction}

In 2001, Lumley and Yaglom published an article `A Century of Turbulence' \cite{Lumley01}. After  they had made it clear that neither author had been involved with turbulence for the entire century, their conclusions were bleak. We may quote from their Abstract, as follows:
\begin{quote}
\emph{`This field does not appear to have a pyramidal structure, like the best of physics. We have very few great hypotheses. Most of our experiments are exploratory experiments. What does this mean?'}
\end{quote}
They go on to answer their own question: 
\begin{quote}
\emph{`We believe that it means that, even after 100 years, turbulence studies are still in their infancy.'}
\end{quote}

The passage of another two decades does not seem to have altered that conclusion, and even if we restrict our attention to isotropic turbulence --- arguably the most fundamental branch of the subject --- this is still characterised by various unresolved issues. For instance, we have the endlessly contentious problems posed by topics like the scaling of two-time correlations and the free decay of the total kinetic energy. In the latter case, there is often a failure to appreciate that the three distinct problems posed by the mathematical physics problem of free decay; the numerical simulation of free decay; and the decay of grid-generated turbulence are all, in principle, not quite the same problem. There is also a widespread belief that the infinite Reynolds number limit is the same thing as setting the viscosity to zero, accompanied it would appear by a belief in the failure of the continuum description of the fluid concerned. But, above all, there is the idea that the Komogorov `$-5/3$' spectrum is subject to intermittency corrections. From a fundamental view this is difficult to understand because Kolmogorov's theory \cite{Kolmogorov41a} was expressed in terms of the \emph{mean} dissipation, which can hardly be affected by intermittency. 

The trouble seems to be that Kolmogorov's theory (K41, for brevity), despite its great pioneering importance, was an incomplete and indeed inconsistent theory. It was formulated in real space; where, although the energy transfer process can be loosely visualised from Richardson's idea \cite{Richardson22} of a cascade, the concept of such a cascade is not mathematically well defined. Also, having introduced the inertial range of scales, where the viscosity may be neglected, he characterised this range by the viscous dissipation rate, which is not only inconsistent but incorrect\footnote{In his exegesis of Kolmogorov's theories, Batchelor \cite{Batchelor47} referred to `the constant rate at which each set (\emph{sic}) passes to the next smaller neighbour'; but, in general, discussion and criticism of K41 continues to use the term `dissipation'.}. An additional complication, which undoubtedly plays a part, is that his theory was applied to turbulence in general. The basic idea was that the largest scales would be affected by the nature of the flow, but a stepwise cascade would result in smaller eddies being universal in some sense. That is, they would have much the same statistical properties, despite the different conditions of formation. In order to avoid uncertainties that can arise from this rather general idea, we will restrict our attention to stationary, isotropic turbulence in this article.

To make a more physical picture we have to follow Obukhov and work in $k$-space, with the Fourier transform $\mathbf{u}(\mathbf{k},t)$ of the velocity field $\mathbf{u}(\mathbf{x},t)$.  This procedure was introduced by Taylor, in order to allow the problem of isotropic turbulence to be formulated as one of statistical mechanics, with the Fourier components acting as the degrees of freedom. In this way,  Obukhov identified the conservative, inertial flux of energy through the modes as being the key quantity determining the energy spectrum in the inertial range. It follows that, with the input and dissipation being negligible, the flux must be constant (i.e. independent of wavenumber) in the inertial range, with the extent of the inertial range increasing as the Reynolds number was increased. This was later recognized by Osager \cite{Onsager45}. Later still, this property became widely known and for many years has been referred to by theoretical physicists as \emph{scale invariance}\footnote{Scale invariance is a general mathematical property and can refer to various things in turbulence research. It simply means that something which might depend on an independent variable, in either real space or wavenumber space,  is in fact constant.} It should be emphasised that the inertial flux is an average quantitiy, as indeed is the energy spectrum, and any intermittency effects present, which are characteristics of the instantaneous velocity field, will inevitably be averaged out. Of course, in stationary flows the inertial transfer rate is the same as the dissipation rate, but in non-stationary flows it is not.

This is not intended to minimise the importance of Kolmogorov's pioneering work. It is merely that we will argue here that one also needs to consider Obukhov's theory \cite{Obukhov41}, with possibly also a later contribution from Onsager \cite{Onsager45}, in order to have a complete theoretical picture. In effect this seems to have been the view of the turbulence community from the late 1940s onwards. Discussion of turbulent energy transfer and dissipation in isotropic turbulence was almost entirely in terms of the spectral picture.  It was not until the extensive measurements of higher-order structure functions by Anselmet \emph{et al.} \cite{Anselmet84} that the real-space picture became of interest, along with the concept of \emph{anomalous exponents}. To lay a foundation for the present work, we will first state the problem in a more succinct fashion and then consider the joint Kolmogorov-Obukhov picture in more detail.

 \subsection{A short statement of the problem}
By restricting our attention to stationary, isotropic
turbulence, we rule out effects due to mean shear, system
rotation, density stratification, and so on. This leaves us with a stark
choice: deviations from Kolmogorov's predictions for the energy spectrum
(or second-order structure function) must be due either to the Reynolds
number being finite (K41 is based on an assumption of very large
Reynolds numbers)\footnote{Strictly, we should also bear in mind the possibility of the persistence of input/forcing effects.} or to the effects of internal intermittency, as was
suggested later on by Kolmogorov in 1962.

Over the last few decades a veritable industry has grown up, based on
the search for so-called \emph{intermittency corrections}. Recently it has been dominated by multi-scale or multifractal models of turbulence; but this topic now seems less popular. Such
activity finds a sympathetic audience, because many people seem to see
the K41 picture as being counter-intuitive, when one considers aspects of
turbulence such as vortex-stretching, localness,
intermittency and the taking of averages.

Running counter to this belief in `intermittency corrections' (or,
increasingly, `anomalous exponents') which has been dominant in recent
times, there is a growing view (see references 
\cite{Bowman96}-\nocite{Barenblatt98a}\nocite{Esser99}\nocite{Qian00}\nocite{Gamard00}\nocite{Lundgren02}\cite{McComb09a})
that K41 is an asymptotic theory, valid in the limit of infinite
Reynolds number. In this school of thought, any deviations from K41 are
due to finite viscosity. As a result, opinion in the turbulence community is deeply divided on
this fundamental issue. Here we will begin by giving a succinct statement of the situation.

As is well known, in 1941, Kolmogorov \cite{Kolmogorov41a,Kolmogorov41b}
gave two different derivations of his famous result for the
second-order structure function\footnote{These are often referred to as K41A and K41B, respectively.}:
\begin{equation}
S_2 = C \varepsilon^{2/3} r^{2/3},
\label{rtt}
\end{equation}
for $l<r<L$, where $l$ is a measure of the scale at which viscous
effects begin to dominate (\emph{i.e.} the internal scale), $L$ is a measure
of the large scales of the system (\emph{i.e.} the external scale) and the prefactor takes
the value $C\simeq 2$. As is equally well
known, the corresponding result for the energy spectrum in wavenumber is
\begin{equation}
E(k) =\alpha \varepsilon^{2/3} k^{-5/3},
\label{kft}
\end{equation}
where the prefactor $\alpha$ is widely known as the \emph{Kolmogorov
constant} and takes a value of about $\alpha = 1.6$. As pointed out earlier, this result may be attributed to Obukhov \cite{Obukhov41} and Onsager \cite{Onsager45}. Despite this, for our present purposes we will follow standard practice and refer to both (\ref{rtt}) and (\ref{kft}) as K41.

Shortly after this work was published, it was criticised by Landau (see
the footnote on page 126 of \cite{Landau59}). Kolmogorov
\cite{Kolmogorov62} interpreted this criticism as a need to treat the
dissipation rate as a variable; and, working with its average taken over
a sphere of radius $r$, concluded that the right hand side of equation
(\ref{rtt}) should be multiplied by a factor $(L/r)^{\mu}$, where $\mu$
is often referred to nowadays as an \emph{intermittency correction}. We will refer to this later theory as K62.

That development gave rise to further attempts by other workers to
obtain a value for $\mu$. As a result, for many years K41 has had a
question mark hanging over its status as a theory of inertial-range
turbulence. From our present point of view, the question may be posed as: is there an intermittency correction to the Kolmogorov spectrum?

\subsection{The Kolmogorov-Obukhov theory of turbulence}

For completeness, we should begin by mentioning that Kolmogorov also used the K\'{a}rm\'{a}n-Howarth equation, which is the energy balance equation connecting the second- and third-order structure functions, to derive the so-called `$4/5$' law for the third-order structure function \cite{Kolmogorov41b}. This procedure amounts to a \emph{de facto} closure, as the time-derivative is neglected (an exact step in our present case) and the term involving the viscosity vanishes in the limit of infinite Reynolds number. This is often described as \emph{`the only exact result in turbulence theory'}; but increasingly it is being said, perhaps more correctly, to be `the only \emph{asymptotically} exact result in turbulence'.

As part of this work \cite{Kolmogorov41b}, Kolmogorov also assumed that the skewness was constant; and this provided a relationship between the second- and third-order structure functions which recovered the `$2/3$' law. It is interesting to note that Lundgren \cite{Lundgren02} used the method of matched asymptotic expansions to obtain both the `$4/5$' and `$2/3$' laws, without having to make any assumption about the skewness. His work also offered a way of estimating the extent of the inertial range in real space.

The K\'{a}rm\'{a}n-Howarth equation is local in the independent variables and therefore does not describe an energy cascade. In contrast, the Lin equation (which is just its Fourier transform) shows that all the degrees of freedom in turbulence are coupled together. It takes the form, for the energy spectrum $E(k, t)$, in the presence of an input spectrum $W(k)$:
\begin{equation}
\frac{\partial E(k,t)}{\partial t} = W(k)+ T(k,t)-  2\nu_{0}k^{2}E(k, t),
\label{lin}
\end{equation}
where $\nu_{0}$ is the kinematic viscosity and the transfer spectrum $T(k,t)$ is 
given by
\begin{eqnarray}
T(k,t) & = & 2\pi k^{2}\int d^{3}j\int d^{3}l\,\delta
(\mathbf{k}-\mathbf{j}-\mathbf{l})M_{\alpha\beta\gamma}(\mathbf{k})
\nonumber \\
& \times & \left\{C_{\beta\gamma\alpha}
(\mathbf{j},\mathbf{l},\mathbf{-k};t)
-C_{\beta\gamma\alpha}(\mathbf{-j},\mathbf{-l},\mathbf{k};t)
\right\},
\end{eqnarray}
with
\beq
M_{\alpha\beta\gamma}(\mathbf{k})=-\frac{i}{2}\left[k_{\beta}
P_{\alpha\gamma}(\mathbf{k})+k_{\gamma}P_{\alpha\beta}(\mathbf{k})\right],
\label{M}
\eeq
and the projector $P_{\alpha\beta}(\mathbf{k})$ is:
\beq
P_{\alpha\beta}(\mathbf{k})=\delta_{\alpha\beta}-
\frac{k_{\alpha}k_{\beta}}{|\mathbf{k}|^{2}}, 
\eeq
where $\delta_{\alpha\beta}$ is the Kronecker delta, and the third-order
moment $C_{\beta\gamma\alpha}$ here takes the specific form:
\beq
C_{\beta\gamma\alpha}(\mathbf{j},\mathbf{l},\mathbf{-k};t)=\langle
u_{\beta}(\mathbf{j},t)u_{\gamma}(\mathbf{l},t)u_{\alpha}(\mathbf{-k},t)
\rangle.
\eeq
At this stage we will also define the flux of energy $\Pi(\kappa,t)$ due to inertial transfer through the mode with wavenumber $k=\kappa$. This is given by:
\beq
\Pi(\kappa,t) = \int_{\kappa}^{\infty}\,dk\,T(k,t).
\label{tp}
\eeq
Further discussion and details may be found  in Section 4.2 of the book \cite{McComb14a}.

We now have a rather simple picture. The form of the input spectrum should be chosen to be peaked near the origin, such that higher wavenumbers are driven by inertial transfer, with energy being dissipated locally by the viscosity. We can define the rate at which stirring forces do work on the system by:
\beq
\int_0^\infty \, W(k)\, dk = \varepsilon_W.
\eeq
Obukhov's idea of the constant  inertial flux can be expressed as follows. As the Reynolds number is increased, the transfer rate, as given by equation (\ref{tp}), must also increase and reach a maximum value which, in turn must be equal to the viscous dissipation. Thus we introduce the symbol $\varepsilon_T$ for the maximum inertial flux as:
\beq
\varepsilon_T = \Pi_{\mbox{max}},
\eeq
and for stationary turbulence at sufficiently high Reynolds number, we have the limiting condition:
\beq
\varepsilon = \varepsilon_T = \varepsilon_W.
\eeq
Thus the loose idea of a local cascade involving eddies in real space is replaced by the precisely formulated concept of scale invariance of the inertial flux in wavenumber space.

Our main purpose in the present work is to examine this concept of scale invariance in wavenumber space and also the effect on the intermittency of the averaging process. The paper is organised as follows:
\begin{description}
\item[Section 2] We begin by describing our numerical methods and summarising the simulations which we carried out.
\item[Section 3] The  use of a statistical ensemble, in addition to the usual shell averaging, is discussed with particular reference to isotropy, the evaluation of errors and the resolution of both small and large scales.
\item[Section 4] A detailed assesment is made of the onset of scale invariance of the energy flux and also the Kolmogorov spectrum, for a range of wavenumbers.
\item[Section 5] We make use of flow visualization in real space to assess the effect of ensemble-averaging on the internal intermittency.
\item[Section 6] In order to provide a context for our results, we discuss various recent investigations, both theoretical and experimental, which support the K41 theory in the limit of infinite Reynolds numbers.
\end{description}

\section{The numerical simulation}

Over the last decade or more, we have been using direct numerical simulation along with analytical methods to study outstanding problems and unresolved issues in isotropic turbulence. This has involved carrying out simulations of the Navier-Stokes equation, based on an initially Gaussian field, with a prescribed initial spectrum, and allowing a turbulent velocity field to develop with time for both free decay and forced simulations.

The initial spectrum was taken to be a standard form given by:
\begin{equation}
E(k,0) = C_1k^{C_2}\exp({-C_3k^{C_4})}, 
\label{inspect}
\end{equation}
where the constants $C_1-C_4$ have to be chosen. As we require the initial spectrum to be peaked near the origin, so that the excitation of higher wavenumbers is purely by nonlinear transfer, it is usual to be influenced by the low-$k$ expansion of the spectrum, thus:
\begin{equation}
E(k,0) = Ak^2  + Bk^4 + \mathcal{O}(k^6).
\label{lowkexp}
\end{equation}For decades there has been controversy over whether the spectrum should start off as $k^2$ or $k^4$, so like others in the field we have used both forms. However, more recently it has been shown that $A=0$ is an exact result \cite{McComb16a}, and accordingly we now only consider trial spectra that begin as $k^4$. 

We began by studying freely decaying turbulence, and showed that the so-called Taylor dissipation surrogate was in fact a better surrogate for the nonlinear energy transfer and only became equal to the dissipation when the Reynolds number was high enough for spectral scale invariance to be present \cite{McComb10b}. Later we studied the onset of Navier-Stokes turbulence in free decay from arbitrary initial conditions and proposed new onset criteria for this \cite{McComb18b}. For forced stationary turbulence, we showed that correcting for systematic errors led to the canonical result $n=2/3$ for the exponent of the second-order structure function as the asymptotic result for large Reynolds numbers \cite{McComb14b}. We have also shown that in stationary turbulence the dimensionless dissipation rate asymptotes to a universal value at large Reynolds numbers \cite{McComb15a}. In this work, the results of the simulation supplemented an analysis which was asymptotically exact in the limit of large Reynolds numbers. Lastly, we found that at very low Reynolds numbers, a forced simulation undergoes a symmetry-breaking phase transition to a self-organised state \cite{McComb15b}. Our general approach has also been extended to magnetohydrodynamic turbulence, e.g. see \cite{Berera14}, \cite{Linkmann15a}.

Our simulations have been well validated by means of extensive and detailed
comparison with the results of other investigations. Further details of the 
performance of our code  may be found in 
the thesis by Yoffe \cite{Yoffe12}, along with  a 
direct comparison with the freely-available pseudospectral
code {\tt{hit3d}} \cite{Schumakov07,Schumakov08}. The Taylor-Green vortex was simulated as a benchmarking problem and our results were in good agreement with those of Brachet \emph{et al} \cite{Brachet83}.
Furthermore our data
reproduces the characteristic behaviour for the plot of the dimensionless 
dissipation rate $\Ceps$ against $\Rl$
\cite{McComb14a}, and agree closely with other representative results in 
the literature, such as the work by Wang, Chen, Brasseur and Wyngaard 
\cite{Wang96}, Cao, Chen and Doolen \cite{Cao99}, Gotoh, Fukayama and Nakano 
\cite{Gotoh02}, Kaneda, Ishihara, Yokokawa, Itakura and Uno 
\cite{Kaneda03}, Donzis, Sreenivasan and Yeung \cite{Donzis05} and Yeung, 
Donzis and Sreenivasan \cite{Yeung12}, 
although there are some differences in forcing methods.

\subsection{Details of the numerical method}

We used a pseudospectral direct numerical simulation (DNS), with full de-aliasing implemented by
truncation of the velocity field according to the two-thirds rule \cite{Orszag71}. Time
advancement for the viscous term was performed exactly using an integrating
factor, while the non-linear term was stepped forward in time using Heun's method
\cite{Heun00},
which is a second-order predictor-corrector routine. Each
simulation was started from a Gaussian-distributed random field with a
specified energy spectrum, which followed $k^4$ for the low-$k$
modes. Measurements were taken after the simulations had reached a
stationary state.
The system was forced by negative damping, with the Fourier transform
of the force $\vec{f}$ given by
\begin{align}
 \vec{f}(\vec{k},t) &=
      (\varepsilon_W/2 E_f) \vec{u}(\vec{k},t) \quad
\text{for} \quad  0 < \lvert\vec{k}\rvert < k_f ; \nonumber \\
  &= 0   \quad \textrm{otherwise},
\label{forcing}
\end{align}
where $\vec{u}(\vec{k},t)$ is the instantaneous velocity field (in
wavenumber space). The highest forced wavenumber, $k_f$, was chosen to
be $k_f = 2.5k_{min}$, where $k_{min}=2\pi/L_{box}=1$ is the lowest 
resolved wavenumber. As $E_f$ was the total energy contained in the forcing
band, this ensured that the energy injection rate was $\varepsilon_W =
\textrm{constant}$. It is worth noting that no method of energy
injection  employed in the numerical simulation of isotropic turbulence
is experimentally realizable. The present method  of negative
damping has also been used in other investigations
\cite{Jimenez93,Machiels97a,Yamazaki02,Kaneda03,Kaneda06}, albeit not necessarily
such that $\vep_W$ is maintained constant (and note the
theoretical analysis of this type of forcing by Doering and Petrov
\cite{Doering05}). Also, note that the correlation between the force and the
velocity is restricted to the very lowest wavenumbers.

For each Reynolds number studied, we used the same initial spectrum and
input rate $\varepsilon_W$. The initial spectrum took the form:
\begin{equation}
E(k,0) = C_1k^{4}\exp(-C_3k^{2}), 
\label{spectin}
\end{equation}
which is (\ref{inspect}) with $C_2=4$ and $C_4=2$. The other constants were given the values: $C_1 = 0.001702$ and $C_3=0.8$. 
The only initial condition changed from one run (i.e. value of the Reynolds number) to another was the
value assigned to the kinematic viscosity. Once the initial transient
had passed, the velocity field was sampled every half a large-eddy
turnover time, $\tau = L/U$, where $L$ denotes the average integral scale and
$U$ the rms velocity.
The ensemble populated with these sampled
realizations was used, in conjunction with the usual shell averaging, to
calculate statistics. 

Simulations were run using lattices of size $64^3$,\
$128^3,\ 256^3,\ 512^3$ and $1024^3$, with corresponding Reynolds
numbers ranging from $R_\lambda = 10.6$ up to $335.2$. 
The smallest
wavenumber was $k_\text{min} = 2\pi/L_\text{box} = 1$ in all
simulations, while the maximum wavenumber satisfied $k_\text{max}\eta
 \geqslant 1.30$ for all runs except that in which $R_\lambda = 335.2$ which satisfied
$k_\text{max}\eta \geqslant 1.01$, where $\eta$ is the Kolmogorov dissipation
lengthscale.\footnote{Note that a subsequent run at $2048^3$ achieved $R_\lambda = 435.2$ and satisfied $k_\text{max}\eta \geqslant 1.30$. This was reported in \cite{McComb14b}, but was not used in the work presented here.} 
The integral scale, $L$, was found to lie between
$0.17 L_{\text{box}}$ and $0.23 L_{\text{box}}$.

It can be seen in Figure 2 of McComb, Hunter and Johnston \cite{McComb01a} that a small-scale
resolution of $k_{max}\eta > 1.6$ is desirable in order to capture the relevant
dissipative physics. Evidently, this would restrict the attainable Reynolds
number of the simulated flow, and the reference suggests that
$k_{max}\eta \geqslant 1.3$ would still be acceptable (containing $\sim 99.5\%$
of dissipative dynamics \cite{Yoffe12}).
In contrast, at $k_{max}\eta \simeq 1$ a non-negligible part of the dissipation
is left out of account. Many high resolution simulations of isotropic turbulence
 try to attain Reynolds numbers as high as possible and thus opt for minimal
resolution requirements.
In this paper the simulations have been
conducted following a more conservative approach, where the emphasis has
been put on better resolving the physical processes, thus necessarily compromising to some extent
on Reynolds number.
Large-scale resolution has only relatively recently received attention in the
literature. As mentioned above, the largest scales of the flow are smaller
than a quarter of the simulation box size. Further discussion is given later on in Section 3.4, where we consider the dissipation requirements in more detail.

\subsection{Summary of simulations carried out}

We begin by briefly summarising our previous discussion before providing the relevant definitions, and the details of the simulations.

The time evolution of forced isotropic turbulence was simulated for a
range of Taylor-Reynolds numbers, $8.70 \leq \Rl \leq 335.2$. At each of the Reynolds numbers studied,  the system was initialised as a Gaussian random
field, using an initial spectrum with $k^4$ as its low $k$ behaviour, and allowed to 
reach a steady-state solution of
the Navier-Stokes equations for shell-averaged quantities. Once this initial transient period had passed,
the velocity field was sampled at intervals corresponding to one half of a large eddy turnover time, $\tau = L/u$,
to create a set of realisations making up a statistical ensemble. As well as having been shell-averaged, the energy and transfer
spectra were also averaged over this ensemble and used to calculate the
statistics of the velocity field.

\subsubsection{Notation and definitions}
The various parameters calculated either during the simulation, or from the spectra after the simulation, include:
\begin{description}
 \item [Total energy] This is found by integrating the energy spectrum over all $k$:
   \begin{equation}
    E(t) = \int dk\ E(k,t). 
   \end{equation}

 \item [Root-mean-square (rms) velocity]{A characteristic velocity scale is found from the total energy, since the total energy is proportional to the velocity squared:
 \begin{equation}
  E(t) = \tfrac{1}{2}\langle u^2(\vec{x},t)\rangle = \tfrac{1}{2} \left[ \langle u_x^2(\vec{x},t)\rangle + \langle u_y^2(\vec{x},t)\rangle + \langle u_z^2(\vec{x},t)\rangle \right] \ .
\end{equation}
 By assuming isotropy, we have $\langle u_x^2(\vec{x},t)\rangle = \langle u_y^2(\vec{x},t)\rangle = \langle u_z^2(\vec{x},t)\rangle = u^2$ so that $E(t) = \tfrac{3}{2}u^2$, or
\begin{equation}
 u(t) = \sqrt{\tfrac{2}{3} E(t)} \ .
\end{equation}
}
 \item [Dissipation spectrum]{The dissipation spectrum has the form
 \begin{equation}
 D(k,t) = 2\nu_0 k^2 E(k,t)\ ,
\end{equation}
so is readily found from the energy spectrum. 
}
 \item [Dissipation rate]{The dissipation rate is the integral of the dissipation spectrum:
 \begin{equation}
 \varepsilon(t) = \int dk\ D(k,t)  \ .
\end{equation}
}

 \item [Integral scale]{This gives a characteristic length-scale of the system based on large-scale structures. It was initially introduced with model fits to the correlation function $f(r) \sim e^{-r/L}$.  It is defined in Fourier space as
 \begin{equation}
 L(t) = \frac{3\pi}{4E(t)} \int dk\ \frac{E(k)}{k}.
\end{equation}

}
 \item [Taylor micro-scale]{Another important length-scale, this time characterising the small-scale structures of the system.  It is found as:
 \begin{equation}
 \lambda(t) = \left(\frac{10 \nu_0 E(t)}{\varepsilon(t)}\right)^{1/2} = \left(\frac{15 \nu_0 u^2(t)}{\varepsilon(t)}\right)^{1/2} \ .
\end{equation}
}
 \item [The Reynolds numbers]{ In general it is defined as:
 \begin{equation}
 Re = \frac{Ul}{\nu_0} \ ,
\end{equation}
where $U$ and $l$ are some characteristic (possibly time-dependent) velocity- and length-scales and $\nu_0$ is the kinematic viscosity. In this investigation we use the integral-scale Reynolds number $R_L = uL/\nu_0$ and the Taylor-Reynolds number $R_\lambda = u\lambda/\nu_0$. }
 \item [The Kolmogorov scale]{The Kolmogorov length-scale gives the approximate scale at which viscous effects become important and is given by:
 \begin{equation}
  \eta(t) = \left(\frac{\nu_0^3}{\varepsilon(t)}\right)^{1/4} \ .
\end{equation}

}
 \item [Longitudinal velocity derivative skewness]{Also referred to as simply the skewness, the longitudinal velocity derivative skewness is one of the most sensitive parameters in quantifying turbulence. In real space, it is defined as:
\begin{equation}
 S(t) = \frac{\left\langle (\partial_{1} u_1(\vec{x},t))^3\right\rangle}{\left\langle (\partial_{1} u_1(\vec{x},t))^2 \right\rangle^{3/2}} \ ,
\end{equation}
where $\partial_1 = \partial/\partial x_1$, or in Fourier space as:
\begin{equation}
 S(t) = \frac{2}{35} \left( \frac{\lambda(t)}{u(t)} \right)^3 \int dk\ k^2 T(k,t) \ .
\end{equation}
It should be noted that pseudospectral methods have access to both of these methods, and there is often a discrepancy between what should be equivalent results. 
}
 \item[Structure functions]{Structure functions are found in configuration space by considering the correlations of the difference between two points. The $n^{th}$-order longitudinal structure function is defined as:
\begin{equation}
 S_n(r) = \left\langle \Big[ \delta \vec{u}(\vec{r}) \cdot \hat{\vec{r}}\Big]^n \right\rangle = \left\langle \Big[\big(\vec{u}(\vec{x}+\vec{r},t)-\vec{u}(\vec{x},t)\big)\cdot \hat{\vec{r}}\Big]^n \right\rangle \ .
\end{equation}
}

 \item[Dissipative wavenumber]{The dissipation wavenumber $k_d$ is the reciprocal of the Kolmogorov microscale. To quantify how well resolved a computation is, we consider the lowest wavenumber $k_{\textrm{diss}}$ such that
 \begin{equation}
  \int_0^{k_{\textrm{diss}}} dk\ 2\nu_0 k^2\ E(k,t) \geq 0.995\varepsilon \ .
 \end{equation}
 That is, the wavenumber up to which 99.5\% of the dissipation is accounted for. This should satisfy $k_{\textrm{diss}} < \kmax$ for the simulation to be well resolved.}
\end{description}

\subsubsection{Preliminary results}

\begin{figure}[H]
 \begin{center}
  \subfigure[Fluctuation of energy $E$ and dissipation rate $\varepsilon$.]{
   \includegraphics[width=0.70\textwidth,trim=3px 35px 15px 40px,clip]{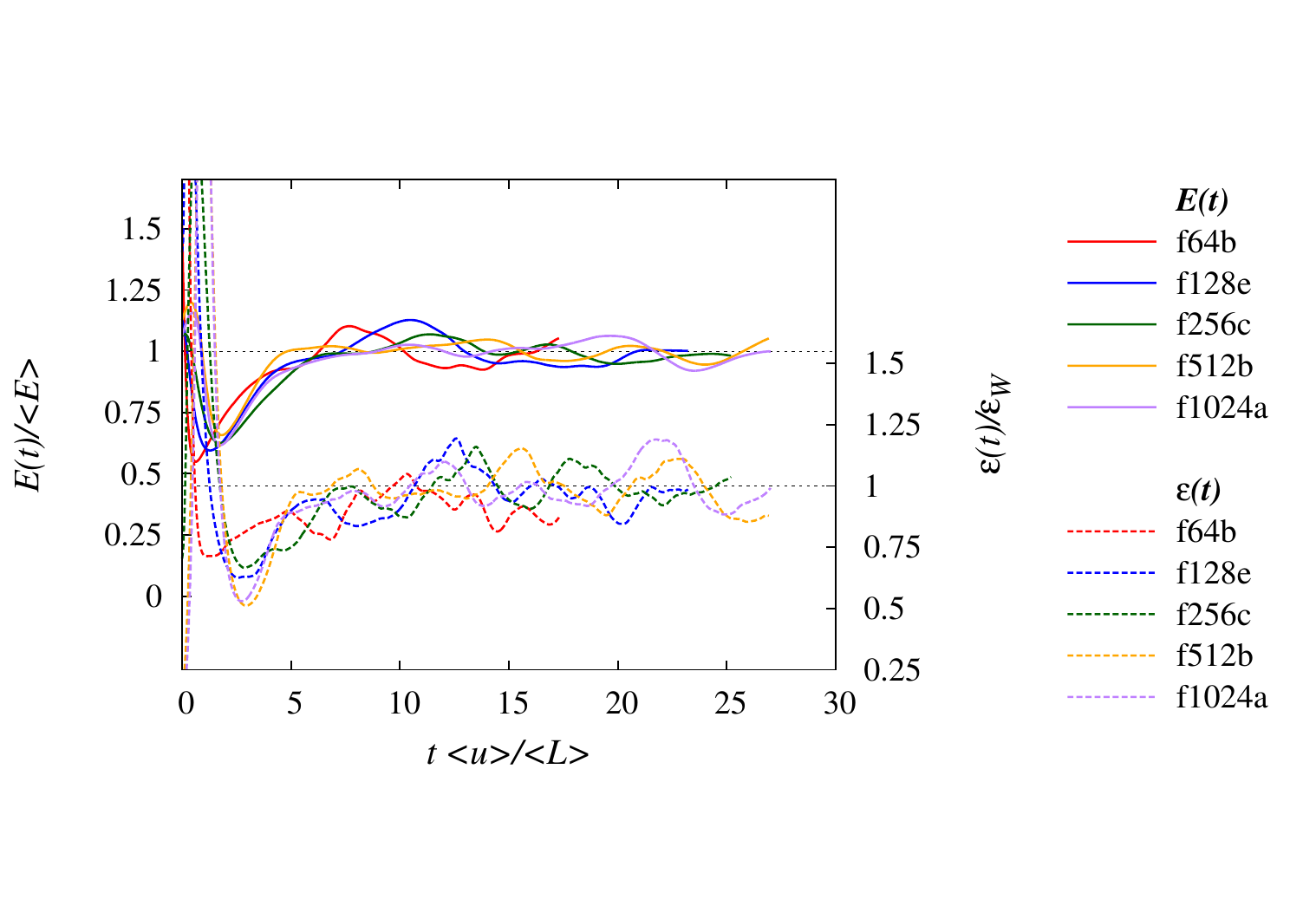}
  }
  \subfigure[Fluctuation of the integral length-scale $L$ and the Taylor microscale $\lambda$.]{
   \includegraphics[width=0.70\textwidth,trim=3px 35px 15px 40px,clip]{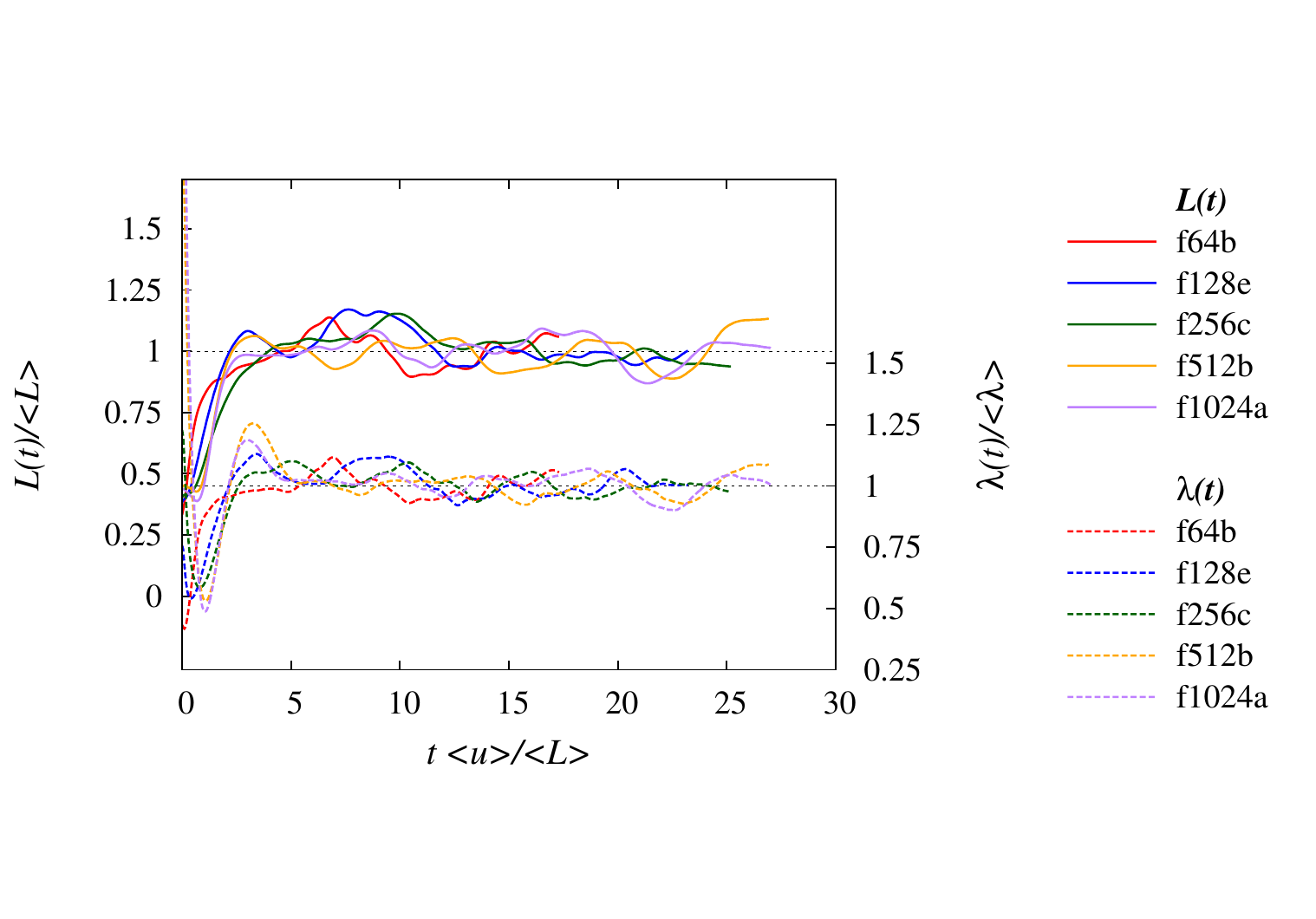}
  }
 \subfigure[Fluctuation of skewness factor $S$ and the transfer power $\Pi(0,t)$.]{
   \label{sfig:forced_time_series_S-Pi0}
   \includegraphics[width=0.70\textwidth,trim=3px 35px 13px 40px,clip]{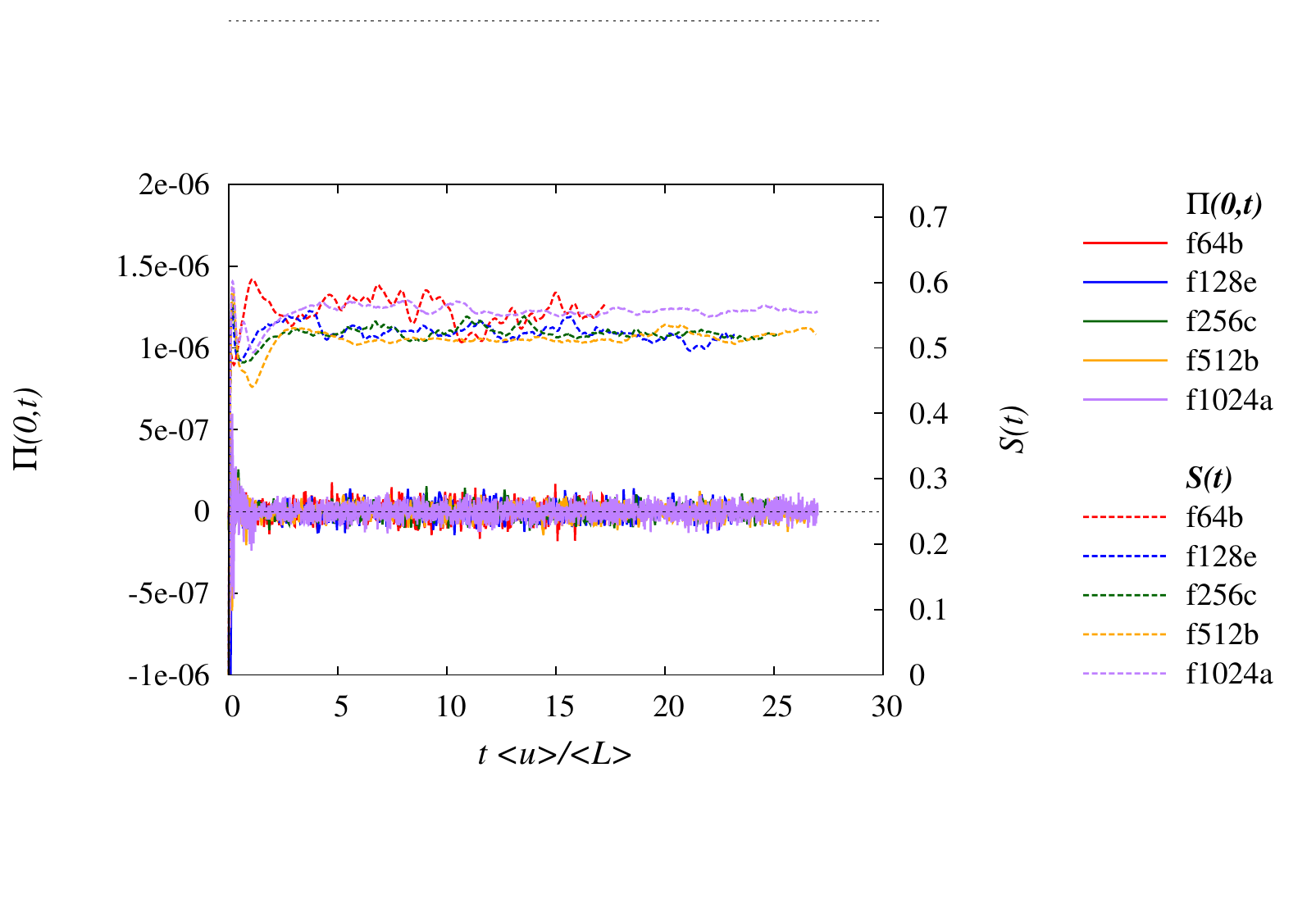}
  }
 \end{center}
 \caption{Time variation of key parameters for forced turbulence. Parts
 (a) and (b) are scaled by their steady-state mean value.}
 \label{fig:forced_time_series}
\end{figure}

Table \ref{tbl:summary_stats} summarises
the mean values of the most important statistical quantities in our simulations. Figure \ref{fig:forced_time_series} shows the evolution of some of these key parameters as
the simulation progresses from its Gaussian initial condition to the steady
state. The quantities have been scaled by their ensemble-averaged mean value. Note that each sub-figure shows two variables, one of which is plotted using continuous lines and the other using dotted lines. The left hand ordinate shows values for the continuous lines, whereas the right hand ordinate is for the dotted lines. The key for both types of line may be found to the right hand side of the figure.
 
As can be seen, after $t \sim 10
\langle L \rangle/\langle u \rangle$ most simulations have settled to their
evolved state. The figures also highlight the fact that stationarity is a
statistical concept. Fluctuations around the mean are expected, and
present in the system, but they should vanish under further (time or ensemble) averaging. An analogy can be drawn with the canonical ensemble in statistical mechanics: the turbulent system fluctuates about mean values determined by the rate at which the system is driven by the energy input.

The integral over the transfer spectrum, $\Pi(0,t)$, is shown in figure \ref{fig:forced_time_series}(c)
and can be seen to fluctuate around zero. The time-averaged values,
given in table \ref{tbl:summary_stats}, show $\Pi(0)$ to be consistently of
order $10^{-8}$ or smaller, indicating that the non-linear term is
conserving energy. We shall expand on the properties of the statistical ensemble in the next section.

\begin{table}[tbp!]
\begin{center}
\begin{tabular}{r|llllllllllc}
ID & $R_L$ & $R_\lambda$ & $u$ & $L$ & $\lambda$ & $\varepsilon$ & $\varepsilon_T$ & $S$ & $k_{\textrm{diss}}$ & $k_d$ & $\Pi(0)\ (\times 10^{-9})$ \\
\hline
\hline
\texttt{f64d}   & 10.6   & 8.70  & 0.441 & 2.163 & 1.777 & 0.083 & 0.026 & 0.566 & 5   & 3   & 62.9  \\
\texttt{f64c}   & 12.8   & 9.91  & 0.440 & 2.041 & 1.578 & 0.081 & 0.031 & 0.615 & 6   & 4   & 5.20  \\
\texttt{f64a}   & 19.0   & 13.9  & 0.485 & 1.956 & 1.435 & 0.086 & 0.037 & 0.583 & 7   & 5   & -42.9 \\
\texttt{f64b}   & 39.5   & 24.7  & 0.523 & 1.512 & 0.943 & 0.092 & 0.060 & 0.554 & 13  & 10  & -64.0 \\
\texttt{f128a}  & 82.7   & 42.5  & 0.581 & 1.442 & 0.733 & 0.094 & 0.079 & 0.540 & 22  & 18  & -73.9 \\
\texttt{f128b}  & 88.2   & 44.0  & 0.578 & 1.374 & 0.686 & 0.096 & 0.083 & 0.533 & 24  & 19  & 60.1  \\
\texttt{f128c}  & 101.4  & 48.0  & 0.586 & 1.383 & 0.655 & 0.096 & 0.084 & 0.535 & 26  & 21  & -46.5 \\
\texttt{f128d}  & 105.7  & 49.6  & 0.579 & 1.279 & 0.600 & 0.098 & 0.088 & 0.531 & 29  & 23  & 20.7  \\
\texttt{f128e}  & 158.6  & 64.2  & 0.607 & 1.307 & 0.529 & 0.099 & 0.092 & 0.529 & 38  & 30  & -5.10 \\
\texttt{f256a}  & 284.6  & 89.3  & 0.600 & 1.185 & 0.372 & 0.098 & 0.095 & 0.522 & 64  & 50  & -58.1 \\
\texttt{f256b}  & 360.1  & 101.3 & 0.607 & 1.187 & 0.334 & 0.099 & 0.096 & 0.521 & 76  & 59  & 37.5  \\
\texttt{f256c}  & 432.6  & 113.3 & 0.626 & 1.243 & 0.326 & 0.100 & 0.099 & 0.525 & 80  & 65  & -69.7 \\
\texttt{f512a}  & 1026   & 176.9 & 0.626 & 1.181 & 0.204 & 0.102 & 0.100 & 0.537 & 162 & 129 & 11.8  \\
\texttt{f512b}  & 1373   & 203.7 & 0.608 & 1.129 & 0.167 & 0.099 & 0.098 & 0.518 & 168 & 168 & -30.9 \\
\texttt{f512c}  & 81.5   & 41.8  & 0.581 & 1.403 & 0.720 & 0.097 & 0.082 & 0.535 & 22  & 18  & -26.6 \\
\texttt{f512d}  & 146.5  & 60.8  & 0.589 & 1.243 & 0.516 & 0.098 & 0.093 & 0.525 & 38  & 30  & 38.0  \\
\texttt{f512e}  & 287.8  & 89.4  & 0.605 & 1.189 & 0.369 & 0.101 & 0.096 & 0.525 & 65  & 51  & -75.7 \\
\texttt{f512f}  & 436.3  & 113.0 & 0.620 & 1.267 & 0.328 & 0.096 & 0.096 & 0.535 & 83  & 64  & 22.1  \\
\texttt{f512g}  & 785.2  & 153.4 & 0.626 & 1.255 & 0.245 & 0.098 & 0.095 & 0.541 & 132 & 99  & 70.9  \\
\texttt{f1024a} & 2415   & 276.2 & 0.626 & 1.158 & 0.132 & 0.100 & 0.100 & 0.557 & 323 & 247 & -4.40 \\
\texttt{f1024b} & 3535   & 335.2 & 0.626 & 1.130 & 0.107 & 0.102 & 0.102 & 0.541 & 337 & 337 & -34.3 \\ 
\end{tabular}
\caption{Summary of the mean statistics for the simulations.}
\label{tbl:summary_stats}
\end{center}
\end{table}

\clearpage

\section{Properties of the ensemble}

Shell-averaging was used to obtain time-varying
statistics such as the fluctuation of total energy or the dissipation rate.
 These quantities were presented as time
series. For stationary turbulence, once it reaches an evolved state, rather than
run multiple simulations, an ensemble can be generated by sampling the
field at various times. If the sample time between realisations is longer
than the typical correlation time scales of the system, we can consider the
times to be uncorrelated realisations of the flow. From this new ensemble,
we can calculate a single mean value for each of various parameters of the
stationary flow, along with their associated error.

First, we must discard the transient data collected while our system evolved
from its initial condition into a stationary solution of the Navier-Stokes
equation. Typically, this takes around 10 eddy turnover times. The remaining
data is then sampled every $\Delta t$ and used to calculated a mean value.
Here, $\Delta t$ should be of the order of $\tau = L/u$, the eddy turnover time, and in practice we found that $\Delta t = \tau/2$ was sufficient. For the simulations in this
work, we collected data for at least 15$\tau$ after the transition period.
The ensemble-averaged value for the parameter $A$ was then calculated as
\begin{equation}
 \langle A \rangle = \frac{1}{T} \sum_{t_i \in \mathbb{T}} A(t_i) \ ,
\end{equation}
where T is the number of realisations in our ensemble, $\mathbb{T}$. 
An estimate of the error is given by the standard
deviation,
\begin{equation}
 \sigma_A^2 = \langle A^2 \rangle - \langle A \rangle^2 \ ,
\end{equation}
although we occasionally refer to the \emph{standard error} on the mean,
denoted $\hat{\sigma}$, where
\begin{equation}
 \hat{\sigma} = \frac{\sigma}{\sqrt{T}} \ .
\end{equation}

In order to establish the reliability of our code, we compared our results for certain key parameters with representative values in the field, as described in the following subsections.

\subsection{The compensated energy spectrum}
\begin{figure}[tb!]
 \centering
  \includegraphics[width=0.8\textwidth]{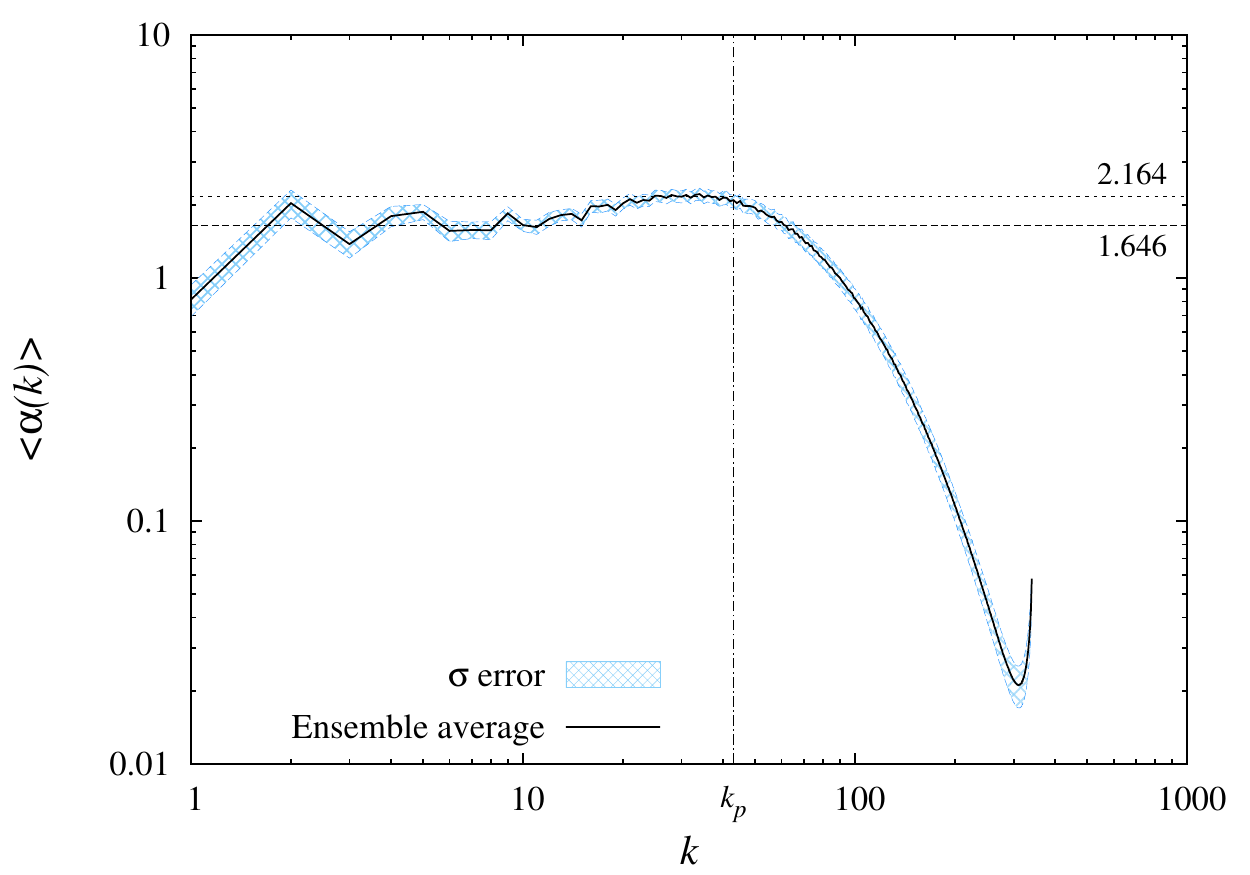}
 \caption{Ensemble-averaged compensated energy spectrum for $R_\lambda \sim
 280$. (-- -- --) Averaged value of $\alpha$. ($\cdots$)
 Anomalous plateau. (-- $\cdot$ --) $k_p \sim 43$.}
 \label{fig:kol_const_E}
\end{figure}

Rearranging the Kolmogorov (K41) energy spectrum in terms of a
wavenumber dependent $\alpha(k)$ gives the \emph{compensated} energy
spectrum,
\begin{equation}
 \alpha(k) = \varepsilon^{-2/3} k^{5/3} E(k) \ .
\end{equation}
Regions in
which this spectrum is flat thus take the Kolmogorov form, $k^{-5/3}$, with
$\alpha = \textrm{constant}$. Figure \ref{fig:kol_const_E} shows the
compensated energy spectrum for an $N = 1024$ simulation. The spectrum has
been ensemble-averaged with $\Delta t = \tau$, allowing us to plot an estimate
of the error.

As noted by Yeung and Zhou \cite{Yeung97}, there appear to be two
plateaus: one at lower $k$ and one at medium $k$. In the paper, the authors
show how the location of the inertial range has been misidentified in
many numerical simulations, causing the value of $\alpha$ to be
overestimated. They present arguments for the plateau at lower $k$
corresponding to the actual inertial range. This is based on the peak of the
dissipation spectrum coinciding with the higher plateau, hence it cannot
correspond to inertial behaviour. This is also observed in our data, with
the peak of the dissipation spectrum at $k_p \sim 43$, indicated in figure
\ref{fig:kol_const_E}. Ishihara, Gotoh and Kaneda \cite{Ishihara09}
also provide a discussion of this mis-identification.

\begin{figure}[tb!]
 \centering
  \includegraphics[width=0.8\textwidth]{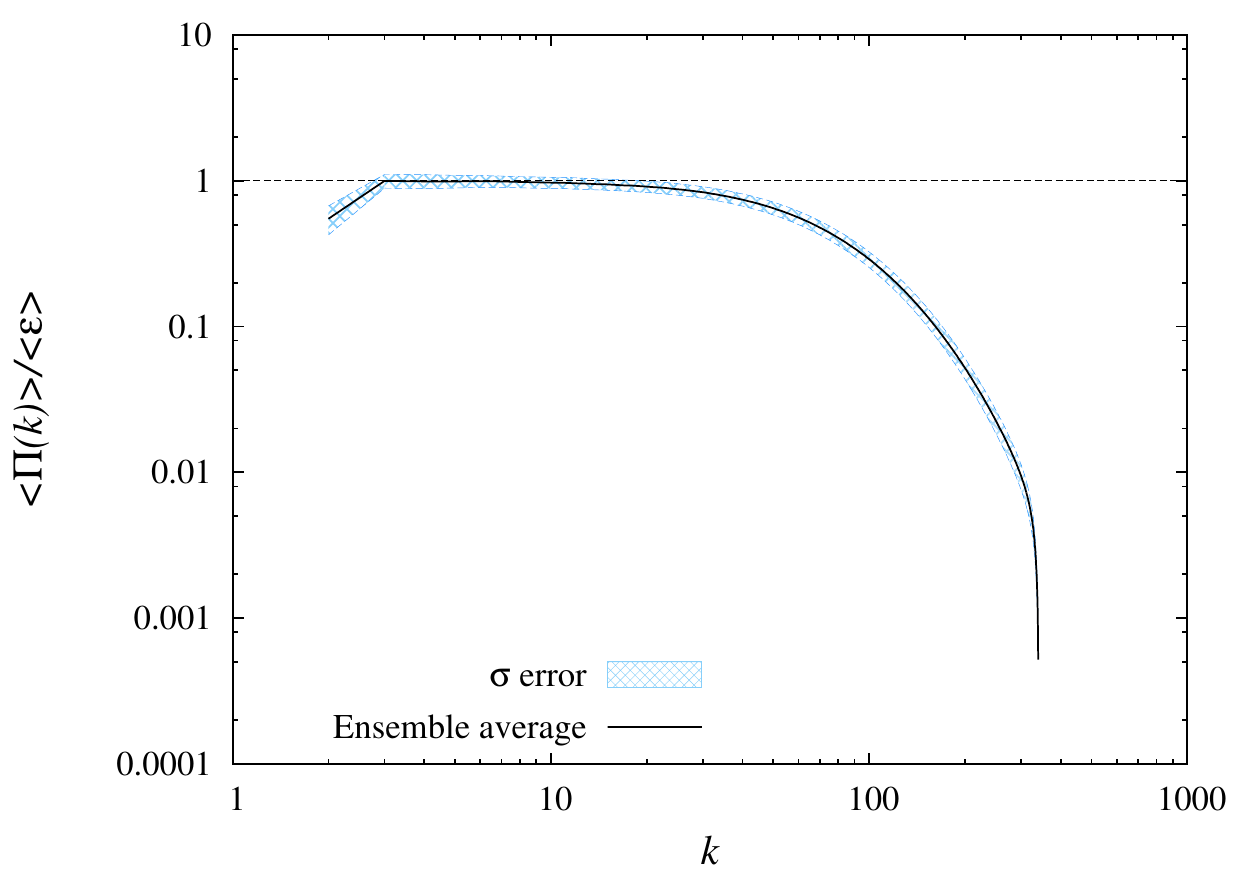}
 \caption{Ensemble-averaged scaled transport power spectrum for $R_\lambda \sim 280$.}
 \label{fig:kol_const_P}
\end{figure}

To make an estimate for the value of the constant $\alpha$, we resort to the
scaled transport power spectrum. As discussed earlier, in the inertial subrange of wavenumbers energy is
transferred at the dissipation rate, such that the flux through a wavenumber $k$
satisfies
\begin{equation}
 \Pi(k,t) = \varepsilon(t) \ ,
\end{equation}
making $\langle \Pi(k) \rangle/\varepsilon$ a simple test for an inertial
range. In figure \ref{fig:kol_const_P}, this can be seen to be unity for the
range $3 \leq k \leq 7$, corresponding to the lower $k$ plateau in figure
\ref{fig:kol_const_E}. To obtain a mean value for this plateau, we averaged
over the range to find the value
\begin{equation}
 \alpha = 1.646 \pm 0.144 \ .
\end{equation}
This value is highlighted in figure \ref{fig:kol_const_E} by the dashed
line, along with the value corresponding to the `anomalous' plateau of 2.164
(dotted line).

Ishihara \etal\ \cite{Ishihara09} found $\alpha = $ 1.5 -- 1.7 in
their high-$R_\lambda$ simulations, placing our result within their range.
In fact, studying the data found in Gotoh and Fukayama
\cite{Gotoh01}, one finds the value for their most similar Reynolds
number, $R_\lambda = 284$, to be 1.64, in good agreement with the
above. They quote an average value of $\alpha = 1.65 \pm 0.05$, and our own
result agrees with this within one error unit. Yeung and Zhou \cite{Yeung97}
found a value of 1.62 for $R_\lambda = 240$. Note that the Kolmogorov
constant can be measured from one- or three-dimensional energy spectra using
the relation $\alpha = (55/18) C_1$, where $C_1$ is measured from
one-dimensional spectra \cite{Yeung97}. Comparison can then also be
made to the experimental value obtained by Sreenivasan
\cite{Sreenivasan95} of $C_1 = 0.53 \pm 0.055$ which gives $\alpha =
1.62 \pm 0.17$. Mydlarski and Warhaft \cite{Mydlarski96} found the
experimental value $C_1 = 0.51$, giving $\alpha = 1.56$.
Further values for comparison obtained using direct mumerical simulation and large-eddy simulation can be found in
\cite{Yeung97,Gotoh01,Wang96}. 


\subsection{Longitudinal velocity derivative skewness}\label{subsec:skewness}
The skewness was computed in both real and Fourier space to obtain the 
values
\begin{equation}
 S_x = 0.551 \pm 0.015 \qquad\qquad \textrm{and} \qquad\qquad S_k = 0.557 \ ,
\end{equation}
respectively.
The Fourier-space result has been calculated using the ensemble-averaged
transfer spectrum, and as such it is difficult to associate an error with
it. However, agreement with the real-space result is good.

This can be compared to other stationary simulations such as Ishihara \etal\
\cite{Ishihara09}, who obtained $S \sim 0.5$, or Machiels
\cite{Machiels97a} who quoted a result of $S = 0.51$ for $R_\lambda
\simeq 190$. Vincent and Meneguzzi \cite{Vincent91} found a value of
$S = 0.5$ for $R_\lambda \sim 150$, which is the same as Kerr
\cite{Kerr85} for $R_\lambda < 80$. Gotoh, Fukayama and Nakano
\cite{Gotoh02} performed a series of simulations on $512^3$ and
$1024^3$ grids. For $R_\lambda = 284$, the closest Reynolds number to that
used here, they found $S = 0.531$. The average value of their $R_\lambda =
284$ and 381 runs gives $S = 0.553$. Jim\'enez, Wray, Saffman and Rogallo
\cite{Jimenez93} obtained $S = 0.525$ for $R_\lambda = 168.1$, while Wang,
Chen, Brasseur and Wyngaard \cite{Wang96} found a value of $S =
0.545$ for the largest forced run with $R_\lambda = 195$. Sreenivasan and
Antonia \cite{Sreenivasan97} comment on skewness increasing
monotonically with Reynolds number and present a collection of data from DNS
and experiment to support this. This can also be observed in
\cite{Ishihara07}.

\subsection{Dissipation-scaled energy spectrum}
\begin{figure}[bth!]
 \centering
  \includegraphics[width=0.7\textwidth]{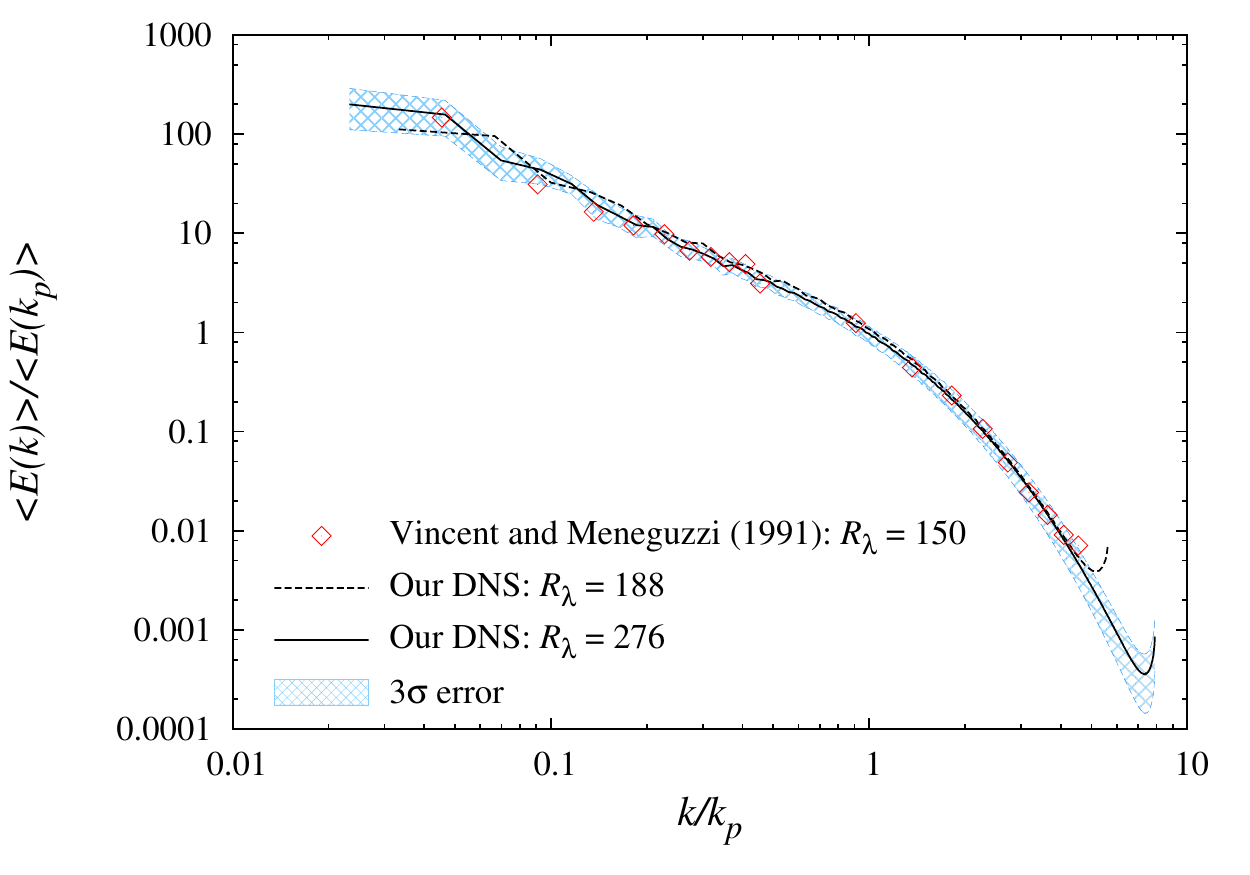}
 \caption{Comparison of dissipation-scaled energy spectra.}
 \label{fig:scaled_E}
\end{figure}

She, Chen, Doolen, Kraichnan and Orszag \cite{She93a} found that
energy spectra taken at various Reynolds numbers collapse when scaled on the
peak of the dissipation spectrum; that is, $k/k_p$ and $E(k)/E(k_p)$. The
authors presented the collapse of DNS for Reynolds numbers $R_\lambda \sim 70$
to 200, along with experimental data. In figure \ref{fig:scaled_E} we plot
our own DNS results for two Reynolds numbers, along with data points from
Vincent and Meneguzzi \cite{Vincent91} for $R_\lambda = 150$ for
comparison. Note that the points have been extracted by hand from their
figure. The data is seen to collapse well. The error shown is that for
$R_\lambda = 276$.

\subsection{Resolution of the small scales}

\begin{figure}[tb]
 \begin{center}
  \includegraphics[width=0.65\textwidth,trim=0 90px 0 80px,clip]{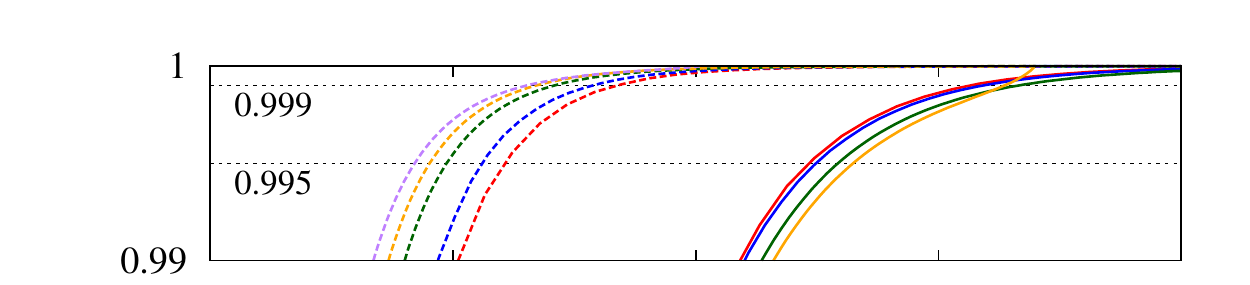}
  \includegraphics[width=0.65\textwidth,trim=0 0 0 10px,clip]{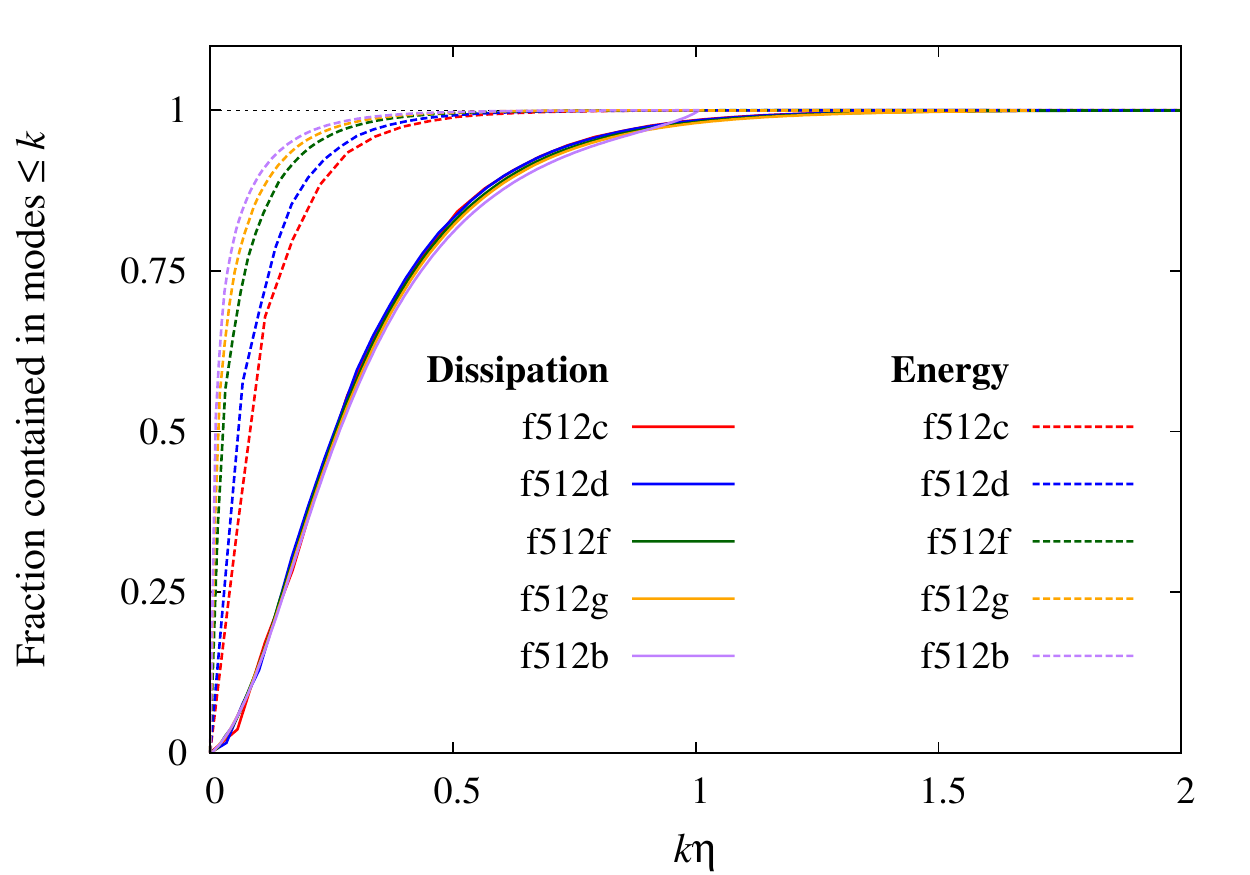}
 \end{center}
 \caption{Location of energy and dissipation rate in highly resolved
 simulations. To retain 99.5\% of dissipative dynamics, one must satisfy
 $\kmax\eta > 1.25$ while 99.9\% requires $\kmax\eta > 1.7$.}
 \label{fig:forced_resolved}
\end{figure}

A general rule for DNS is that one must satisfy $\kmax \eta > 1$, with
$\kmax \eta = 1$ being regarded as only partially resolved. It has been suggested that it
is actually necessary to satisfy $\kmax \eta > 1.5$ to capture the relevant
dynamics \cite{McComb01a}. Therefore, a series of `highly resolved' runs was performed, for
which $\kmax \eta > 1.5$, see runs \frun{f512c-g}. This allowed
us to explore the distribution of energy and dissipation without artefacts
caused due to the system being under-resolved. Figure
\ref{fig:forced_resolved} shows our results. We plot the total energy and
dissipation rate which has been taken into account for up to mode $k$ and normalised by their actual totals,
thus
\begin{equation}
 \frac{1}{\langle E \rangle} \int_{k_{\textrm{min}}}^k dk'\ E(k') \
 \quad\qquad\textrm{and}\quad\qquad \frac{1}{\langle \varepsilon \rangle}
 \int_{k_{\textrm{min}}}^k dk'\ 2\nu_0 k'^2\ E(k') \ .
\end{equation}
We also plot the partially resolved run \frun{f512b} for comparison, which
can be seen to kick up unphysically as it reaches $k\eta = 1$. This also
occurs for run \frun{f512g} as it reaches its cutoff $\kmax\eta = 1.7$. The
energy really is contained in much lower wavenumbers (larger length-scales)
than the dissipative loss. By $k\eta \sim 0.5$ we have already accounted for
virtually all the energy, but only around 75\% of the dissipation rate (see also \cite{McComb01a}). The
additional graphic in figure \ref{fig:forced_resolved} shows a close up of
the final percentile. This highlights two key points: First, if we want to
include 99.5\% of dissipative dynamics, we must use $\kmax\eta \simeq 1.25$.
Whereas, to include 99.9\% requires $\kmax\eta \simeq 1.7$. Second, as the
Reynolds number is increased, energy is contained in progressively lower
$k\eta$ while dissipation is pushed to higher $k\eta$.

\subsection{Isotropy}

Since we wish to simulate isotropic turbulence, it is important to
ensure that the velocity field does indeed satisfy this property. This was
done using the method given in Young \cite{Young99}.

\begin{figure}[tbh!]
 \centering
  \includegraphics[width=0.75\textwidth]{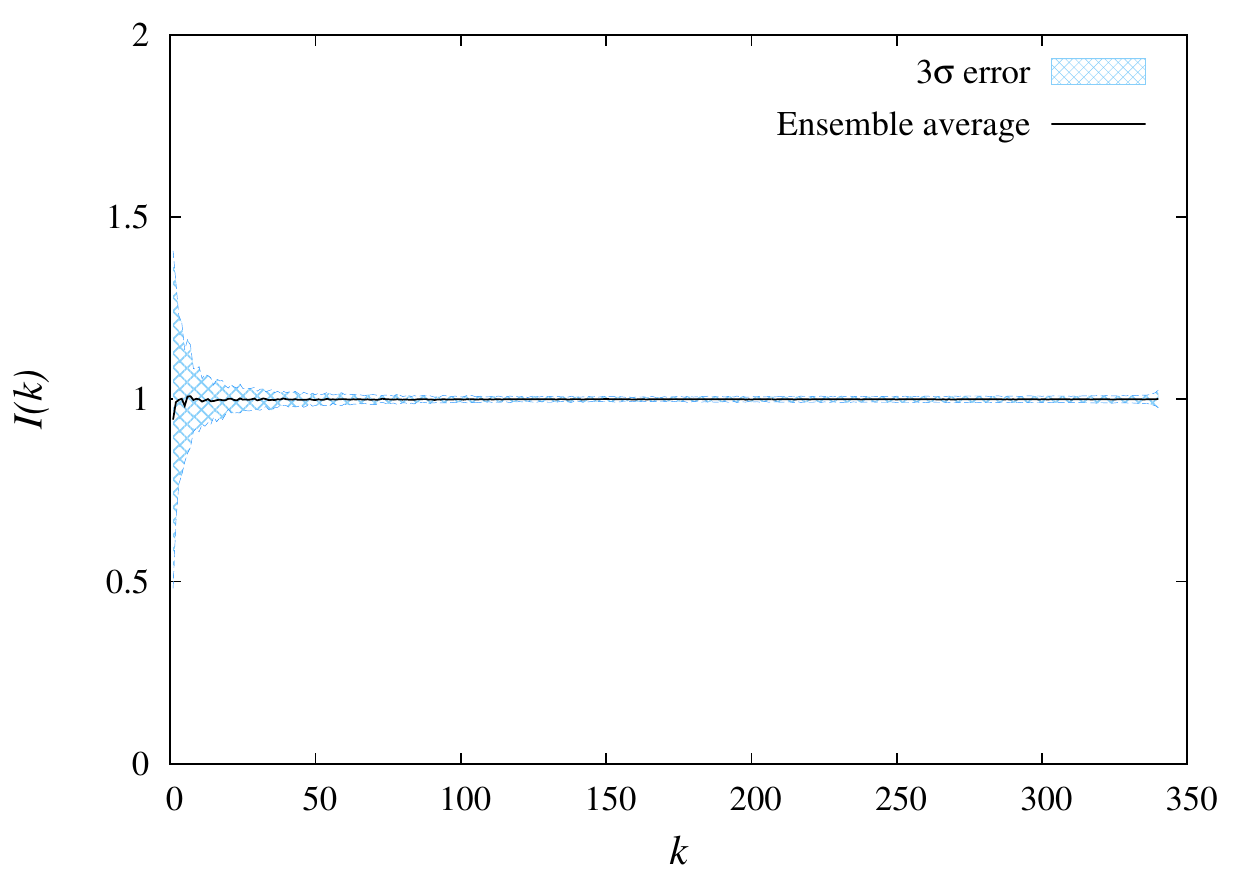}
 \caption{Ensemble averaged isotropy spectrum for an $N = 1024$ lattice.}
 \label{fig:isotropy_spec}
\end{figure}

A random unit vector $\vec{z}(\vec{k})$ which is not parallel to $\vec{k}$
(that is, it satisfies $\vec{z}(\vec{k}) \cdot\hat{\vec{k}} \neq 1$) is
chosen for all wavevectors, and from it we define two mutually orthogonal
unit vectors
\begin{align}
 \vec{e}_1(\vec{k}) = \frac{\vec{k}\times\vec{z}(\vec{k})}{\lvert
 \vec{k}\times\vec{z}(\vec{k}) \rvert} \ , \qquad\qquad \vec{e}_2(\vec{k}) =
 \frac{\vec{k}\times\vec{e}_1(\vec{k})}{\lvert
 \vec{k}\times\vec{e}_1(\vec{k}) \rvert} \ .
\end{align}
These are used to compute the average energy in these two directions,
\begin{equation}
 I_j(k,t) = \left\langle \lvert \vec{e}_j(\vec{k}) \cdot \vec{u}(\vec{k},t) \rvert^2 \right\rangle\ , \qquad j = 1,2 \ ,
\end{equation}
which should be the same for isotropic turbulence. A measure of the degree
of isotropy is, therefore, the ratio
\begin{equation}
 I(k,t) = \sqrt{\frac{I_1(k,t)}{I_2(k,t)}} \ .
\end{equation}
As seen plotted in figure \ref{fig:isotropy_spec}, while individual
realisations fluctuate, the ensemble average is close to 1 for all values of
$k$. The increase in the deviation from unity as one moves towards low $k$
is due to the lower resolution of these shells. Since they contain fewer points
the statistics are not as good.

A representative value can be obtained by averaging over all of Fourier
space. Values for a variety of simulation sizes can be found in table
\ref{tbl:isotropy} and are quite satisfactory, allowing us to conclude that
there is no significant deviation from isotropy in our simulations. The
uncertainty on the scale of the mean, $\sigma / \langle I \rangle$,
decreases as $N$ is increased, since the large $k$ modes are more isotropic
than the low $k$ modes and we are including more of them in the simulation.
\begin{table}[H]
\begin{center}
\begin{tabular}{r||cccc}
$N$ & 128 & 256 & 512 & 1024 \\
\hline
 $\langle I \rangle \pm \sigma$ & $1.002 \pm 0.009$ & $1.005 \pm 0.008$ & $0.9979 \pm 0.0077$ & $1.0002 \pm 0.0034$
\end{tabular}
\vspace{1em}
\caption{Representative values for the total isotropy for various lattice
sizes.}
\label{tbl:isotropy}
\end{center}
\end{table}

\section{Results for scale-invariance and the Kolmogorov spectrum}

\begin{figure}[tb]
 \begin{center}
  \subfigure[Energy spectra scaled with $\eta$]{
   \label{sfig:forced_spec_Ekol}
   \includegraphics[width=0.475\textwidth,trim=0 0 10px 0,clip]{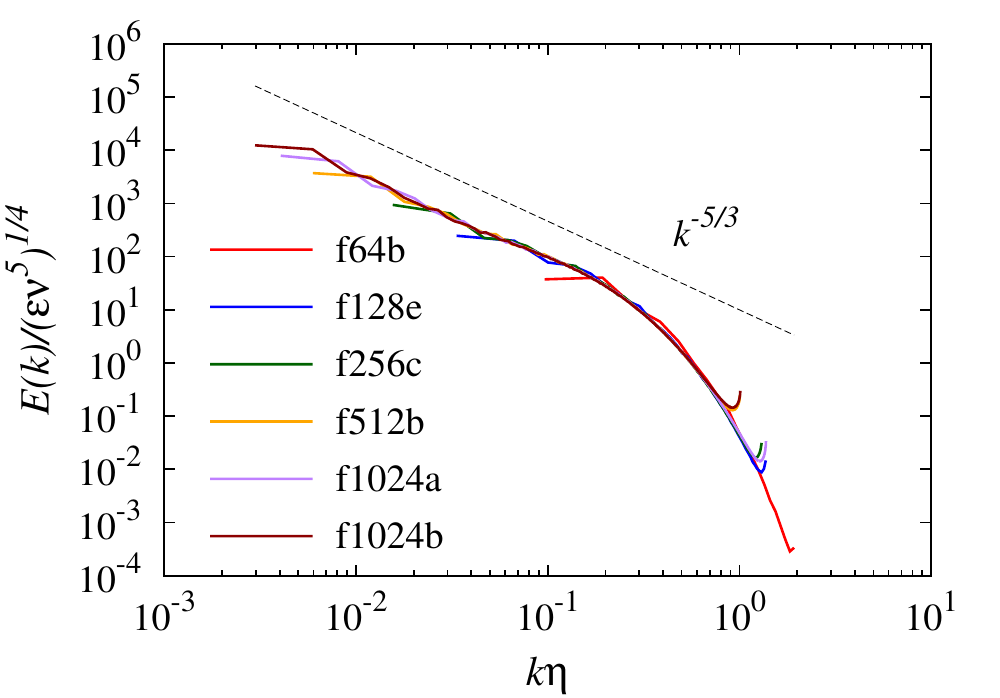}
  }
  \subfigure[Energy spectra scaled with the integral scale]{
   \label{sfig:forced_spec_EL}
   \includegraphics[width=0.475\textwidth,trim=0 0 10px 0,clip]{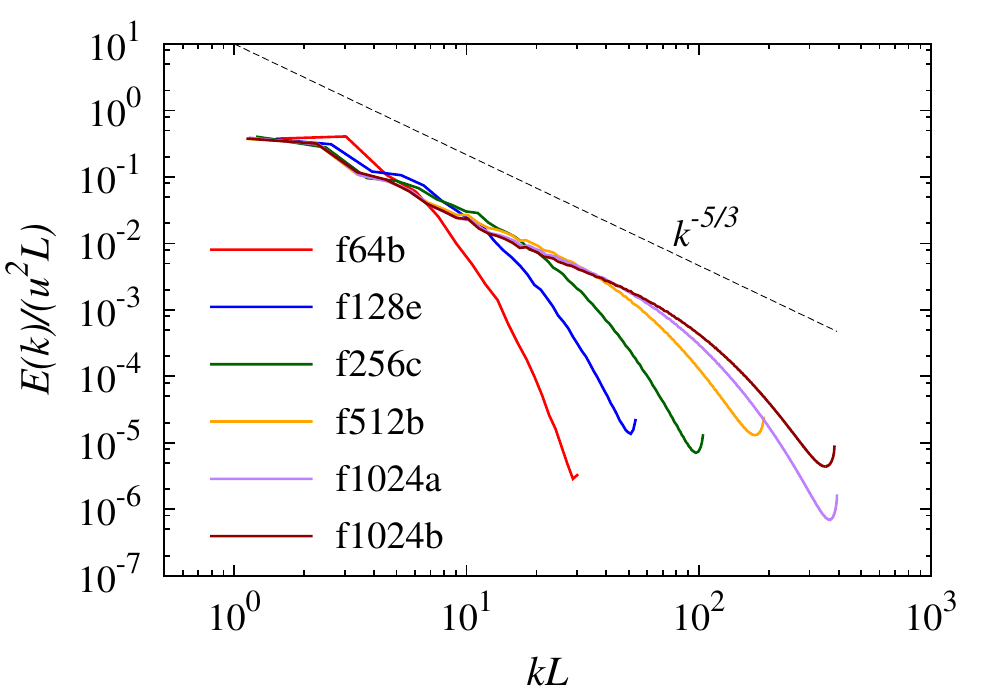}
  }
  \subfigure[The compensated energy spectrum]{
   \label{sfig:forced_spec_Ecomp}
   \includegraphics[width=0.475\textwidth,trim=0 0 10px 0,clip]{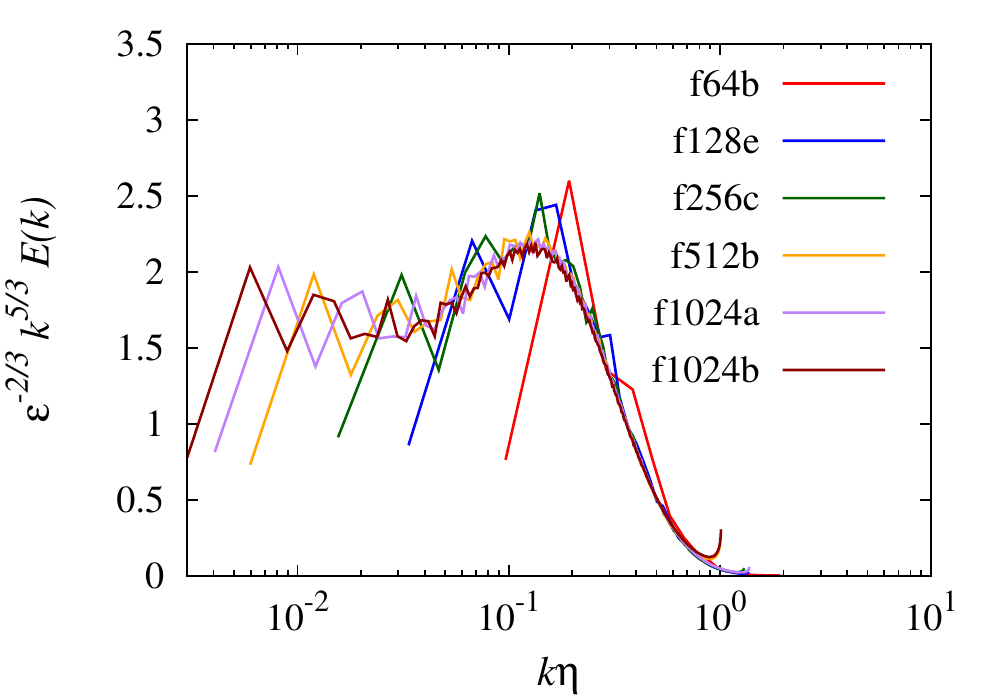}
  }
  \subfigure[The transport power spectrum]{
   \label{sfig:forced_spec_P}
   \includegraphics[width=0.475\textwidth,trim=0 0 10px 0,clip]{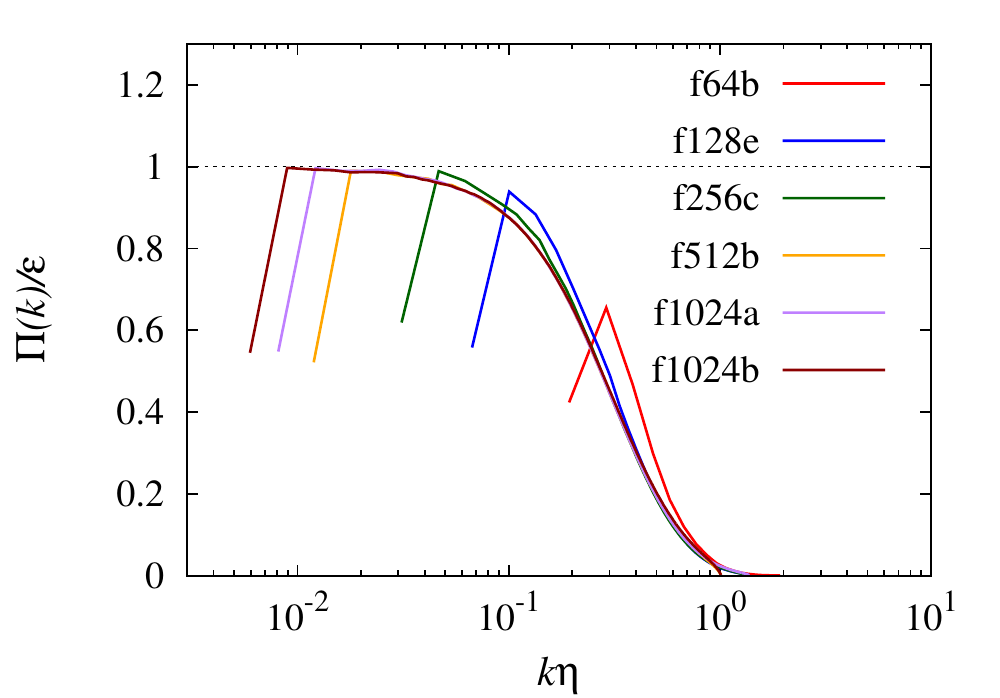}
  }
  \subfigure[Transfer spectra scaled with $\eta$]{
   \label{sfig:forced_spec_T}
   \includegraphics[width=0.475\textwidth,trim=0 0 10px 0,clip]{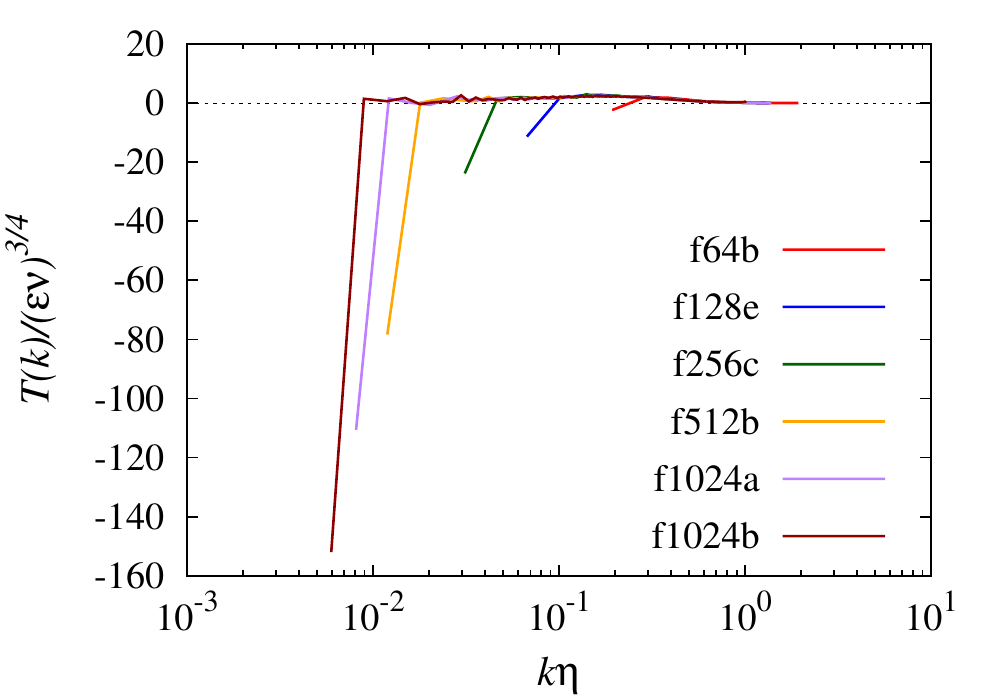}
  }
  \subfigure[Transfer spectra scaled with the integral scale]{
   \label{sfig:forced_spec_Tu3}
   \includegraphics[width=0.475\textwidth,trim=0 0 10px 0,clip]{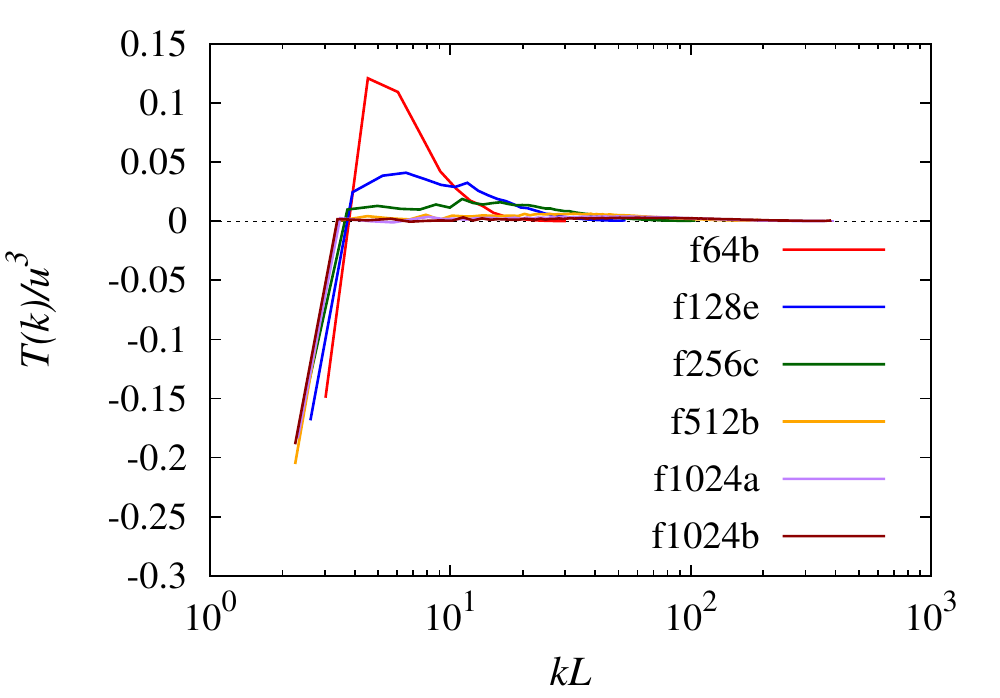}
  }
 \end{center}
 \caption{Energy, transfer and transport power spectra for the range of Reynolds numbers $24.7 \leq \Rl \leq 335.2$.}
 \label{fig:forced_spec}
\end{figure}

The energy spectra taken for a selection of runs are presented in figure
\ref{fig:forced_spec}. Figure \ref{sfig:forced_spec_Ekol} is scaled using
the Kolmogorov length-scale and the appropriate combination of dissipation
range variables $\varepsilon$ and $\nu_0$.
The collapse of data for all runs is very good. The slope of
the data can be seen to be slightly shallower than K41 for a period, hinting
at $-5/3 + \mu$ with $\mu > 0$. This is not in agreement with Kaneda,
Ishihara and Yokokawa, Itakura and Uno \cite{Kaneda03} who found $\mu
\simeq -0.1$ by considering the compensated energy spectrum for the high
Reynolds number simulations performed on the Earth Simulator. This
correction could be Reynolds number dependent and vanish as $Re \to \infty$,
making it a finite Reynolds number effect. An analysis of the Reynolds
number variation of this exponent would help determine whether K41 is an
asymptotic theory or not. Unfortunately, the data obtained here, presented
in figure \ref{sfig:forced_spec_Ecomp}, did not offer a large enough range
to measure this exponent properly. The compensated energy spectra should be
compared to those obtained by Ishihara, Gotoh and Kaneda
\cite{Ishihara09}. Figure \ref{sfig:forced_spec_EL} shows
the energy spectrum scaled using the integral scale, for comparison. The
slope here also looks to be shallower than $k^{-5/3}$.

The scaled transfer spectra are shown in figures \ref{sfig:forced_spec_T}
and \ref{sfig:forced_spec_Tu3}.

\subsection{The Kolmogorov prefactor}
Figure \ref{sfig:forced_spec_Ecomp} shows the compensated energy spectrum.
 Note the pronounced curl up of
the tail for the partially resolved run \frun{f512b}. This is also the case
for \frun{f1024b} (not plotted). The figure shows how a plateau could be
  identified for a low $k\eta$ range for runs with $R_\lambda > 113$. This
  plateau was used to find a value for the Kolmogorov constant, $\alpha$,
  and can be seen to lie around 1.6 -- 1.7. The values measured for the four
  runs for which a plateau could be found are given in table
  \ref{tbl:kol_const}, using a simple average and an error-weighted fit. The
  transport power spectra shown in figure \ref{sfig:forced_spec_P} were used
  to define the fit region. Note that the peak associated with $k\eta \sim
  0.1-0.2$ is not associated with an inertial range. For runs with lower Reynolds number, a plateau
  cannot be identified.

\begin{table}[tb!]
 \begin{center}
  \begin{tabular}{r|cccc}
   ID & $R_\lambda$ & $\alpha$ & $\alpha'$ \\
   \hline\hline
   \frun{f512a}  & 176.9 & $1.663 \pm 0.218$ & $1.632 \pm 0.172$ \\
   \frun{f512b}  & 203.7 & $1.625 \pm 0.166$ & $1.621 \pm 0.165$ \\
   \frun{f1024a} & 276.2 & $1.636 \pm 0.177$ & $1.646 \pm 0.144$ \\
   \frun{f1024b} & 335.2 & $1.643 \pm 0.136$ & $1.625 \pm 0.119$
  \end{tabular}
 \end{center}
 \caption{Measured values of the Kolmogorov constant, as found by identifying
 the range of wavenumbers where $\Pi(k) \simeq \varepsilon$ and averaging
 over those points. The value $\alpha$ is obtained by a simple average over
 the range, whereas $\alpha'$ is calculated using an error-weighted fit on
 the range.}
 \label{tbl:kol_const}
\end{table}

\subsection{Reynolds number dependence of statistics}\label{subsec:forced_Rdep}
\begin{figure}[tb]
 \begin{center}
  \subfigure[Onset of an inertial subrange]{
   \label{sfig:forced_Rvar_eps-pi}
   \includegraphics[width=0.59\textwidth]{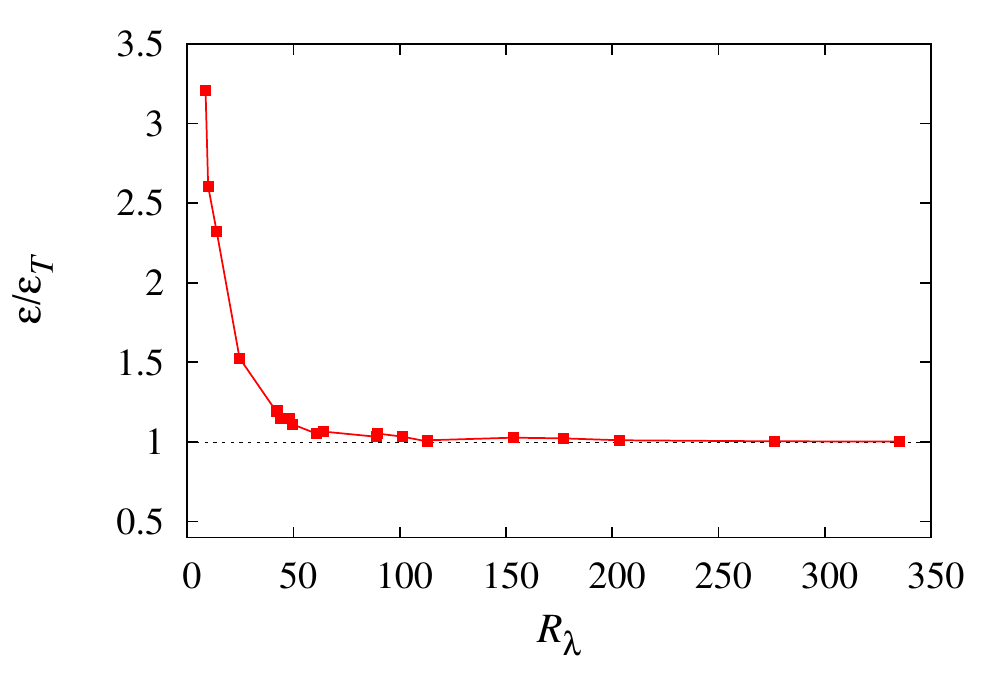}
  }
  \subfigure[Variation of a selection of parameters]{
   \label{sfig:forced_Rvar_params}
   \includegraphics[width=0.59\textwidth]{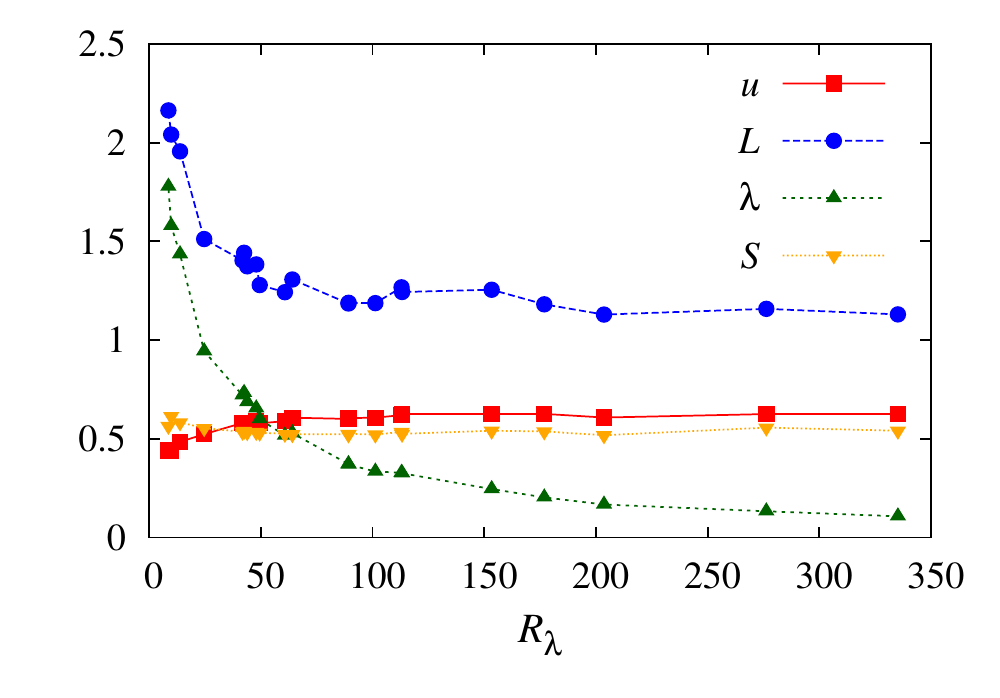}
  }
  \subfigure[The Taylor surrogate, $u^3/L$]{
   \label{sfig:forced_Rvar_eps-pi-surr}
   \includegraphics[width=0.59\textwidth]{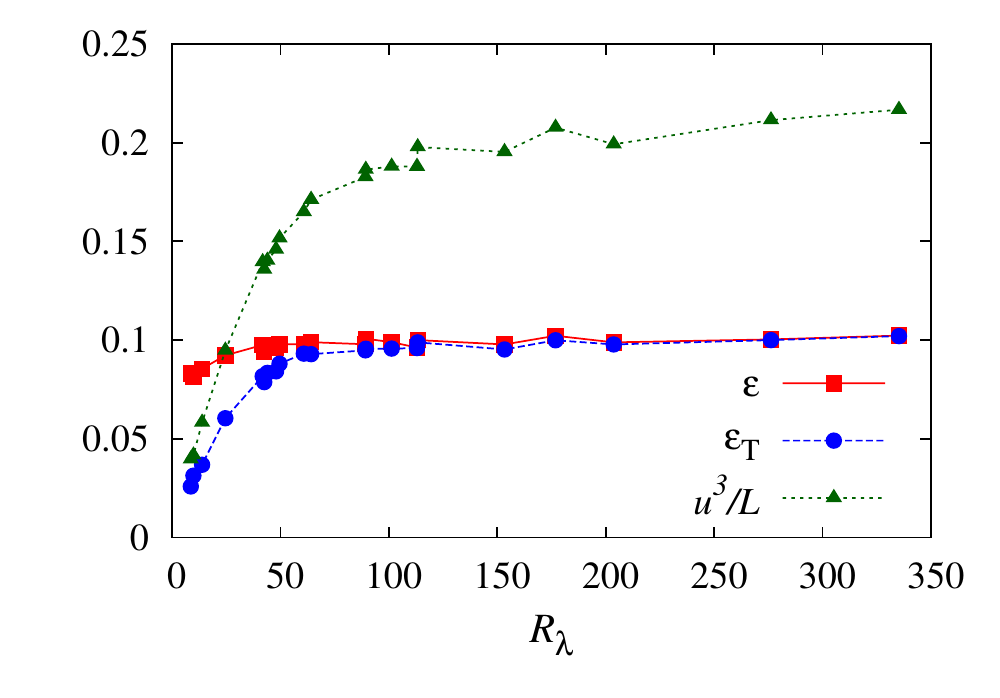}
  }
 \end{center}
 \caption{Variation of some key parameters with increasing Reynolds number.}
 \label{fig:forced_Rvar}
\end{figure}

We now look at how the values of some important parameters vary with
increasing Reynolds number. As mentioned in the previous section,
an indication of the presence of a inertial subrange in a stationary system
is a range of wavenumbers for which the transport power, or flux of energy
through that wavenumber, is equal to the dissipation rate, $\Pi(k) =
\varepsilon$. When this is the case, we find that the maximum transfer rate
$\varepsilon_T = \max \Pi(k) = \varepsilon$. As such, a study of
$\varepsilon_T/\varepsilon$ will give unity for stationary systems in which
the integral and dissipation scales are sufficiently well separated that an
inertial subrange can form. The variation of this quantity with Reynolds
number is presented in figure \ref{sfig:forced_Rvar_eps-pi}. This is in contrast to decaying turbulence, where the
maximum transport can never quite reach the dissipation rate. Note that, for
Reynolds numbers $R_\lambda > 113$, we basically find $\varepsilon_T =
\varepsilon$, thus indicating the presence of an inertial subrange.

Figure \ref{sfig:forced_Rvar_params} shows the Reynolds number variation of
the steady state value of the rms velocity, integral and Taylor
length-scales, and the velocity derivative skewness. We see that the skewness
remains more or less constant, just above 0.5, for the range of Reynolds
numbers available. The length-scales are both seen to decrease as $Re$
increases. However, the integral scale looks like it may have reached a
plateau, whereas the same cannot be said for the Taylor microscale. The rms
velocity initially increases then appears to stay constant. We would expect
the rms velocity to increase as the Reynolds number increases since there
are more modes excited. This may still be the case, but as most of the
energy is located at low wavenumbers the increase is small.

The Taylor surrogate $u^3/L$ is
compared in figure \ref{sfig:forced_Rvar_eps-pi-surr} to the dissipation
rate and inertial flux, $\varepsilon_T$. We see that the
surrogate is better matched to the behaviour of the inertial flux than the
dissipation rate. This is in agreement with the findings of McComb \etal\
\cite{McComb10a}.

\clearpage

\section{Results illustrating the nature of internal intermittency}

Internal intermittency was originally identified in terms of the drop-outs in an anemometer signal (see the monograph by Batchelor \cite{Batchelor53}). But it has long been interpreted as a spottiness in the spatial distribution of quantities like the vorticity or the rate-of-strain. In practice, this means that the phenomenon is seen as an example of a coherent structure. Thus, for many years it has been studied, in the context of coherent structures, by flow visualisation, in both laboratory experiments and numerical experiments (i.e. DNS). Of course in this topic there are many different criteria for establishing the presence of significant events, so we begin with some brief comments on flow-visualisation techniques. Then we discuss the ones that we have used here, before concluding with our results for the behaviour of the intermittency under ensemble-averaging.

\subsection{Visualisation of coherent structures}\label{sec:structures}

Over the years, there have been many attempts to define a vortex in such a
way that they may be identified in flow visualisation, whether that be
experimental data from real flows or DNS data as studied here. The na\"ive
definition of a vortex as a region of high vorticity can be misleading, since
there is no particular value above which vorticity can be universally
regarded as being high. In fact, even in the absence of vortices there can
exist areas of high vorticity in parallel shear flows
\cite{Haller05}. This creates a difficulty in finding unambiguous
criteria which can isolate a unique vortex.

Jeong and Hussain \cite{Jeong95} summarised and compared a selection
of methods available. They highlighted that any criteria should be Galilean
invariant, and found that previous indicators of a vortex, such as
streamlines, isovorticity and minima in the local pressure are not suitable
for use in unsteady flow. Haller \cite{Haller05} provided a
comprehensive review of the definition of a vortex along with a variety of
identification techniques. Despite this, surfaces of isovorticity continue to be used for vortex
identification and can produce good results in the case of isotropic
turbulence.

In this section, we will compare the detection of coherent
structures in visualisations of our DNS data for isotropic turbulence using
isovorticity contours, the $Q$-criterion, and the magnitude of the rate of strain.
 
\subsubsection{Isovorticity}
Surfaces of isovorticity connect regions which have the same magnitude of
vorticity, $\vert\vec{\omega}\vert$. Since the core of a vortex is
associated with high vorticity, with the value progressively dropping as we
move away from the core, these surfaces form structures such as `worms' and
`sheets'. See, for example, the work of Jimenez \emph{et al.} \cite{Jimenez93} and Okamoto \emph{et al.} \cite{Okamoto07}.
From our results, structures identified in the plane $z = 0$ using vorticity are shown in
figures \ref{sfig:structures512_vort} and \ref{sfig:structures1024_vort}, for
two different Reynolds numbers, as part of a comparison with other
identification methods. As can be seen, the magnitude of vorticity shows a
large amount of structure in the plane, and there are several regions of
high vorticity that could be identified as being vortices. Three-dimensional
structures can be seen in figures \ref{sfig:3dstructures256_vort} and
\ref{sfig:3dstructures1024_vort} and show how the vorticity has organised
itself into an entanglement of tubes or `worms', as observed by many other
authors \cite{Ishihara09,Ishihara07,Jimenez93}. These
should be compared to the Gaussian initial condition shown in figure
\ref{sfig:256_Gaussian_vort}, which shows little in the way of organised
structure. Note that the plane with $z=0$ is through the centre of the box.

\subsubsection{The $Q$-criterion}
The $Q$-criterion was originally proposed by Hunt, Wray and Moin
\cite{Hunt88} and is based on the invariants of the deformation
tensor, $A$, whose elements are:
\begin{equation}
 a_{ij} = \frac{\partial u_i}{\partial x_j} \ .
\end{equation}
The eigenvalues, $\Lambda$, of this tensor are found by requiring that
\begin{equation}
 \det (A - \Lambda \unitM) = 0 \ ,
\end{equation}
which in three-dimensions leads to the third-order characteristic equation,
\begin{equation}
 \Lambda^3 - P\Lambda^2 + Q\Lambda - R = 0 \ ,
\end{equation}
with the coefficients:
\begin{align}
 P &= \tr(A); \\
 Q &= \tfrac{1}{2} \Big( \tr(A)^2 - \tr\big( A^2 \big) \Big) ;\\
 R &= \det A \ .
\end{align}
The coefficients are the principle invariants of $A$, since the
eigenvalues do not depend on the choice of basis vectors. We first note that:
\begin{equation}
 \tr A = \frac{\partial u_i}{\partial x_i} = 0,
\end{equation}
for an incompressible fluid, such as that considered here. Next, the
  deformation tensor can be decomposed into its symmetric and antisymmetric
  parts, thus:
\begin{equation}
 S_{ij} = \tfrac{1}{2}\Big( a_{ij} + a_{ji} \Big)
 \qquad\qquad\textrm{and}\qquad\qquad \Omega_{ij} = \tfrac{1}{2}\Big( a_{ij}
 - a_{ji} \Big)\ ,
\end{equation}
which may be recognised as the strain, and vorticity, tensors, respectively.
We can therefore evaluate the trace
\begin{align}
 \tr \big(A^2 \big) &= \tr\big( SS + S\Omega + \Omega S + \Omega\Omega \big), \nonumber \\
 &= \tr\big( SS \big) + \tr\big(S\Omega\big) + \tr\big(\Omega S\big) + \tr\big(\Omega\Omega \big), \nonumber \\
 &= \tr\big( SS^T \big) - \tr\big(S\Omega^T\big) + \tr\big(\Omega S^T\big) - \tr\big(\Omega\Omega^T \big) \ ,
\end{align}
where the last line used the symmetry of $S$ and $\Omega$. Since the trace
has the properties $\tr(AB) = \tr(BA)$ and $\tr(A^T) = \tr(A)$, the two
cross terms cancel, to leave:
\begin{equation}
 Q = \tfrac{1}{2} \Big( \Vert \Omega \Vert^2 - \Vert S \Vert^2 \Big) \ ,
\end{equation}
with the Euclidean matrix norm defined as $\Vert M \Vert^2 =
\tr\big(MM^T\big)$. For the antisymmetric component, we have
$\Vert\Omega\Vert^2 = \tfrac{1}{2} \vert\vec{\omega}\vert^2$, and the value
of $Q$ is calculated as
\begin{equation}
 Q = \tfrac{1}{2} \Big( \tfrac{1}{2}\omega^2 - \Vert S \Vert^2 \Big) \ .
\end{equation}
$Q$ represents the local balance between shear strain rate and vorticity
magnitude, and vanishes at a solid boundary (unlike
$\lvert\vec{\omega}\rvert$) \cite{Jeong95}. When $Q > 0$, the
implication is that the vorticity tensor (quantifying that amount of
rotation) is dominant over the strain-rate tensor (which is related to
dissipation), and so there is a vortex. Figure \ref{sfig:structures512_Q} shows
the $Q$-criterion for a two-dimensional slice through a $512^3$ evolved
velocity field. As can be seen by comparison to
\ref{sfig:structures512_vort} for the vorticity, the $Q$-criterion is more
selective in what it considers to be coherent structures. Figures
\ref{sfig:3dstructures256_Q} and \ref{sfig:3dstructures1024_Q} show the
three-dimensional structures identified using the $Q$-criterion. By
comparison to those obtained using vorticity, we once again see that this
method is stricter with what it considers to be a vortex. Note also that the
`sheet'-like structures obtained using vorticity are no longer present.
Comparison should be made to the Gaussian initial condition plotted in
figure \ref{sfig:256_Gaussian_Q}. See the work by Jeong and Hussain \cite{Jeong95} and by Haller \cite{Haller05}, and the many references therein, for more information and further discussion of these points.

\subsubsection{Rate-of-strain}
\begin{figure}[htb]
 \centering
 \subfigure[Vorticity]{
  \label{sfig:structures512_vort}
  \includegraphics[width=0.475\textwidth,trim=130px 460px 130px 30px, clip]{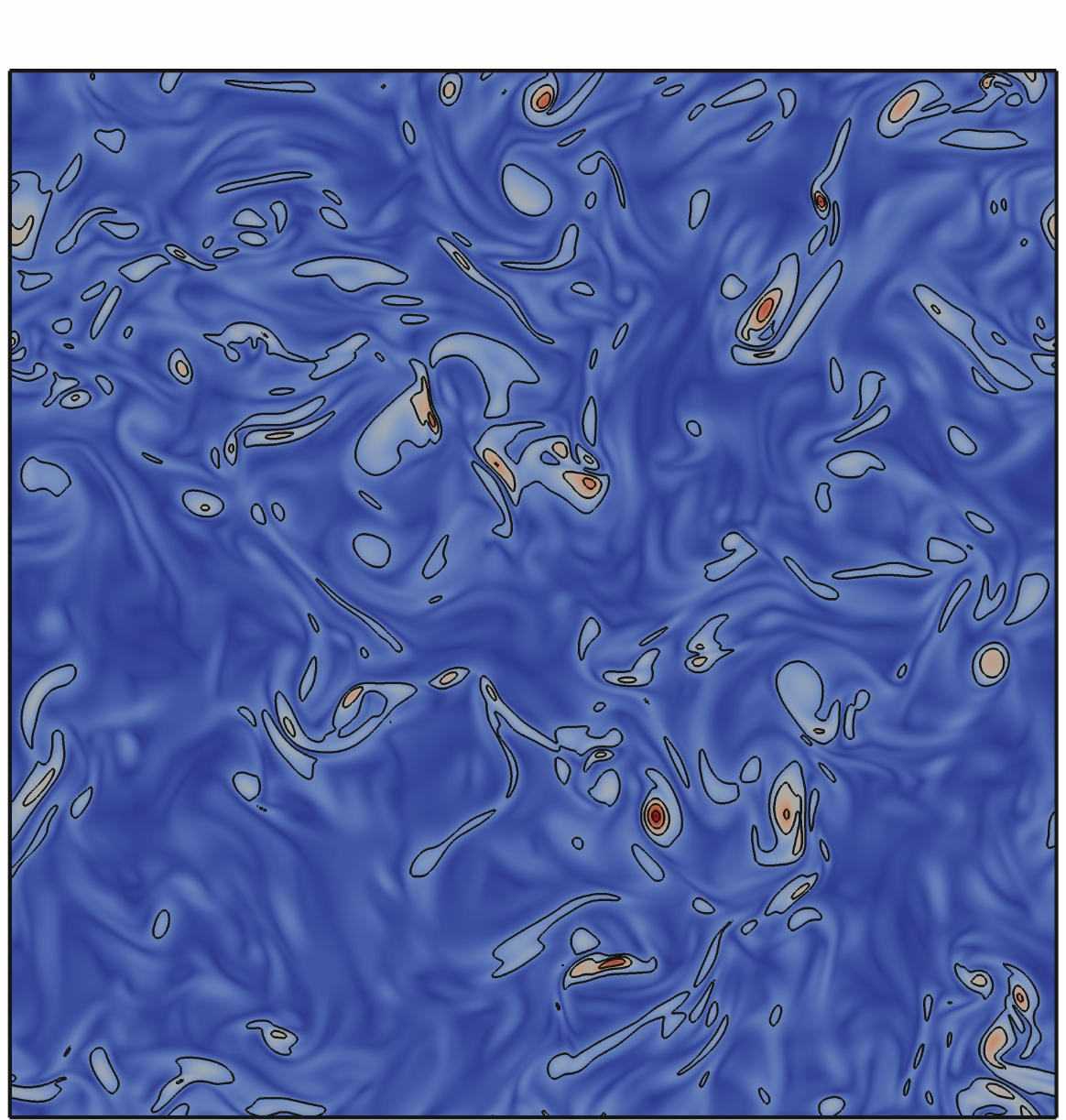}
 }\hfill
 \subfigure[Strain rate]{
  \label{sfig:structures512_strain}
  \includegraphics[width=0.475\textwidth,trim=130px 460px 130px 30px, clip]{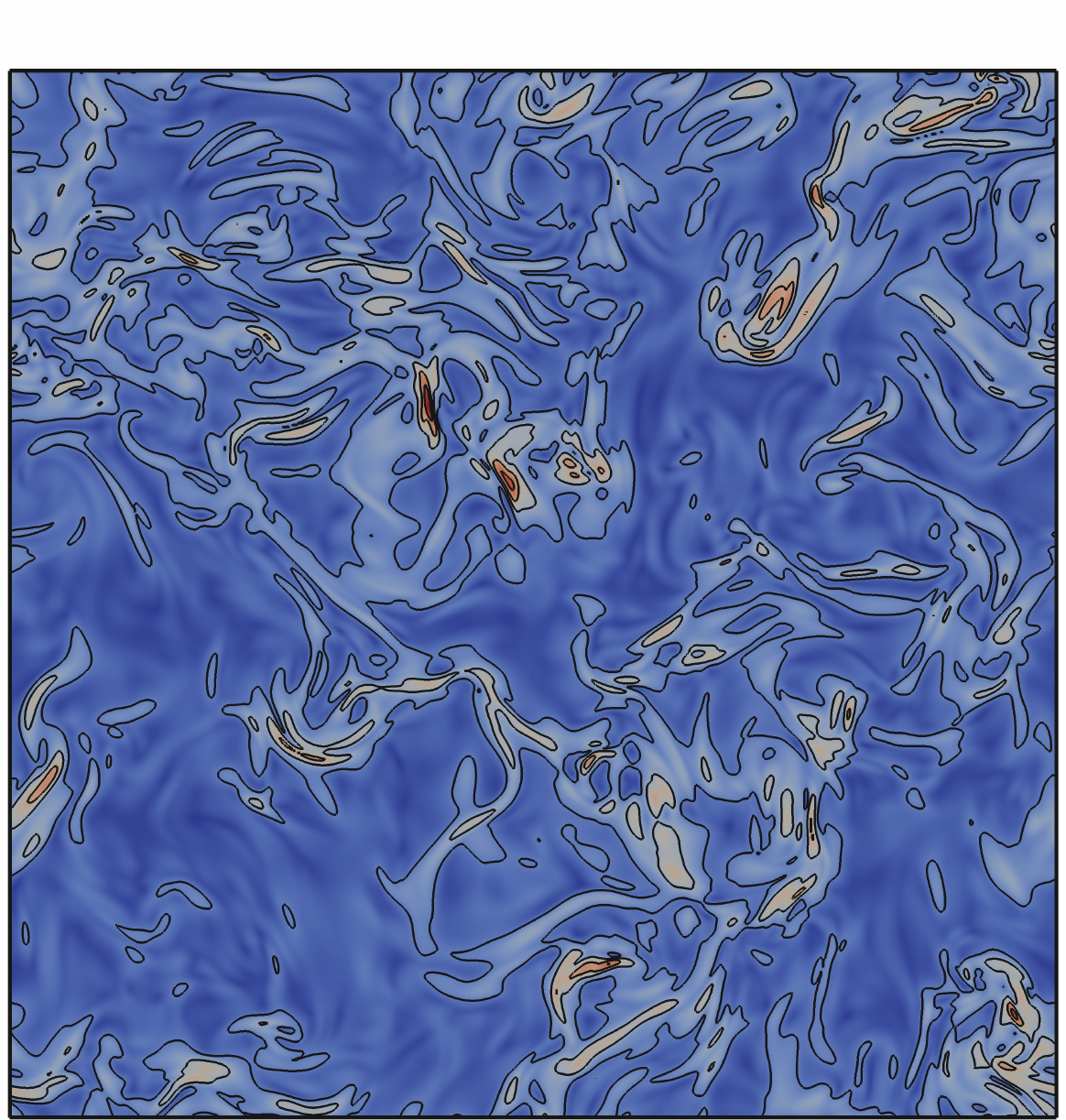}
 } \\
 \subfigure[$Q$-criterion]{
  \label{sfig:structures512_Q}
  \includegraphics[width=0.475\textwidth,trim=130px 460px 130px 30px, clip]{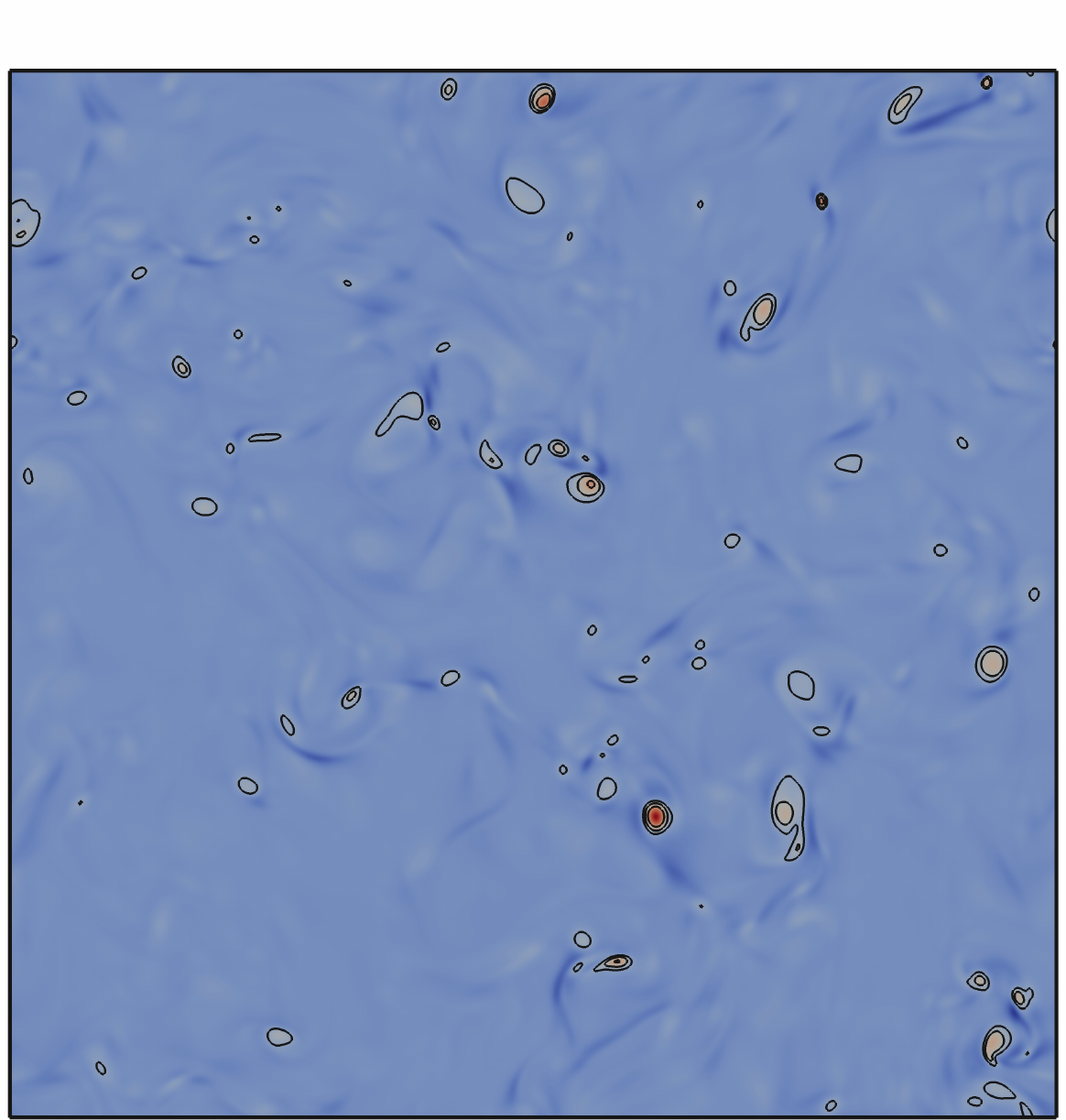}
 }
 \caption{Visualisation of the $z=0$ plane of an $R_\lambda \sim 115$
 evolved velocity field from run \frun{f512f}, using: (a) vorticity,
 $\vmod{\omega}$; (b) magnitude of the strain rate tensor, $\Vert S \Vert$;
 and (c) $Q$-criterion. Contours for a range of values are also plotted.
 Note that the $Q$-criterion identifies far fewer structures. Contours for
 $Q$-criterion all have $Q \geq 0.1 Q_{\textrm{max}}$.}
 \label{fig:structures512}
\end{figure}

\begin{figure}[htb]
 \centering
 \subfigure[Velocity]{
  \label{sfig:structures1024_vel}
  \includegraphics[width=0.475\textwidth,trim=145px 475px 145px 50px, clip]{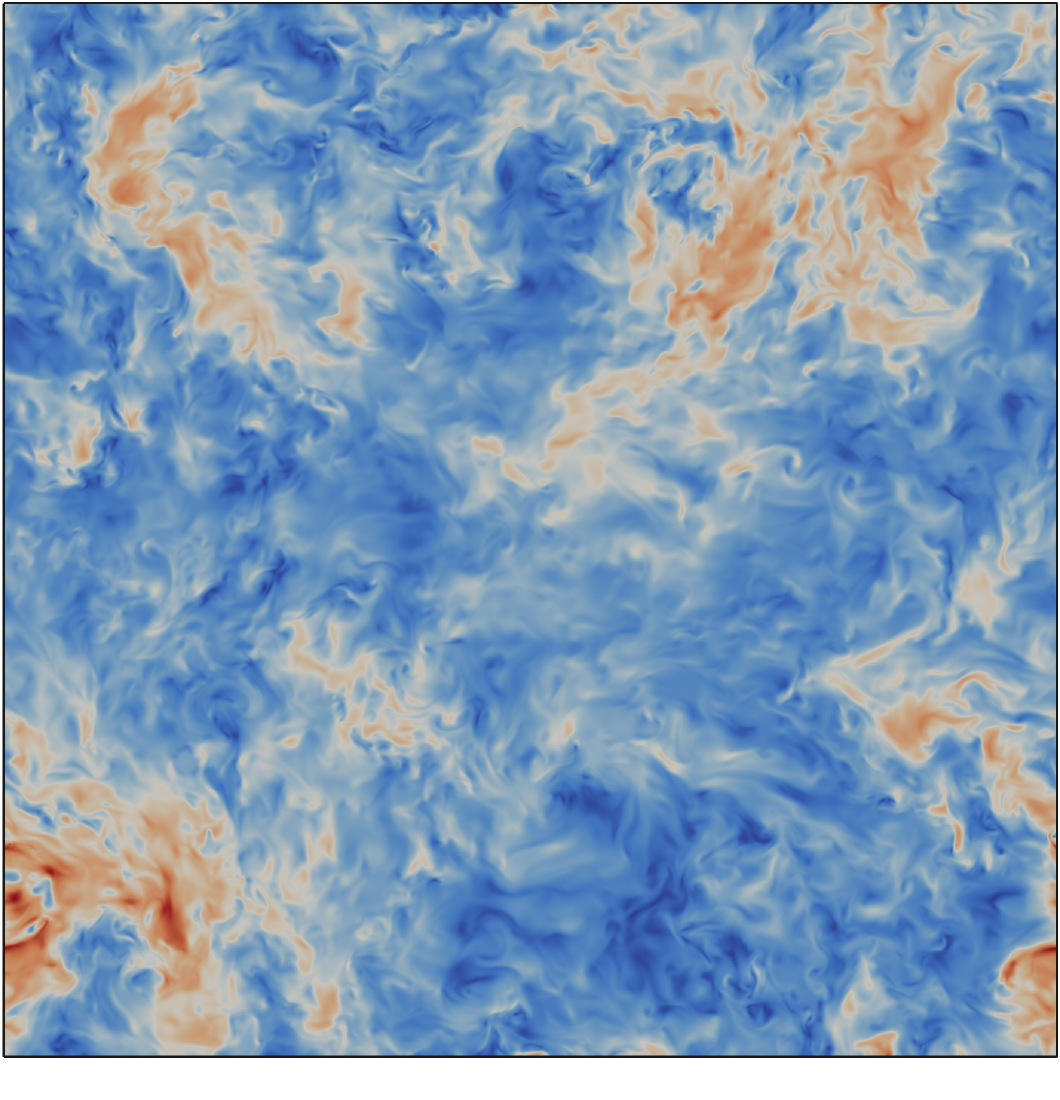}
 }\hfill
 \subfigure[Vorticity]{
  \label{sfig:structures1024_vort}
  \includegraphics[width=0.475\textwidth,trim=145px 475px 145px 50px, clip]{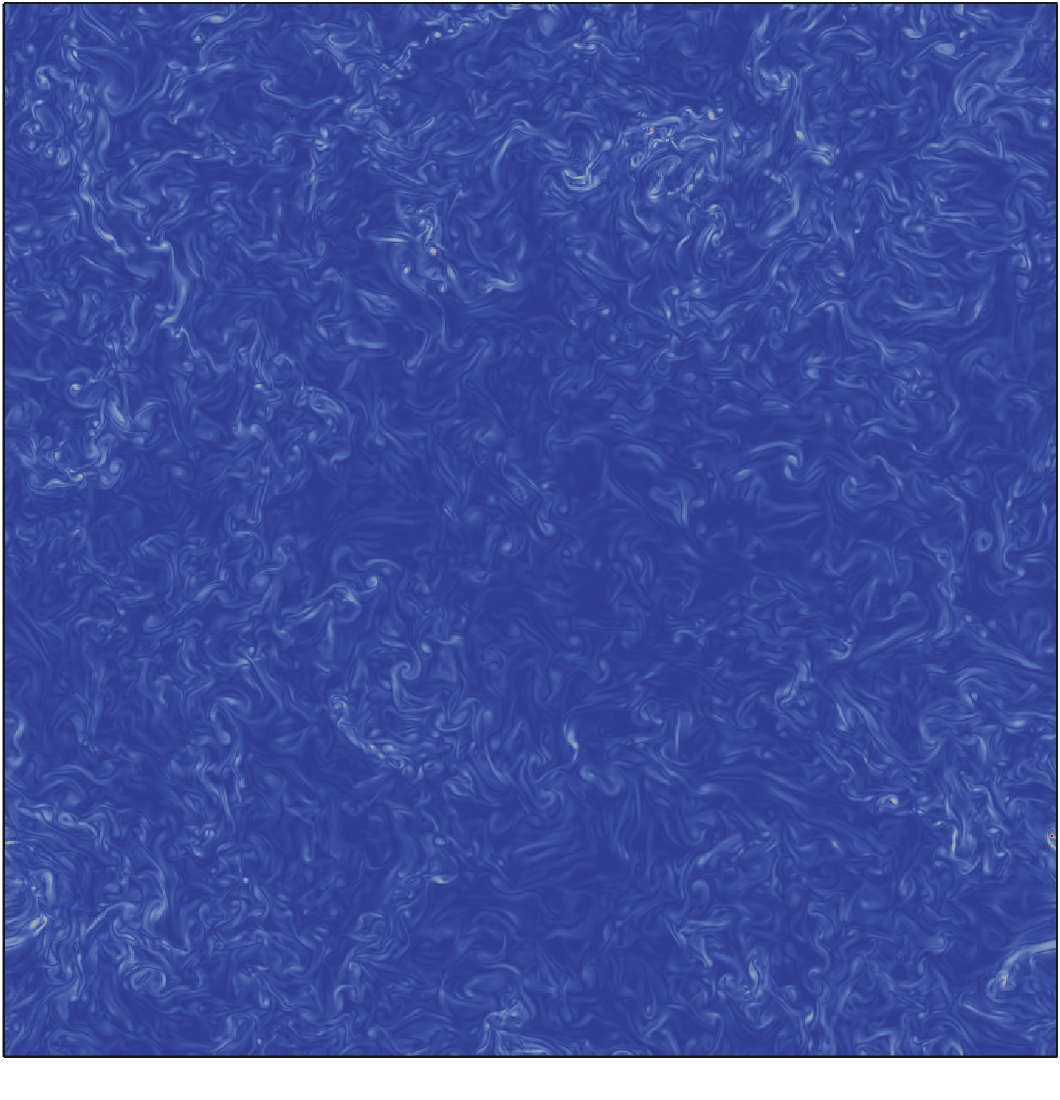}
 } \\
 \subfigure[Strain rate]{
  \label{sfig:structures1024_strain}
  \includegraphics[width=0.475\textwidth,trim=145px 475px 145px 50px, clip]{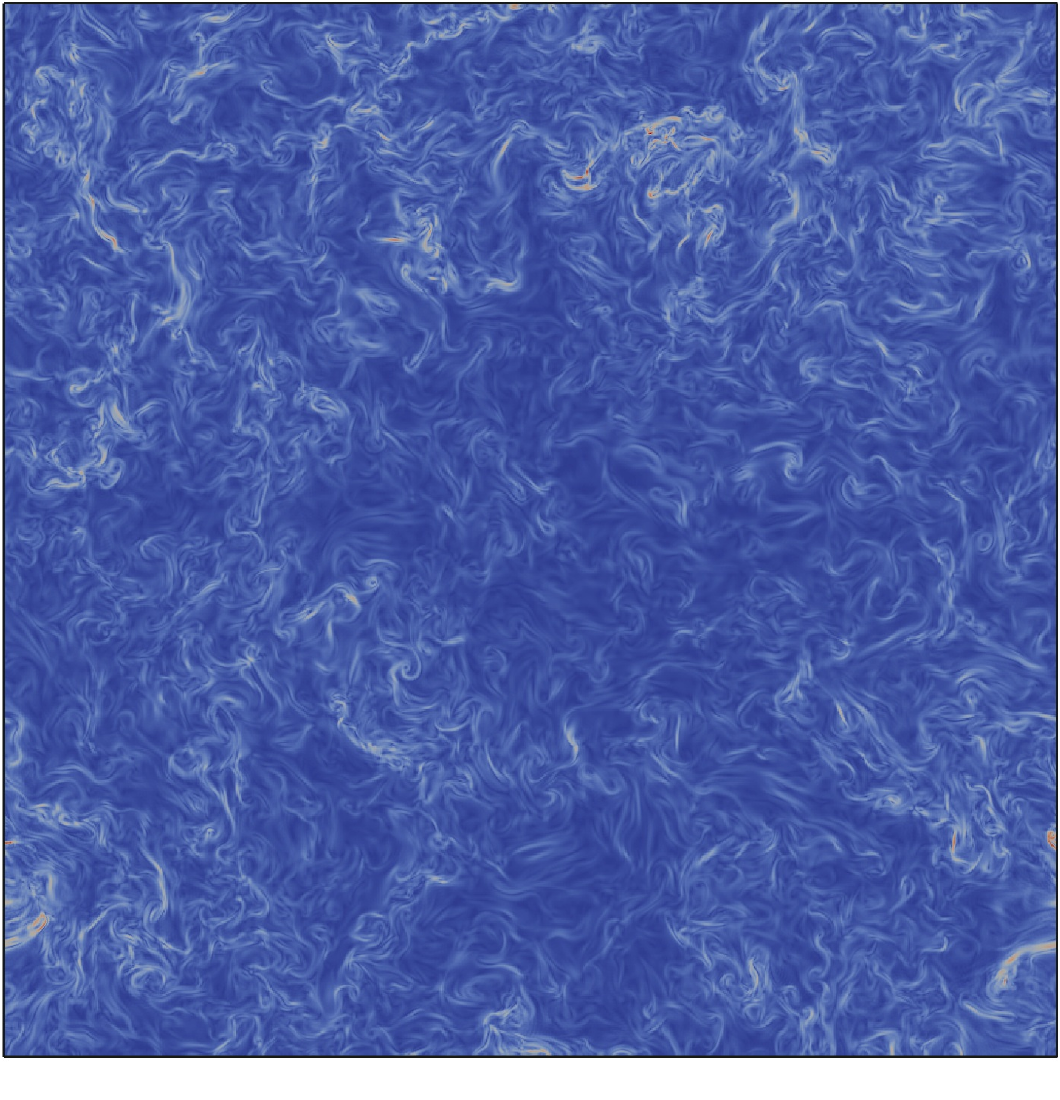}
 }
 \caption{A snapshot of (the $z=0$ plane of) the evolved velocity field from
 run \frun{f1024a}, coloured by: (a) $\vmod{u}$; (b) $\vmod{\omega}$; and
 (c) magnitude of the strain rate tensor, $\Vert S \Vert$. Contours not
 plotted due to the small size of the structures. Magnitude of velocity
 offers little in the way of identifying structures.}
  \label{fig:structures1024}
\end{figure}

\begin{figure}[htb]
 \centering
 \subfigure[Vorticity]{
  \label{sfig:3dstructures256_vort}
  \includegraphics[width=0.475\textwidth,trim=130px 430px 145px 70px, clip]{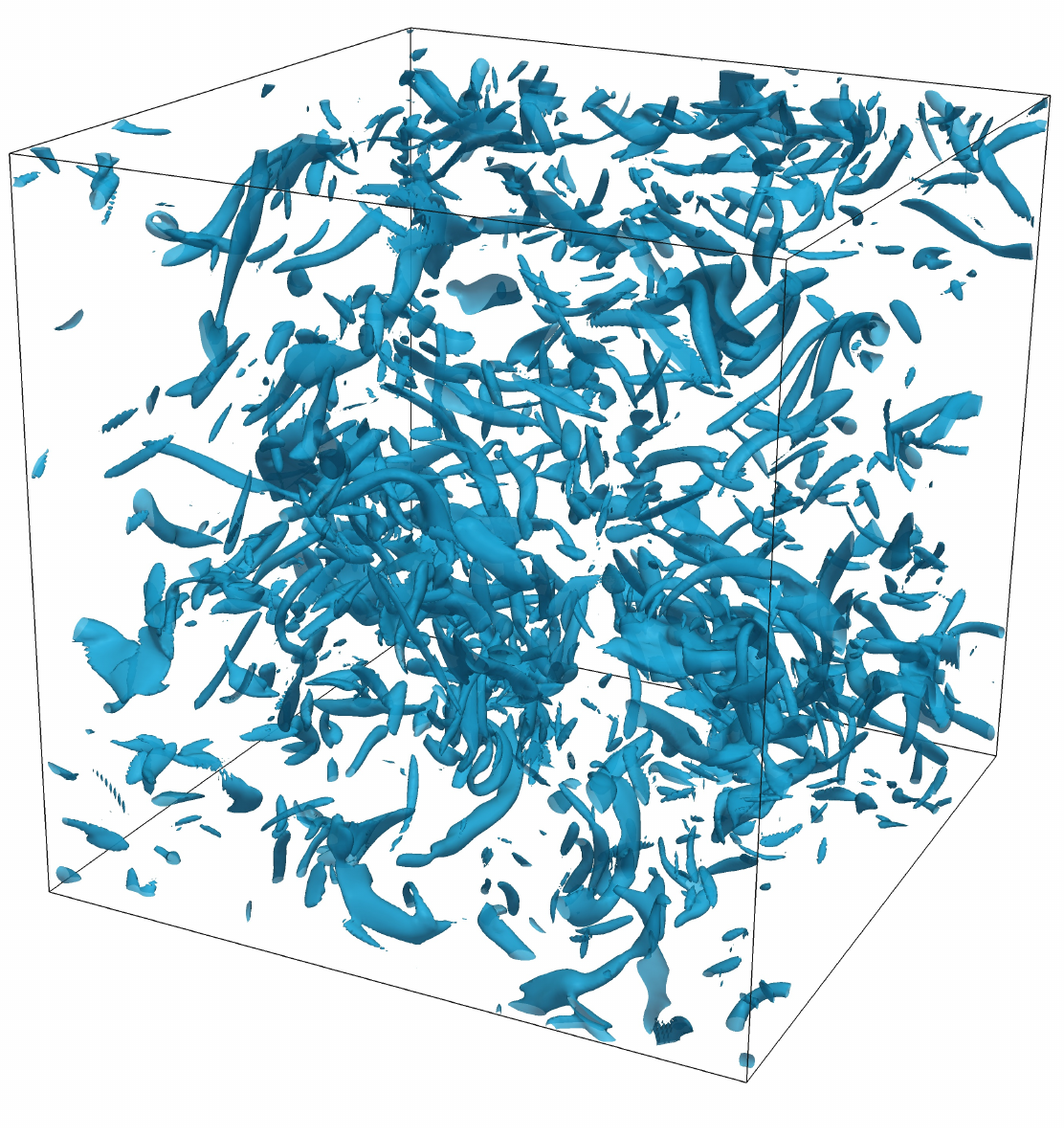}
 }
 \subfigure[Vorticity and rate-of-strain]{
  \label{sfig:3dstructures256_vort_strain}
  \includegraphics[width=0.475\textwidth,trim=130px 430px 145px 70px, clip]{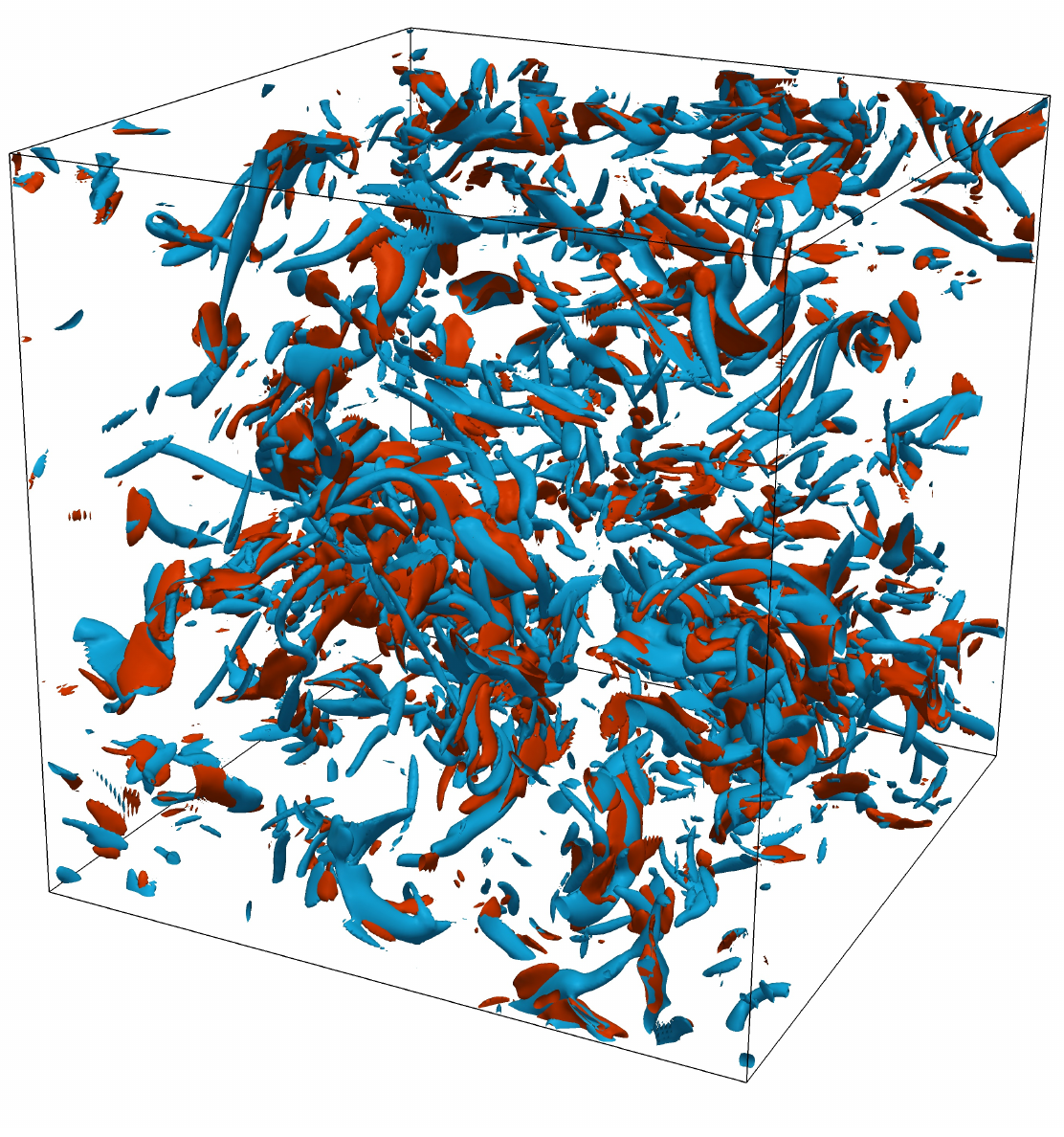}
 }
 \subfigure[Q-criterion]{
  \label{sfig:3dstructures256_Q}
  \includegraphics[width=0.64\textwidth,trim=130px 430px 145px 70px, clip]{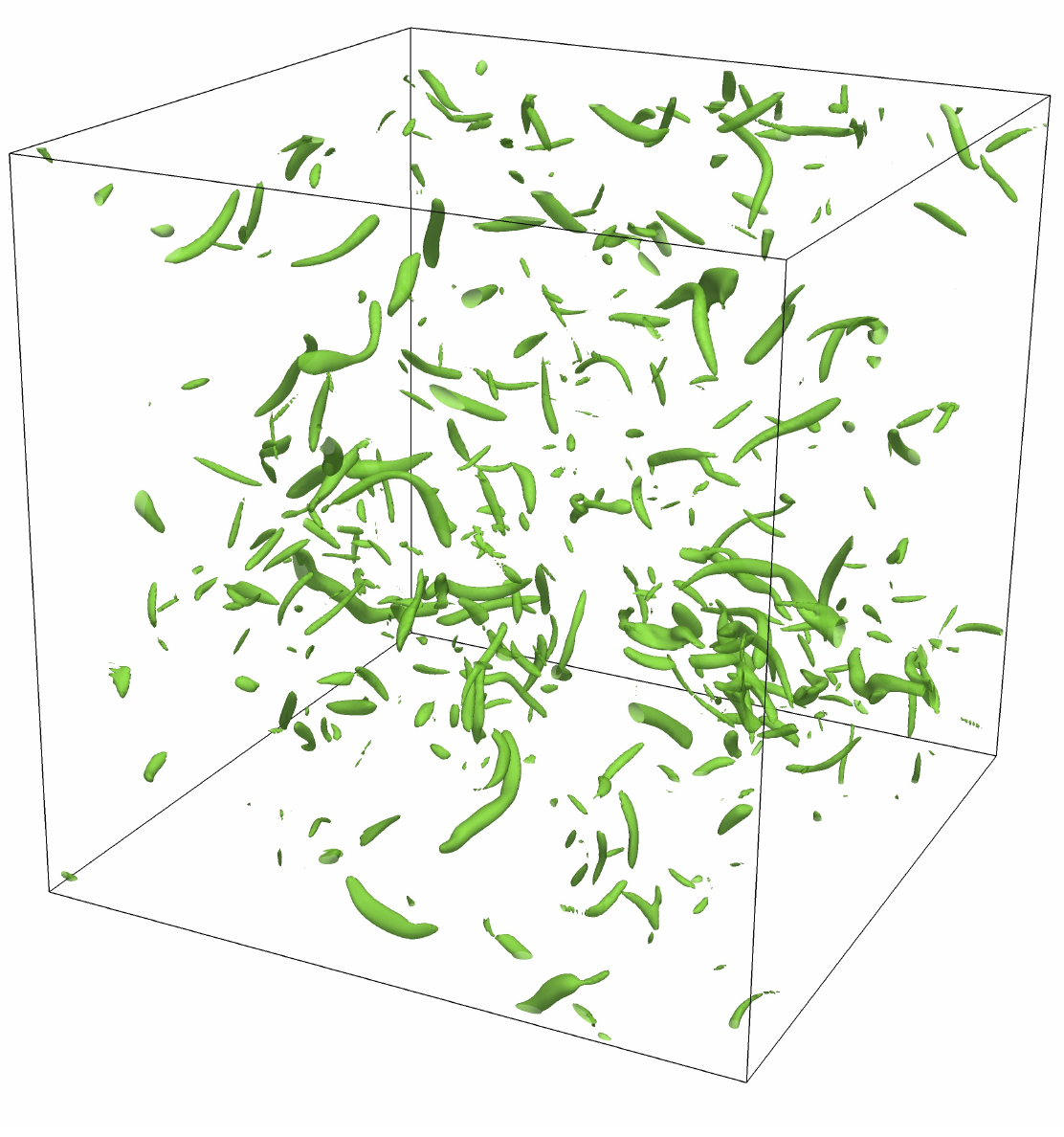}
 }
 \caption{Visualisation of turbulent structures in an $R_\lambda \sim 100$
 evolved velocity field from run \frun{f256b}. Isosurfaces of (a) vorticity
 ($0.25 \omega_{\textrm{max}}$ plotted); (b) vorticity (blue) and strain
 rate ($0.4\Vert S\Vert_{\textrm{max}}$ plotted, red); and (c) Q-criterion
 ($0.1 Q_{\textrm{max}}$ plotted). Regions of high vorticity are seen to be
 correlated with areas of high strain. The $Q$-criterion can be seen to pick
 out fewer structures than just vorticity.}
 \label{fig:3dstructures256}
\end{figure}

\begin{figure}[htb]
 \centering
 \subfigure[Isovorticity]{
  \label{sfig:256_Gaussian_vort}
  \includegraphics[width=0.35\textwidth,trim=130px 430px 145px 70px, clip]{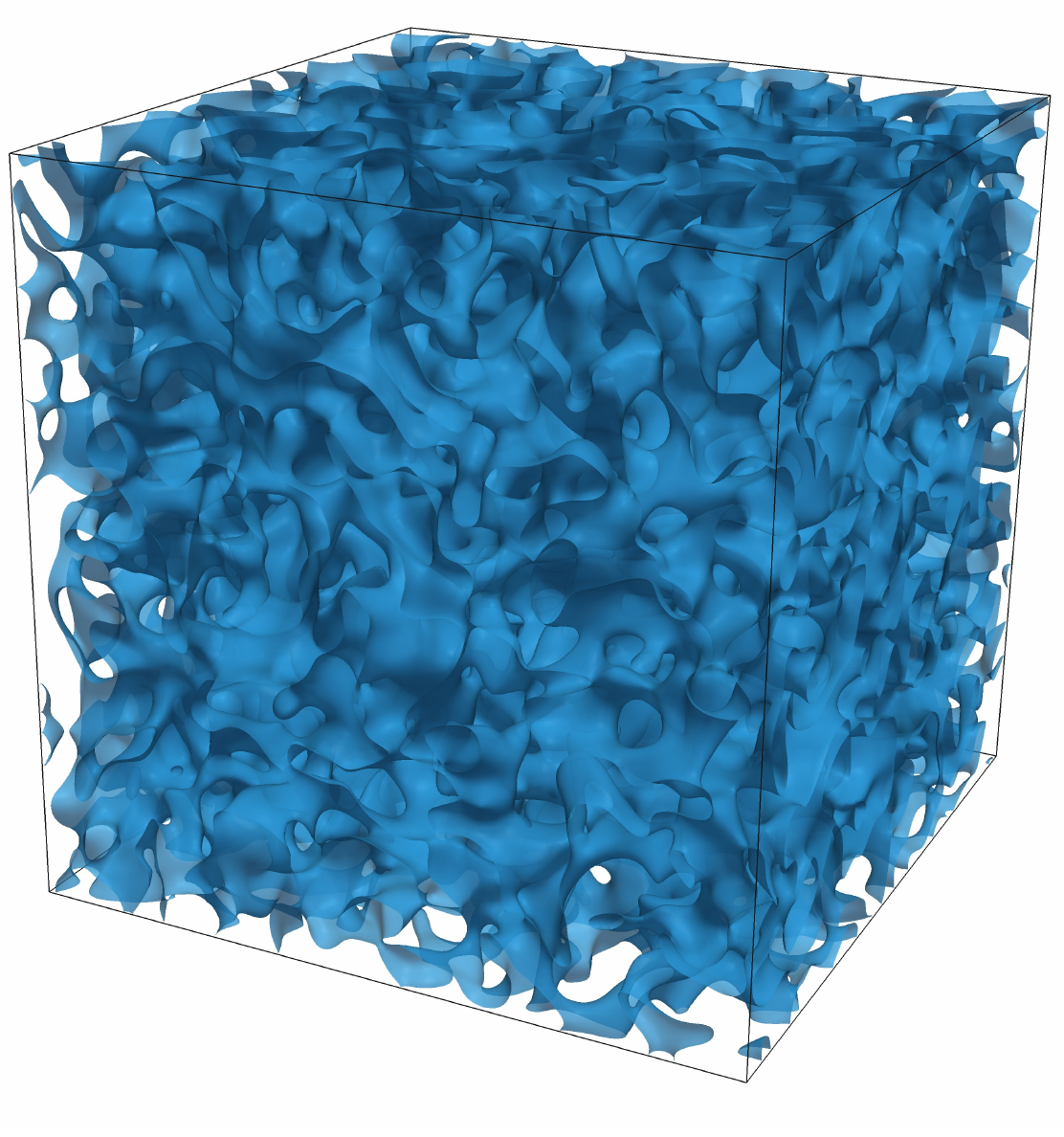}
 }\hspace{0.5in}
 \subfigure[Q-criterion]{
  \label{sfig:256_Gaussian_Q}
  \includegraphics[width=0.35\textwidth,trim=130px 430px 145px 70px, clip]{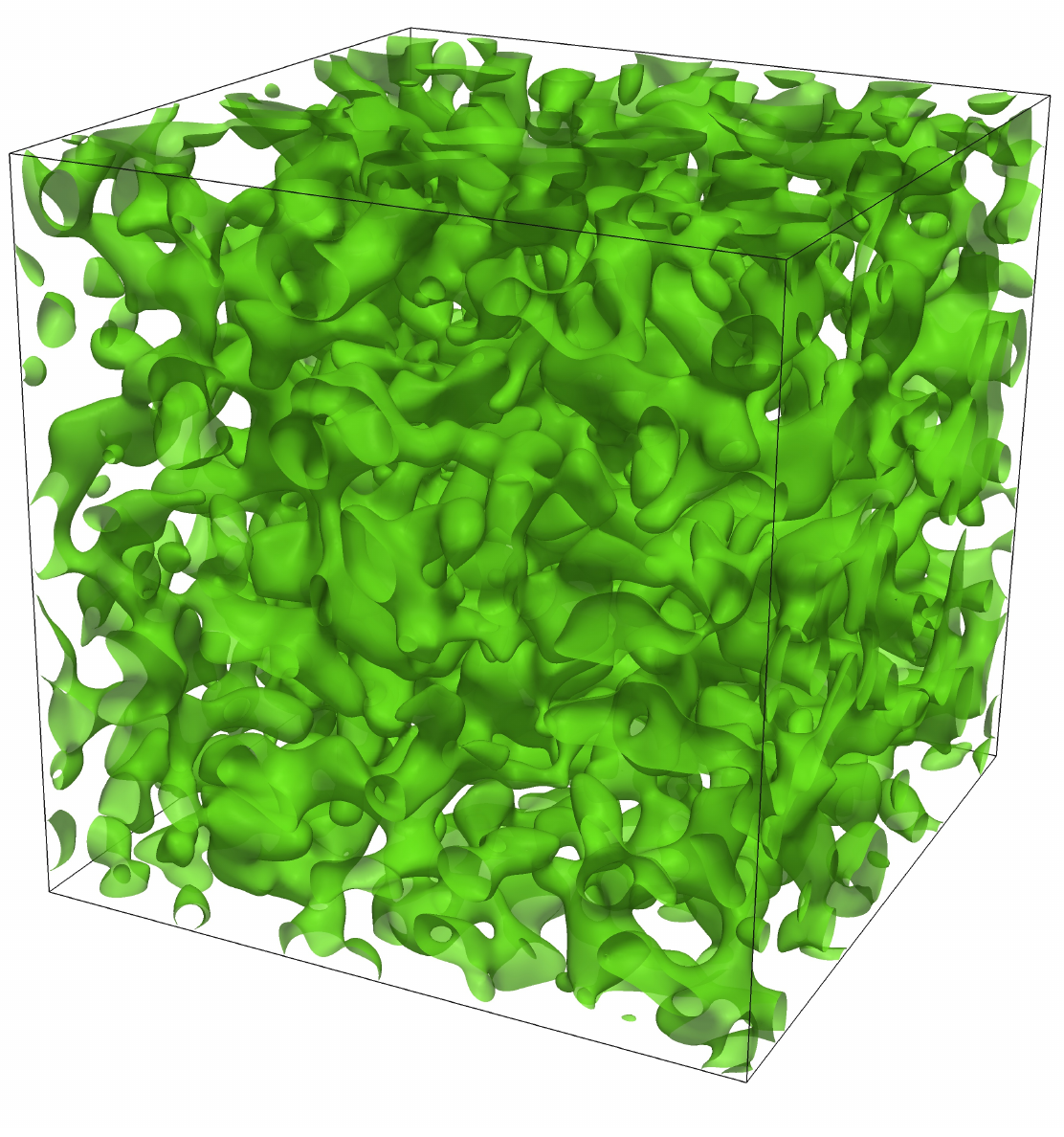}
 }
 \caption{Visualisation of structures in an $N = 256$ initial
 random Gaussian field. There is little evidence of coherent structure. The
 same surfaces have been plotted as figure \ref{fig:3dstructures256} above.}
 \label{fig:256_Gaussian}
\end{figure}

\begin{figure}[htb]
 \centering
 \subfigure[Vorticity]{
  \label{sfig:3dstructures1024_vort}
  \includegraphics[width=0.45\textwidth,trim=165px 465px 185px 60px, clip]{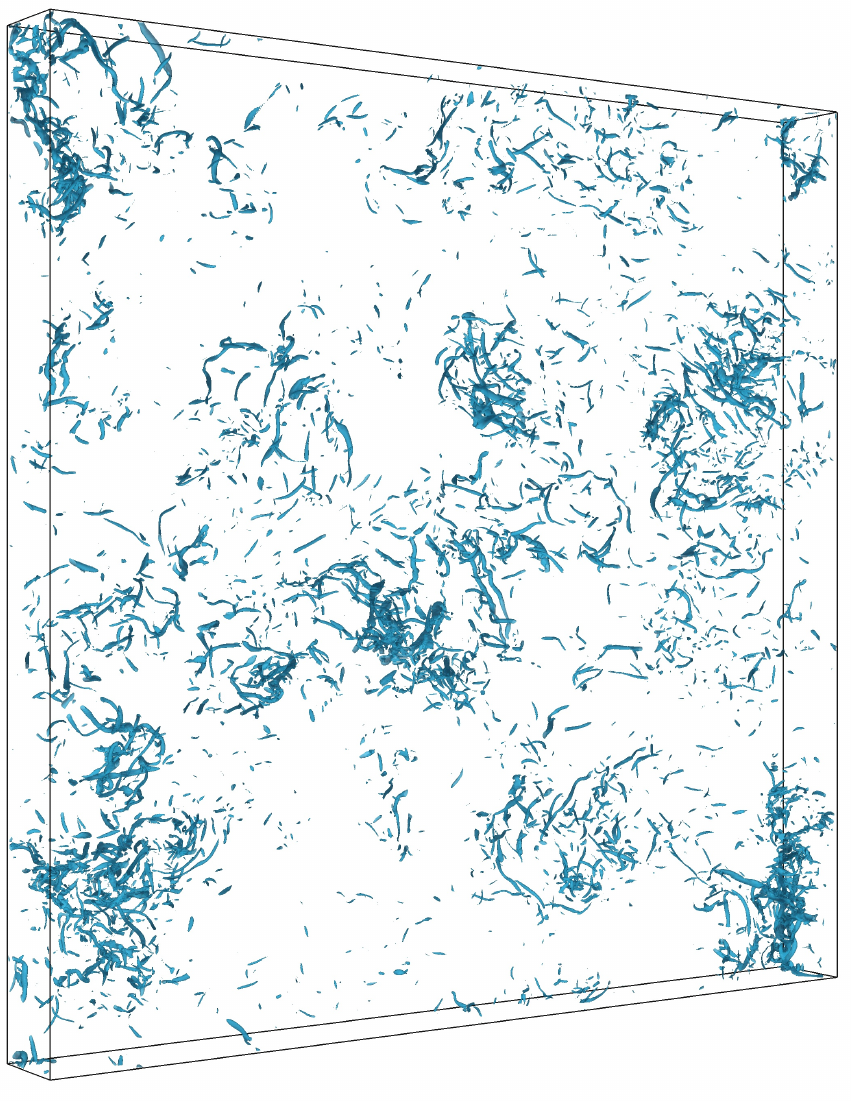}
 }\hfill
 \subfigure[Q-criterion]{
  \label{sfig:3dstructures1024_Q}
  \includegraphics[width=0.45\textwidth,trim=165px 465px 185px 60px, clip]{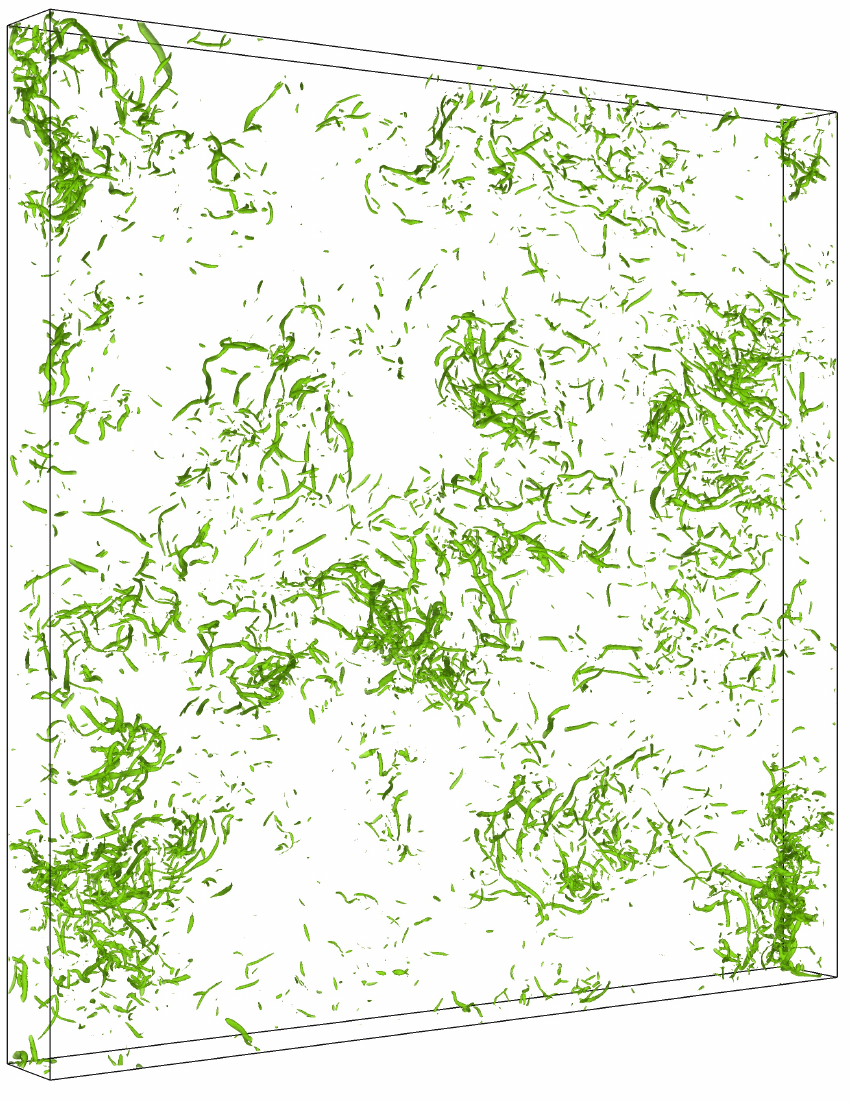}
 }
 \caption{Visualisation of turbulent structures in a $96\times 1024 \times
 1024$ slice (due to memory constraints, the whole volume could not be
 rendered) of an $R_\lambda \sim 335$ evolved velocity field from run
 \frun{f1024b}. Isosurfaces of (a) vorticity ($0.25 \omega_{\textrm{max}}$
 plotted) and (b) Q-criterion ($0.05 Q_{\textrm{max}}$ plotted).}
 \label{fig:3dstructures1024}
\end{figure}

The rate-of-strain tensor $S_{ij}$ defined above can be connected to the dissipation
rate, since
\begin{equation}
 \varepsilon = \frac{\nu_0}{2} \left\langle \left( \frac{\partial
 u_i}{\partial x_j} + \frac{\partial u_j}{\partial x_i} \right)^2
 \right\rangle = 2\nu_0 \big\langle \Vert S \Vert^2 \big\rangle \ ,
\end{equation}
where the average is performed over space. This means that $2\nu_0 \Vert S
\Vert^2$ gives a measure of the \emph{local} dissipation at a point $\vec{x}$.
Since $2\nu_0$ is just a scaling, the magnitude of the strain rate tensor
indicates the strength of the dissipation and allows for the identification
of dissipative structures. We show these in figures
\ref{sfig:structures512_strain} and \ref{sfig:structures1024_strain}: the
former compares contours with the magnitude of vorticity and $Q$-criterion,
discussed below, while the latter shows the structures for a higher Reynolds
number on a larger lattice. Figure \ref{sfig:3dstructures256_vort_strain}
shows the dissipative structures in three-dimensions, indicating that they
are correlated with and attached to the regions of high vorticity, but that
the two criteria are not indistinguishable.

\subsection{Persistence of structure under averaging}

Looking at the snapshot of the velocity field in figure (10a), it can
be seen that there are well-defined structures and a great deal of variation
from point to point. The velocity field is said to be intermittent: there is
a high degree of spatial variation. 

\begin{figure}[tbp!]
 \begin{center}
  \subfigure[$N = 1$]{
   \includegraphics[width=0.38\textwidth,trim=130px 460px 130px 30px, clip]{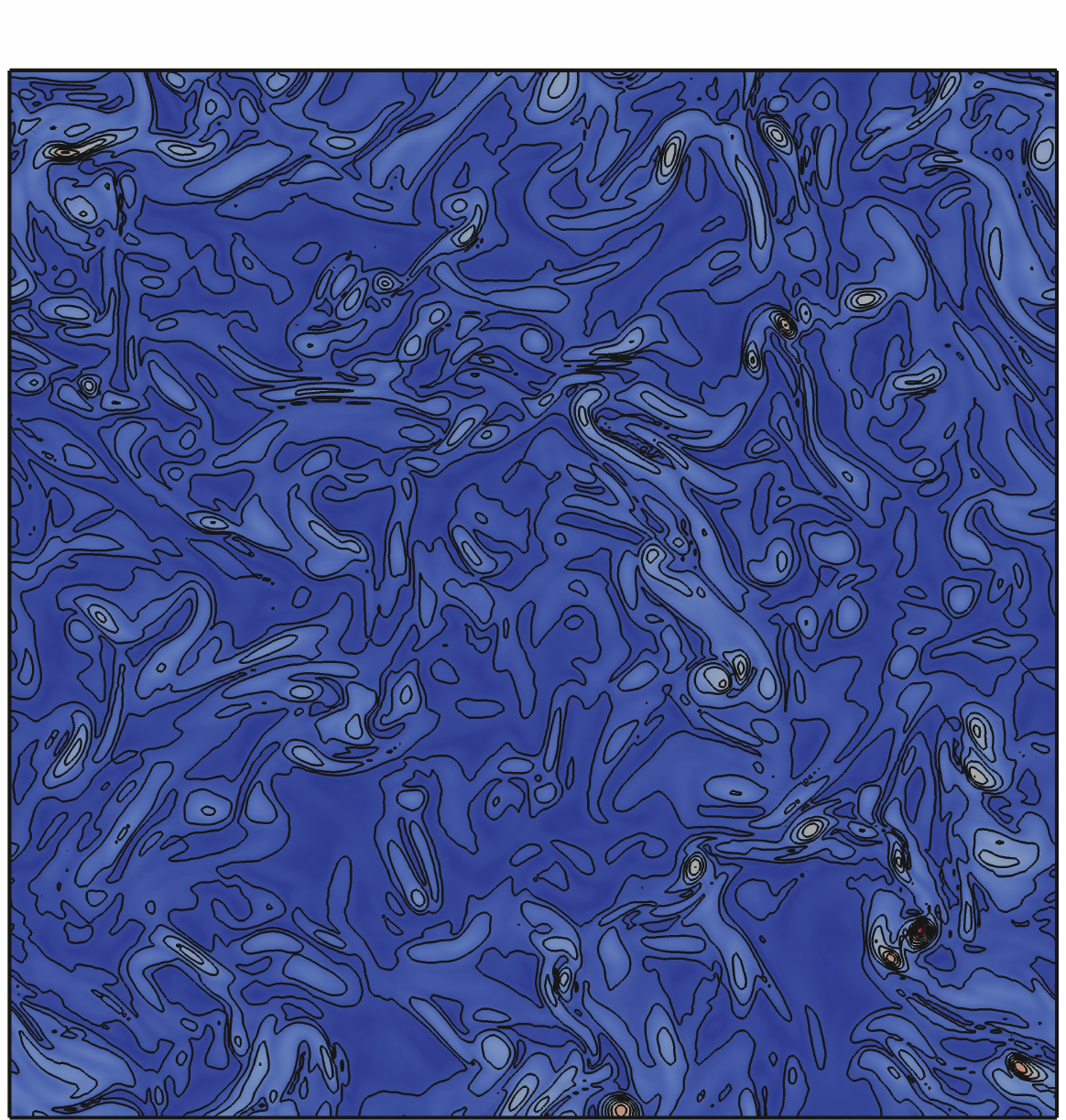}
  }
  \subfigure[$N = 2$]{
   \includegraphics[width=0.38\textwidth,trim=130px 460px 130px 30px, clip]{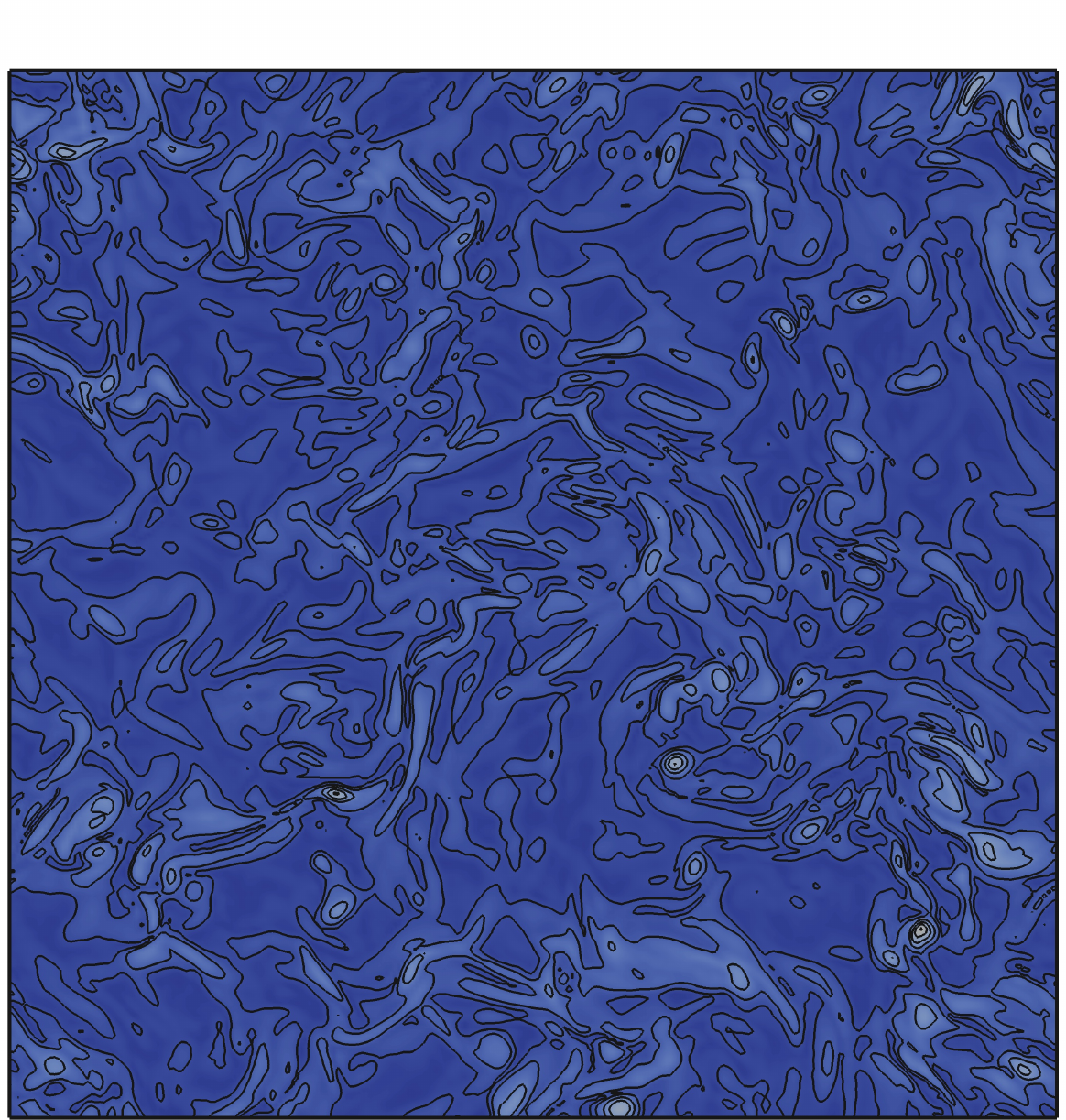}
  }
  \subfigure[$N = 5$]{
   \includegraphics[width=0.38\textwidth,trim=130px 460px 130px 30px, clip]{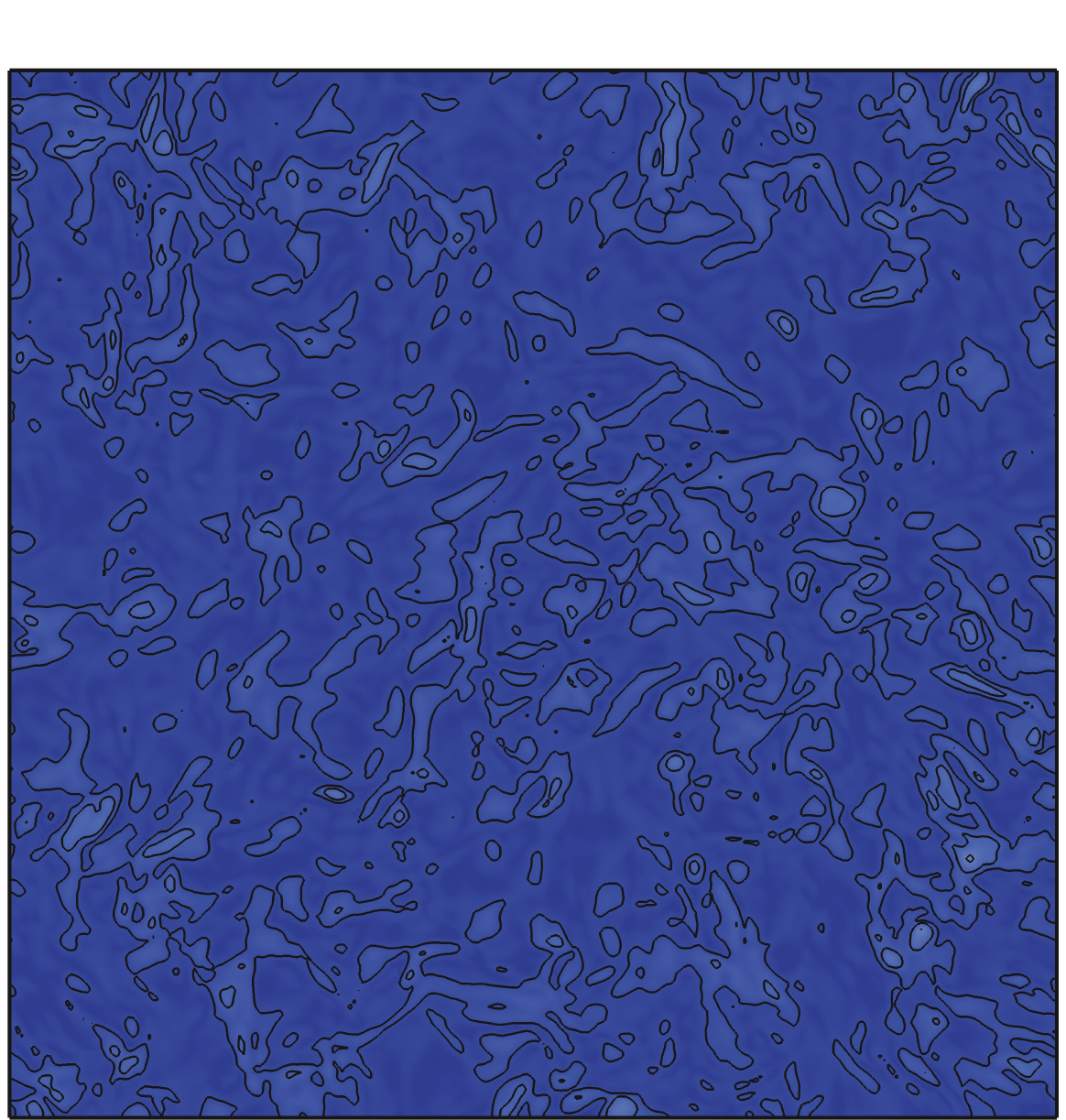}
  }
  \subfigure[$N = 10$]{
   \includegraphics[width=0.38\textwidth,trim=130px 460px 130px 30px, clip]{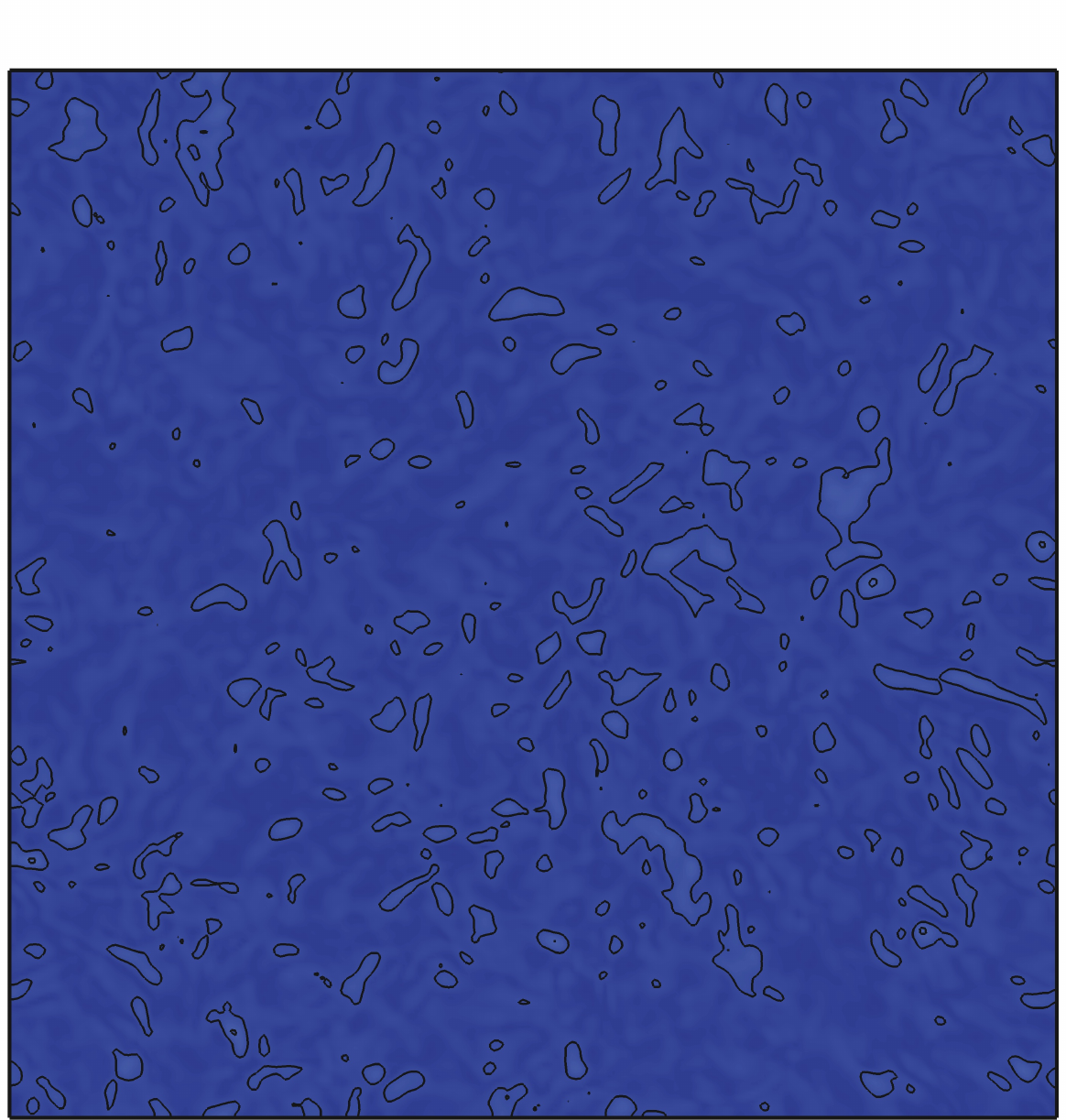}
  }
  \subfigure[$N = 25$]{
   \includegraphics[width=0.38\textwidth,trim=130px 460px 130px 30px, clip]{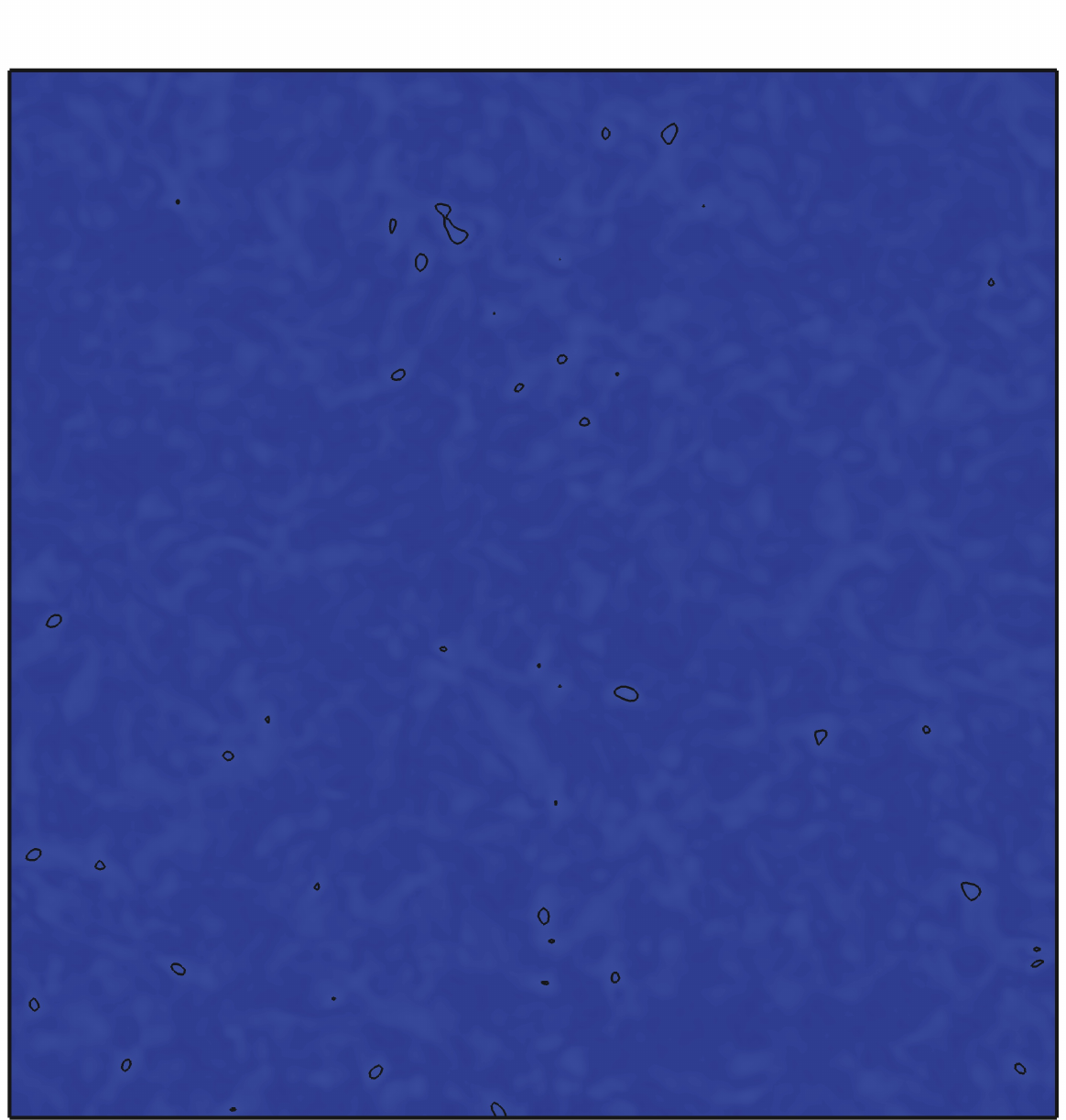}
  }
  \subfigure[$N = 46$]{
   \includegraphics[width=0.38\textwidth,trim=130px 460px 130px 30px, clip]{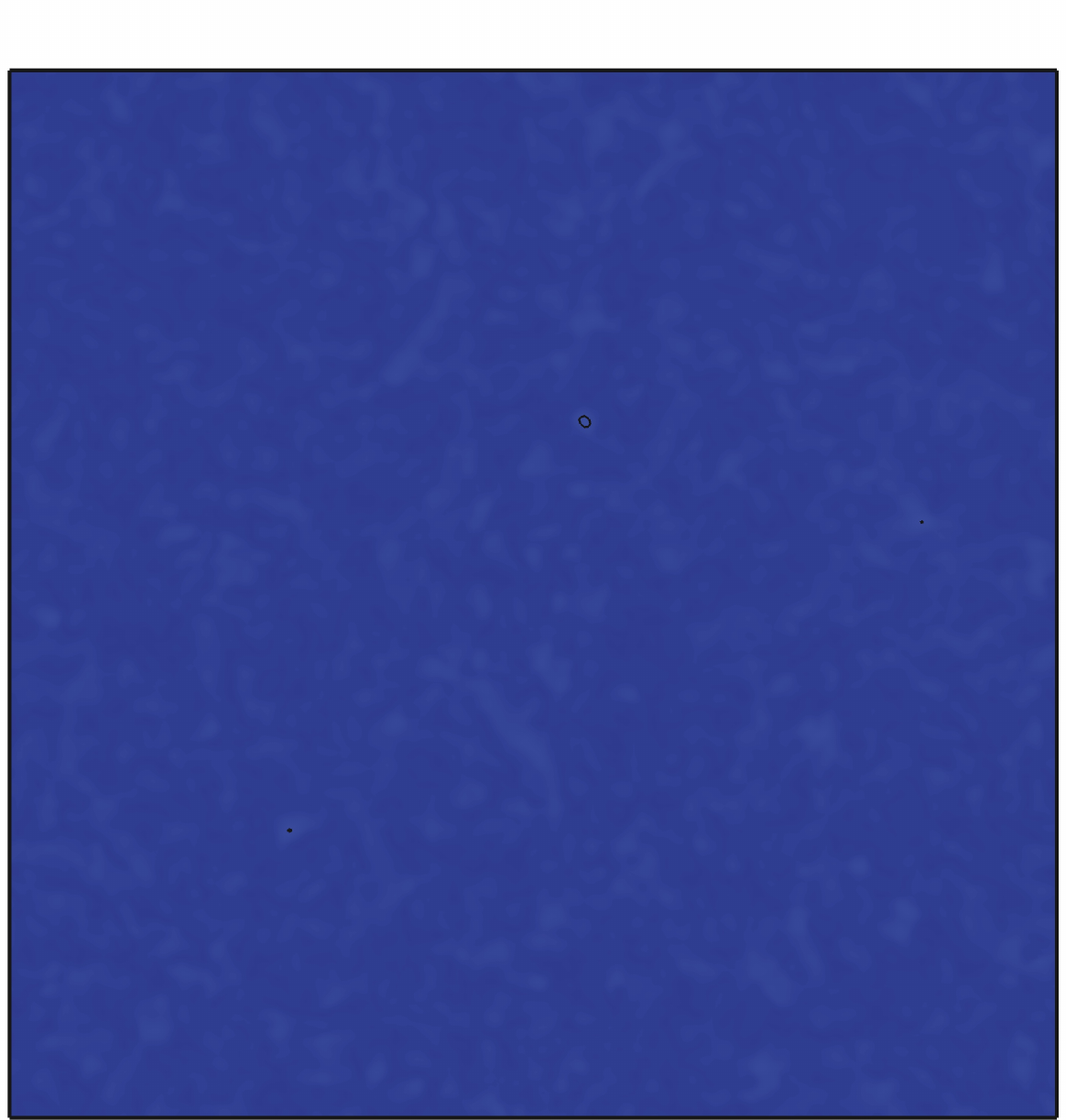}
  }
 \end{center}
 \caption{Contours plotted for 5\%, 10\%, 20\%, 30\%, 40\%, 50\%, 60\%,
 70\%, 80\% and 90\% of $\omega_{\textrm{max}}$. $R_\lambda \sim 100$ on
 $256^3$ run \frun{f256b}.}
 \label{fig:str256_V}
\end{figure}

\begin{figure}[tbp!]
 \begin{center}
  \subfigure[$N = 1$]{
   \includegraphics[width=0.38\textwidth,trim=130px 460px 130px 30px, clip]{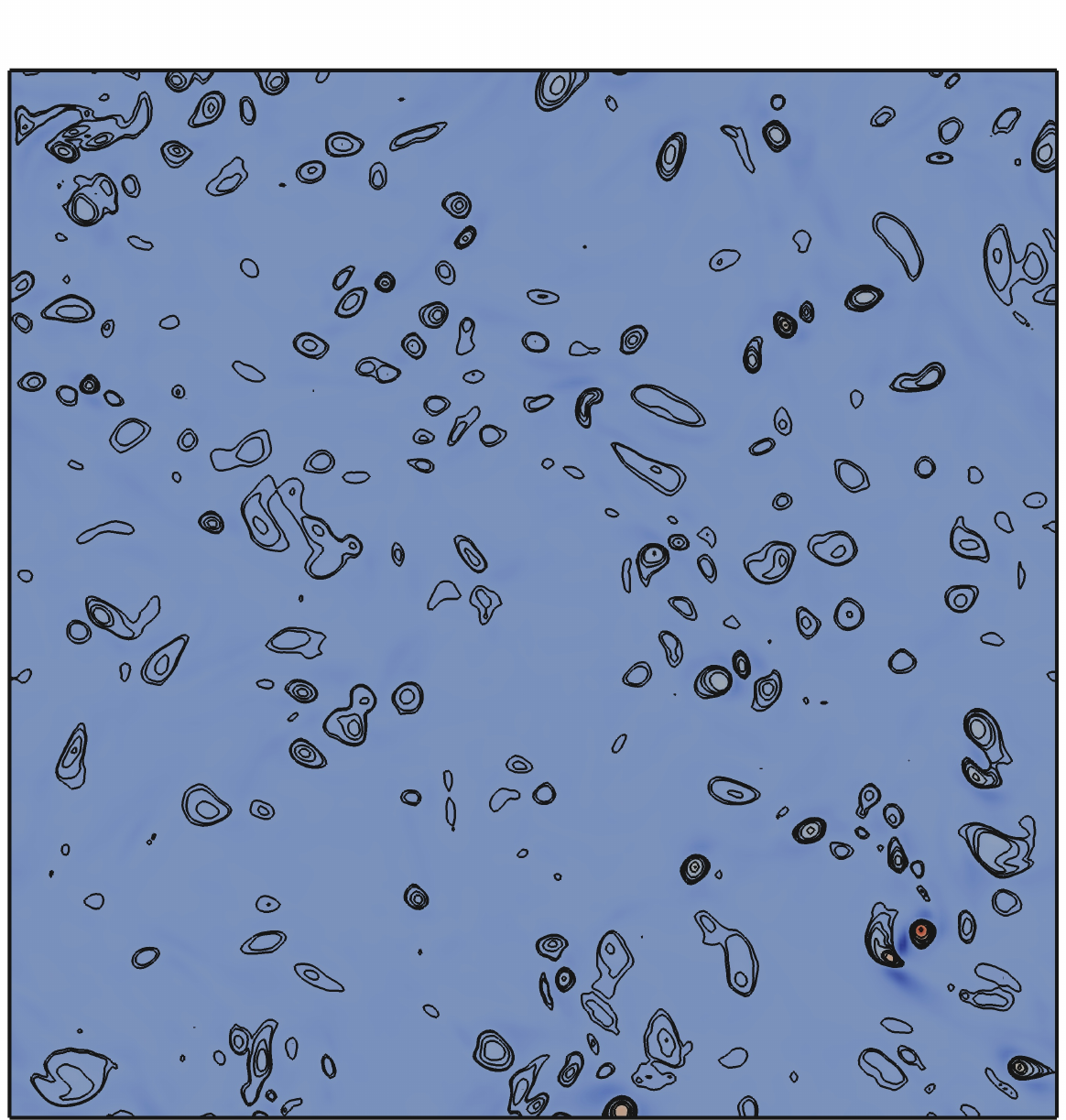}
  }
  \subfigure[$N = 2$]{
   \includegraphics[width=0.38\textwidth,trim=130px 460px 130px 30px, clip]{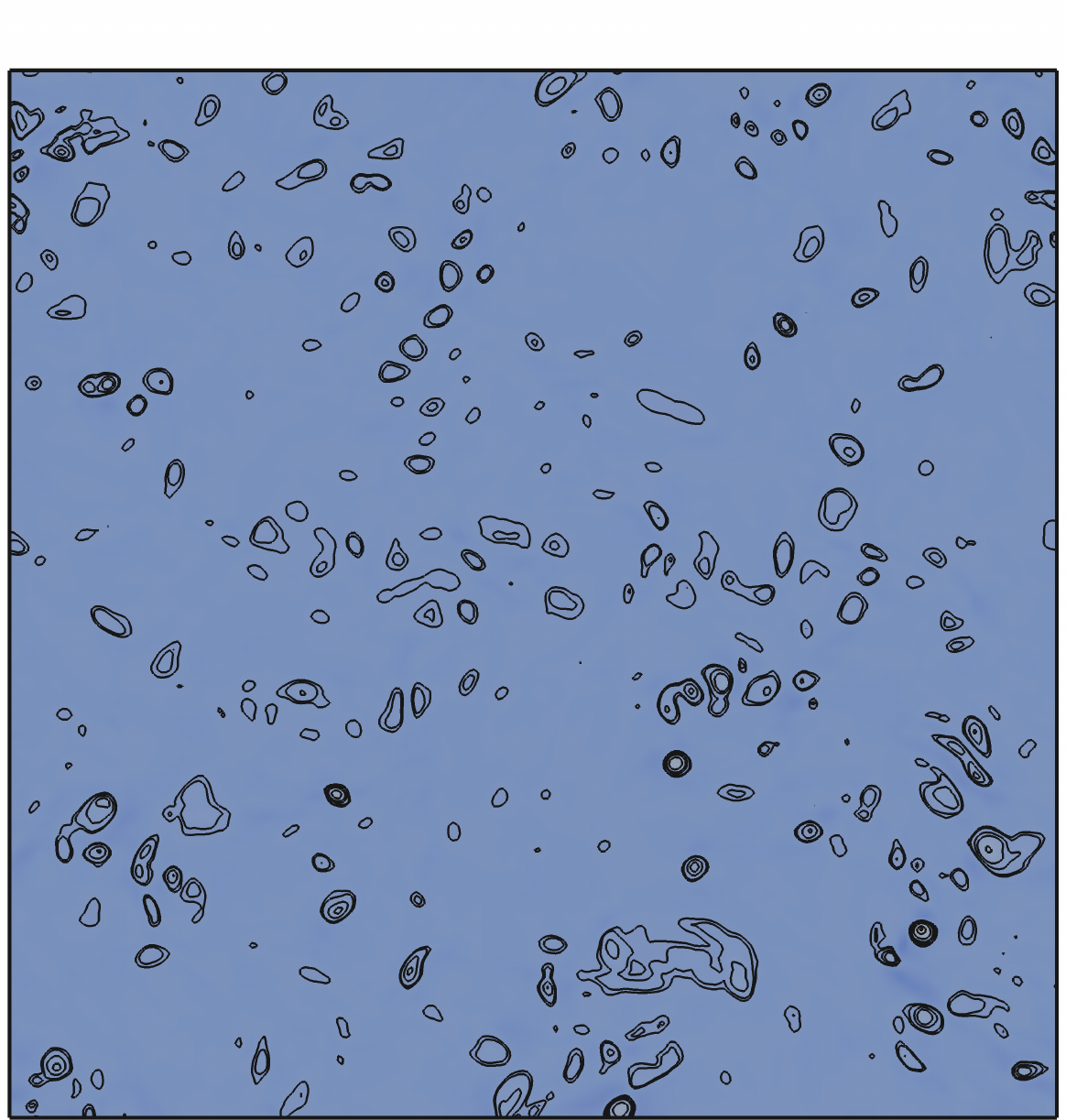}
  }
  \subfigure[$N = 5$]{
   \includegraphics[width=0.38\textwidth,trim=130px 460px 130px 30px, clip]{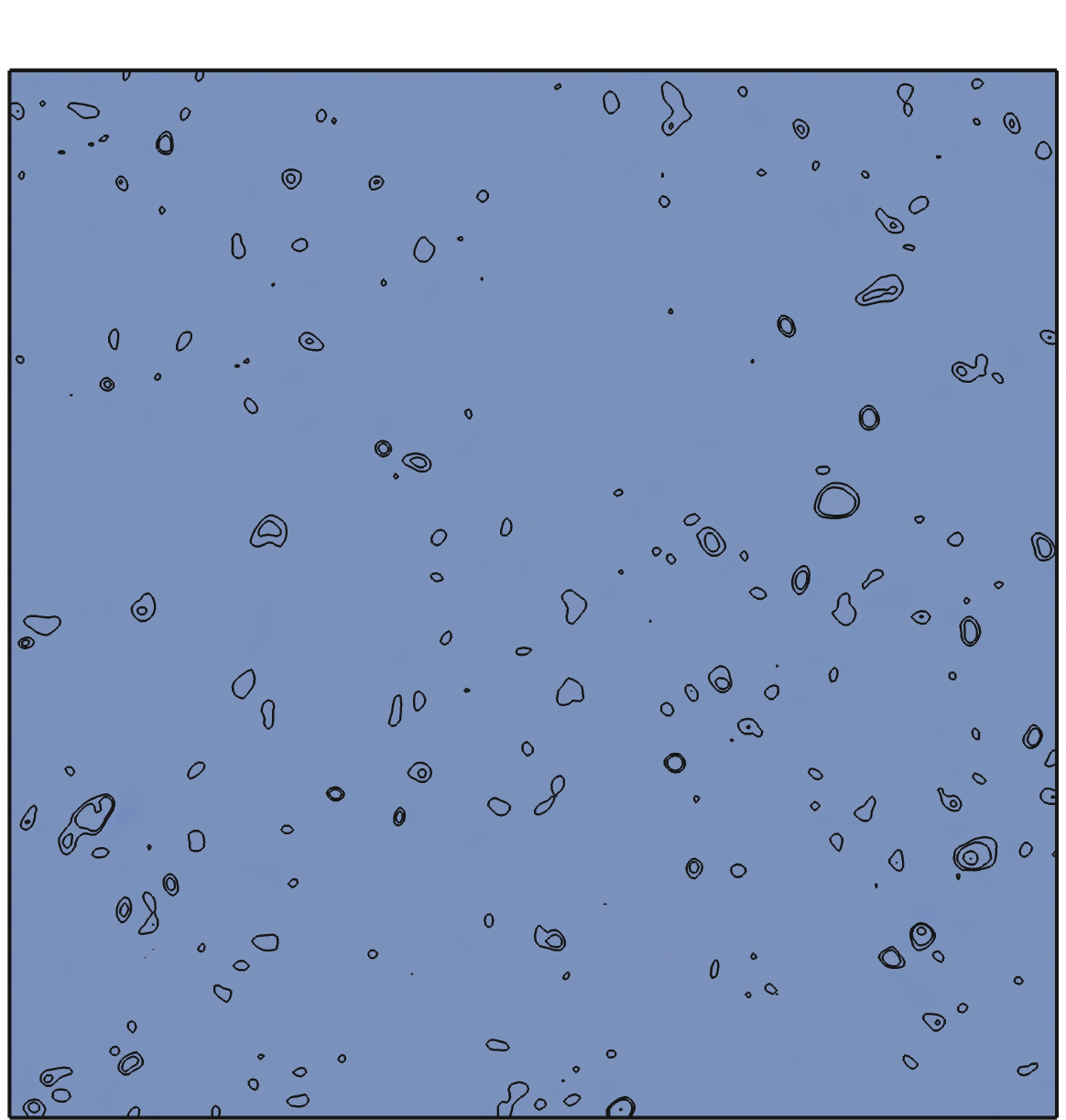}
  }
  \subfigure[$N = 10$]{
   \includegraphics[width=0.38\textwidth,trim=130px 460px 130px 30px, clip]{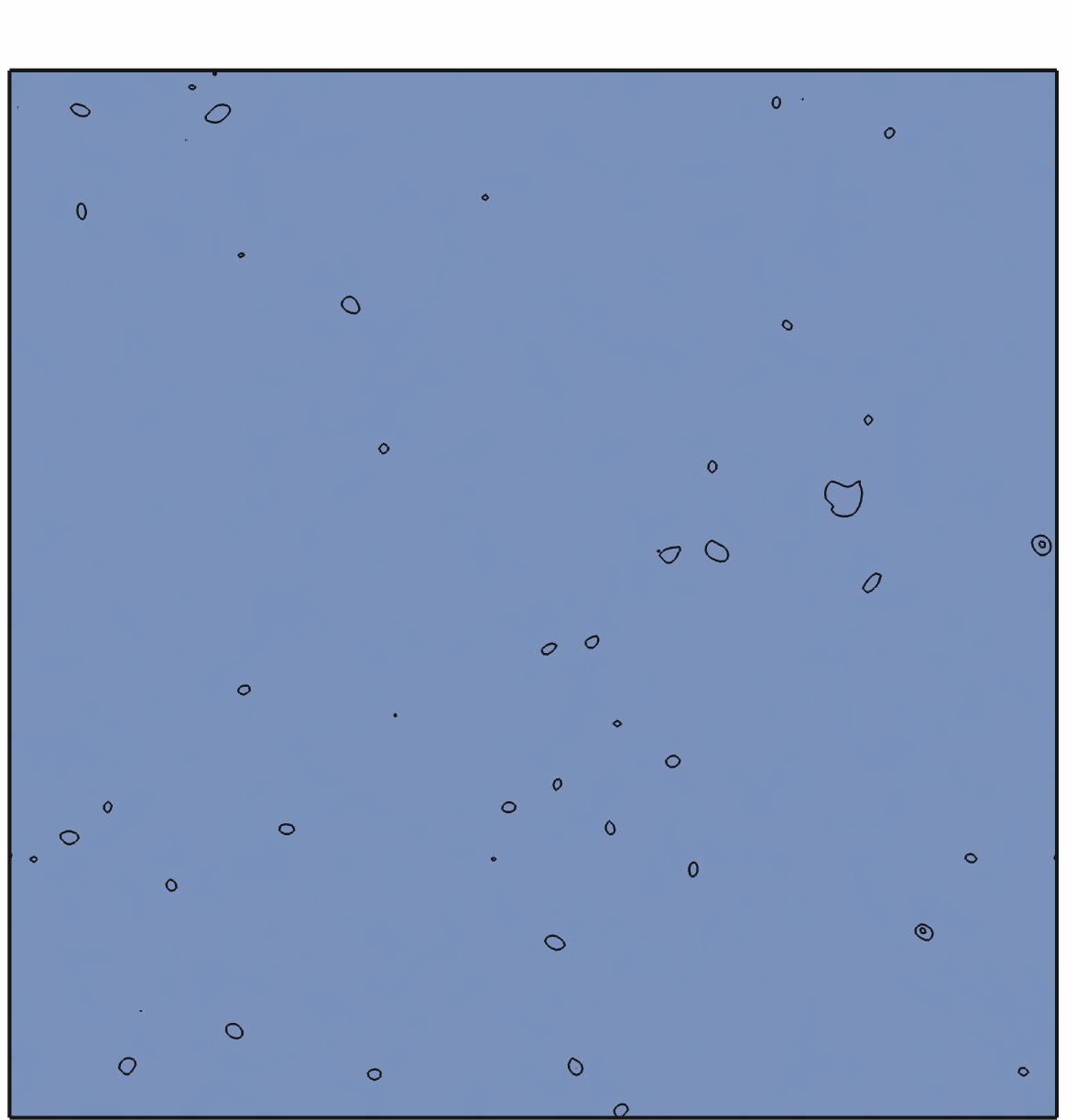}
  }
  \subfigure[$N = 25$]{
   \includegraphics[width=0.38\textwidth,trim=130px 460px 130px 30px, clip]{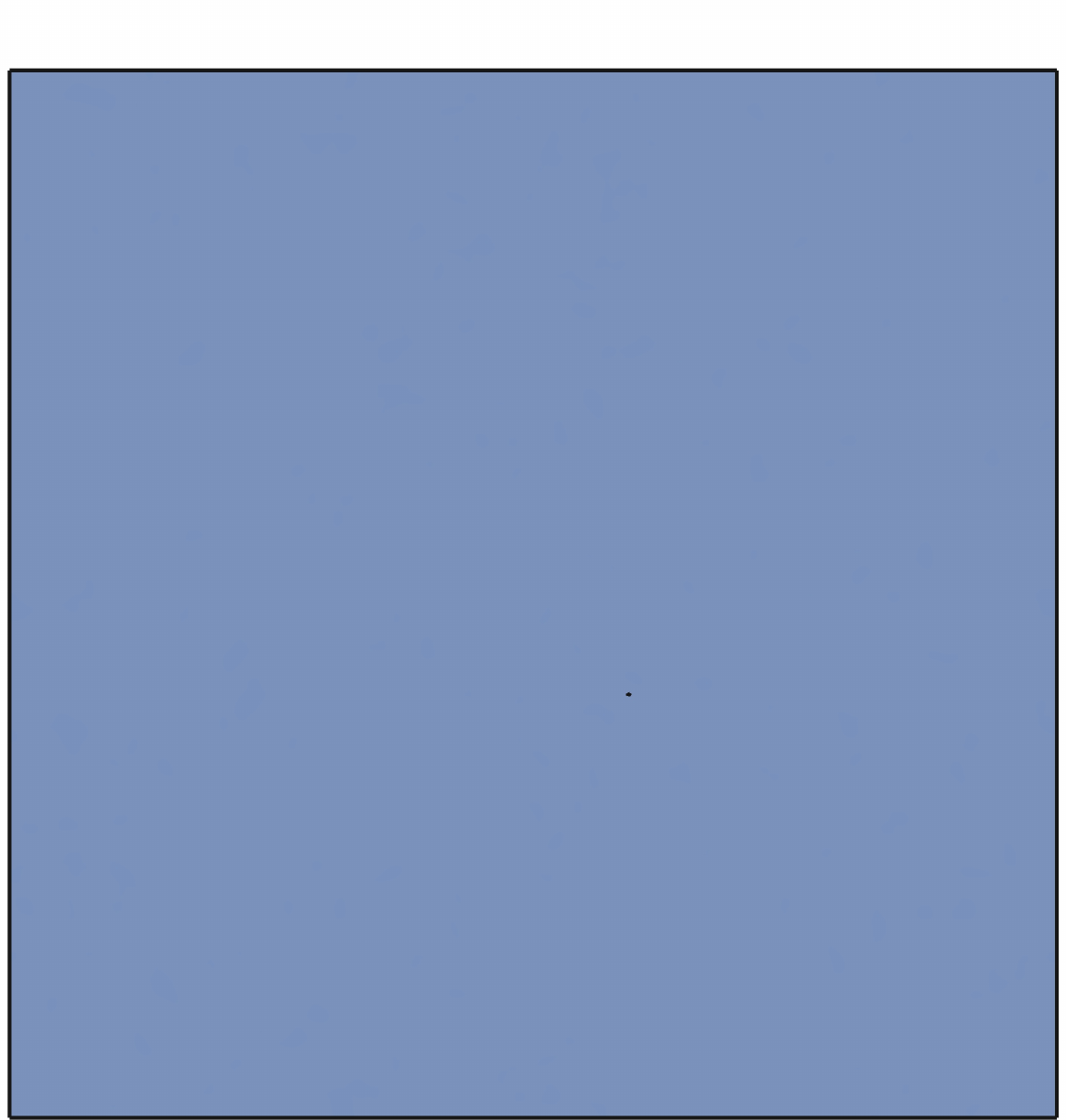}
  }
  \subfigure[$N = 46$]{
   \includegraphics[width=0.38\textwidth,trim=130px 460px 130px 30px, clip]{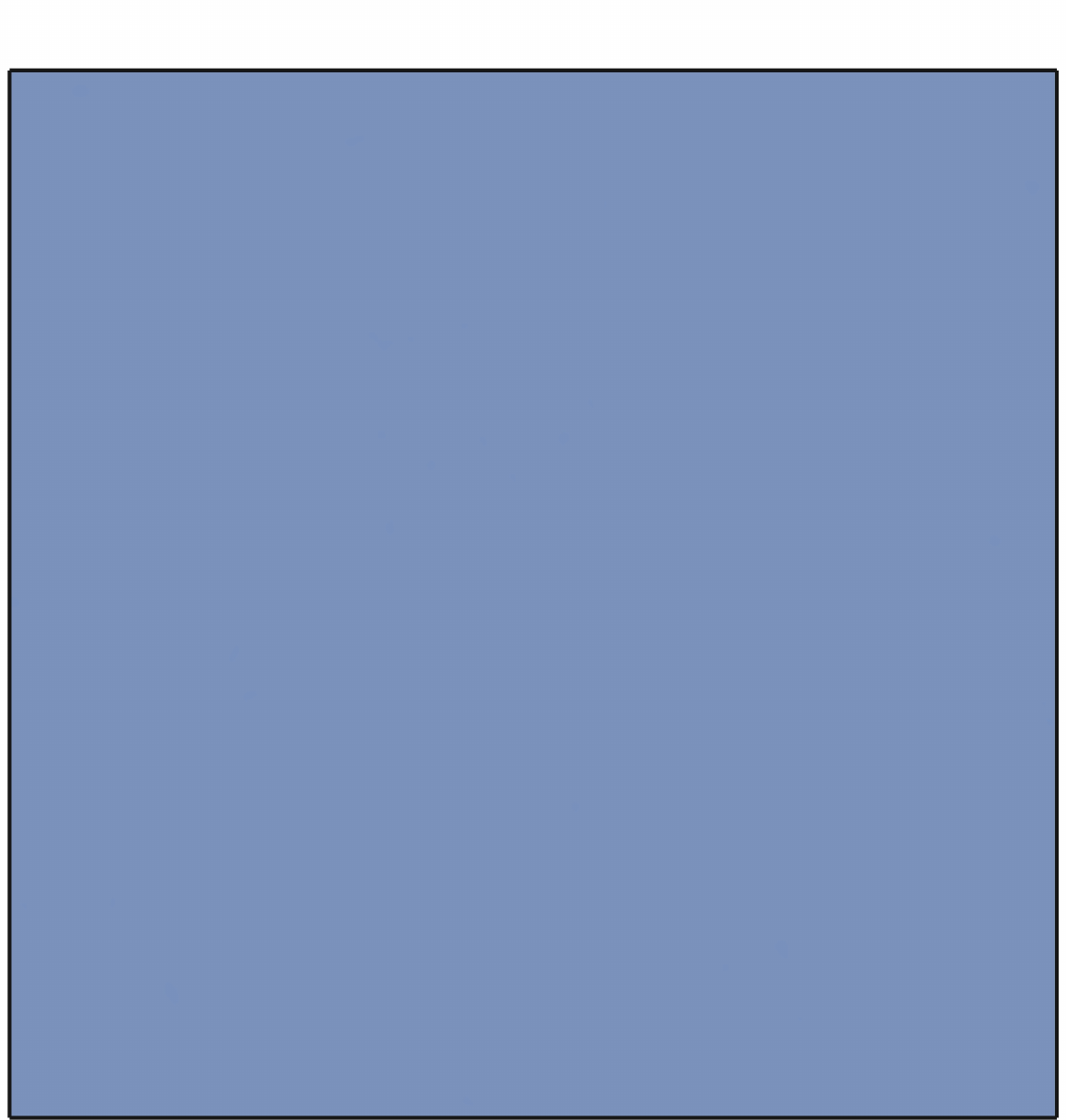}
  }
 \end{center}
 \caption{Contours plotted for 0.5\%, 1\%, 3\%, 5\%, 10\%, 20\%, 25\%, 50\%,
 75\% and 90\% of $Q_{\textrm{max}}$. $R_\lambda \sim 100$ on $256^3$ run
 \frun{f256b}.}
 \label{fig:str256_Q}
\end{figure}

Due to the restriction of isotropy, it is impossible for coherent structure
to exist in homogeneous, isotropic turbulence in anything other than an
instantaneous sense \cite{McComb11}. To test the amount of residual
coherent structure present in a finite ensemble, we ensemble average the
velocity field. The set $\mathbb{S}_N$ contains $N$ realisations of the
velocity field taken from a stationary simulation, sampled at an interval of
one large eddy turnover time. The average is then found as
\begin{equation}
 \left\langle \vec{u}(\vec{x}) \right\rangle_N = \frac{1}{N} \sum_{n = 1}^N
 \vec{u}^{(n)}(\vec{x}) \ ,
\end{equation}
where $\vec{u}^{(n)}(\vec{x})$ is the $n$th member of $\mathbb{S}_N$. The
resultant fields for a sample of $N$ are visualised in figures
\ref{fig:str256_V} and \ref{fig:str256_Q} using vorticity and the
$Q$-criterion, respectively. As can be seen in figure \ref{fig:str256_V},
the case $N = 1$ corresponding to a single realisation displays the expected
mass of structures. When we add in another realisation ($N = 2$) we see a
definite reduction in the amount of structure present; and this becomes even
more dramatic when we move to $N = 5$. By this point, we have no structures
with higher vorticity than our second lowest contour. Proceeding to $N = 10$
and 25 the structures reduce further, until for $N = 46$ we see only two
very small areas with vorticity as high as 5\% of $\omega_{\textrm{max}}$
obtained from the initial single realisation. A similar story is seen in
figure \ref{fig:str256_Q} using the $Q$-criterion, with the difference that
by $N = 46$ there are no structures present with even 0.5\% of
$Q_{\textrm{max}}$ obtained in $N = 1$. Note that there appear to be more
structures in figure \ref{fig:str256_Q} than \ref{sfig:structures512_Q}
above due to the inclusion of significantly lower, less restrictive contour
values.

From this, we conclude that as the ensemble size is increased, there remains
less and less coherent structure in the velocity field. The effect of
intermittency, clearly present in a single realisation, on statistical
quantities disappears as a result of ensemble averaging.

The constraint $\langle \vec{u} \rangle = 0$ does not itself imply that
there is no coherent structure which could remain under averaging, since one
could set up, for example, two counter-rotating vortices. However, these
structures break isotropy, and it is this which prevents their presence
under the ensemble averaging process. This test therefore also assists in
determining the degree to which isotropy is satisfied as the ensemble size
is increased.


\clearpage

\section{Discussion: intermittency effects \emph{versus} finite-Reynolds-number corrections}

In this section, we discuss various contentious issues concerning the Kolmogorov theory that have arisen over the years. These are: difficulties in reconciling the cascade picture with the physics of vortex stretching; the Landau criticism of K41; Kolmogorov's modified theory of 1962; anomalous exponents; and the more recent development of support for K41. We begin with a brief discussion of internal intermittency and highlight what we regard as the two most important points.

\subsection{Internal intermittency}
This type of intermittency was first pointed out by Batchelor and Townsend \cite{Batchelor49}.
At one time, it was referred to as `dissipation-range intermittency', and then more recently as
`fine-scale' or `small-scale' intermittency. Nowadays it seems
to be more usual to call it `internal intermittency'. This is presumably because it is now recognized (mainly from numerical simulations) that this type of intermittency is present at all scales.
In any case, it should be distinguished
from the intermittency associated with a free edge of unconfined shear flows or with the
`bursting process' in duct flows. In both these cases, intermittency is associated with
\emph{structure}, which exists in some average sense. Due to its restrictive symmetries, it
is impossible for structure to exist in isotropic turbulence, in anything other than an instantaneous sense.

There seem to be two crucial difficulties with the idea that the existence of spatial
intermittency implies the need for some corrections to the Kolmogorov (1941) theory. These
are as follows:
\begin{enumerate}
\item The dissipation rate is not the relevant quantity for K41A. The relevant quatity is
the inertial flux of energy. The use of the same symbol $\vep$ (and even the same terminology!) for
both these quantities may have caused some confusion. 
It is for this reason that we have introduced $\vep_{T}$ for flux (and also $\vep_{D}$
for decay rate) \cite{McComb09a}.
\item Internal intermittency is a phenomenon associated with a single realization. It must 
necessarily `average out'; or, in other words, disappear, under any global averaging operation. It 
is not immediately obvious that its existence will affect relationships between globally-averaged
quantities. In the case of K41B, Kolmogorov's starting point was the K\'{a}rm\'{a}n-Howarth equation
and of course this equation is ensemble-averaged and contains the mean rate of dissipation. This
fact was recognised in K62, where it was stated that the `4/5' law for $S_{3}(r)$ was
unaffected by the process of locally averaging the instantaneous dissipation rate $\hat{\vep}$ 
over a sphere of radius $r$. A significant new element in that work was the introduction
of an \emph{ad hoc} expression for the skewness and the abandonment of the assumption of
constant skewness, as previously made in K41B. 
\end{enumerate}

However, despite the lack of any obvious connection between intermittency and the Kolmogorov theory, the use of the term `intermittency effects' became quite widespread. By the 1980s, the exponent $\mu$ was widely referred to as the \emph{intermittency exponent}: see \cite{McComb90a}. More recently the term \emph{anomalous
exponents} has become popular in connection with structure functions of all orders.

\subsection{Cascade or vortex stretching?}

The idea of a cascade, as believed to underpin the Kolmogorov picture,
is often seen to be incompatible with the vortex-stretching behaviour of
turbulence (e.g. see \cite{Saffman68}, \cite{Kraichnan74}). In order to
examine this proposition, we need to understand the meaning of both
these terms and indeed how they relate to the Kolmogorov theories.

The term `cascade' is used to describe a process in which energy is
transferred from large scales to small scales. However, there is no
cascade in real space. This is because there is no inertial flux: the
K\'{a}rm\'{a}n-Howarth equation is a \emph{local} energy balance\footnote{This
is not the case for the Navier-Stokes equation, where the pressure term
is known to be non-local in real space \cite{McComb90a}. But the NSE describes momentum
transfer in a single realization, not ensemble-averaged energy
transfer.}, taken at a position $\mathbf{x}$ or a scale $r$. Of course,
in shear flows, the inhomogeneity leads to a flux of energy from where
it is produced to where it is dissipated. But in isotropic turbulence no such flux
exists. Good discussions of this topic may be found in the books by
Tsinober \cite{Tsinober09} and that of Sagaut and Cambon \cite{Sagaut08}
(who cite the first edition of Tsinober's book).

How then does all this affect K41A and K41B, both of which were formulated in real
space? Taking them in reverse order, K41B does not rely on a cascade. It relies on the
vanishing of the viscous term in the K\'{a}rm\'{a}n-Howarth equation when one takes the limit
of infinite Reynolds numbers at constant dissipation rate. Also, bearing in mind the
local nature of the K\'{a}rm\'{a}n-Howarth equation, one may also apply a suitably chosen 
stirring force, such that its effects are confined to scales greater than some input
scale $r_{I}$. Then the energy balance applies to an \emph{inertial range of scales}
where the detailed effects of dissipation and forcing are not felt. This separation of
input (or energy-containing) scales from viscous dissipation scales offers a sort of
intuitive basis for K41A, which is actually no more than a dimensional analysis. 

It is generally understood that turbulence is characterised by various processes
involving vortical structures, in which the vorticity increases with time in ways that
we tend to associate with vortex stretching. One can think of a vortex tube being
stretched by a velocity gradient, and in the process, conservation of angular momentum
ensuring that the associated kinetic energy is concentrated in ever-smaller regions of
space.

In the past, various models have been proposed to describe this process and these
include arrays of tubes, plane vortex sheets, and so on. Naturally in each case, analytical
tractability is a prime consideration in choosing the model. A brief introduction
to such methods may be found in \cite{McComb90a}. However, in recent years it has
become clear, from numerical simulations of isotropic turbulence, that surfaces of constant vorticity
tend to take the form of randomly-coiling worm-like structures.

When we seek to confront the cascade picture with the vortex-stretching
picture, the difficulty in so doing is two-fold. First, the cascade is
in wavenumber space, whereas the vortex structures are in real space.
Second, the cascade is an ensemble-averaged process, whereas the vortex
structure is an instantaneous phenomenon. It is characteristic of a
single realization and may be eliminated by averaging. Nevertheless, it
is still possible, in very general terms, to identify some apparent
inconsistency between the two pictures, as follows.

Granted that the cascade describes the transfer of energy from small wavenumbers to
large; and intuitively associating the corresponding `scales' in $x$-space with the reciprocal
of the wavenumbers, one can see a major difficulty in reconciling the two
processes \cite{Saffman68}, \cite{Kraichnan74}. That is, when a vortex tube is drawn out in space, the length scale of the extensional field may be expected to be much larger than the diameter of the 
extended tube. This is not immmediately compatible with the idea of localness
of a cascade. Or alternatively, when a vortex tube, or any comparable vortical structure, is stretched,
at least one dimension of the structure remains large (when compared to the 
cross-section) and thus does not appear to support the idea of transfer to a
larger wavenumber (i.e. smaller scales). 

We can offer counter arguments to these points. But before doing
so, let us enter a caveat. This to the effect that we should not rely
too heavily on intuition about Fourier analysis, which might just cope with Fourier transforms
applied to some very simple problem such as waves in linear electric
circuits or physical optics. In contrast, fluid turbulence is not only a highly
nonlinear phenomenon but also involves random amplitudes and phases. In
some many-body problems, the approximate cancellation of phases
justifies a \emph{random phase approximation}; and in other cases the
phases are known to cancel exactly. Indeed, when we form the energy
spectrum in turbulence, the phases cancel exactly by construction. But
this does not alter the fact that making intuitive connections between
phenomena in $x$-space and those in $k$-space is likely to prove a
rather fraught procedure. Having said that, let us now take the two
points in order.

The idea of localness is crucial to the cascade. When we consider the
Fourier-transformed NSE, the first thing we learn is that nonlinear
mixing couples each Fourier mode to every other mode. So it would be
helpful if we had something like a `nearest neighbour' assumption for
interactions \emph{e.g.} as in the Ising model for ferromagnetism). The
situation is made worse in turbulence by the fact that the interactions between modes
of the NSE are triadic. So the concept of `nearest neighbour' is not
available to us and has to be replaced by some idea of `strongly
interacting triads' and `weakly interacting triads'. This topic has
been the subject of numerical investigations. For example, see the results of Domaradzki and Rogallo \cite{Domaradzki90}, Shanmugasundaram \cite{Shanmugasundaram92} and Yeung \emph{et al.} \cite{Yeung95}. The latter investigation is
particularly interesting in our present context as it considers
concurrent real-space and wavenumber-space views. However, interesting
although these matters are, we do not need to rely on them in order to
justify Kolmogorov's picture. It was recognised by Batchelor
\cite{Batchelor71}, at least as early as the 1950s, that the key was
not the transfer spectrum but the flux through mode $k$. This concept was implicit in the work of Obukhov \cite{Obukhov41} and Onsager \cite{Onsager45}.
It follows rigorously from the symmetry of the NSE, that the \emph{local flux}
through mode $k$ is determined by a sum over all contributions $j$ such
that $j\leq k$. This, combined with scale invariance, is the only concept of localness that K41 needs.
As the Reynolds number increases, and the energy-containing and
dissipation ranges move apart, the inertial range becomes that range of
wavenumbers where the flux is approximately constant and equal to the
dissipation.

As regards the second point, one must be aware of identifying a vortical
structure, such as a vortex tube, with a particular Fourier mode. If a
given vortex tube contributes to $u^{2}_{1}(k)$ and $u^{2}_{2}(k)$,
where $k$ is large; but not to the corresponding $u^{2}_{3}(k)$, then
some other vortex tube (or, tubes) must make up the deficit. At the end of the day,
the combined effects of ensemble-averaging and isotropy will ensure that
this is so. We would reiterate that the Kolmogorov spectral picture
involves only average quantities and reasoning from some speculation
about a single realization is likely to prove tenuous at best.

\subsection{The Landau criticism of K41 and problems with averages}

The idea that K41 had some problem with the way that averages were taken has
its origins in the famous footnote on page 126 of the book by Landau and
Lifshitz \cite{Landau59}. This footnote is notoriously difficult to
understand; not least because it is meaningless unless its discussion of
the `dissipation rate $\vep$' refers to the \emph{instantaneous}
dissipation rate. Yet $\vep$ is clearly defined in the text above (see
the equation immediately before their (33.8)) as being the \emph{mean}
dissipation rate. Nevertheless, the footnote ends with the sentence 'The
result of the averaging therefore cannot be universal'. As their
preceding discussion in the footnote makes clear, this lack of
universality refers to 'different flows': presumably wakes, jets, duct
flows, and so on.

We can attempt a degree of deconstruction as follows. We will use our own notation, and to this end we introduce the instantaneous structure function $\hat{S}_2(r,t)$, such that $\langle \hat{S}_2(r,t) \rangle =S_2(r)$. Landau and Lifshitz consider the possibility that $S_2(r)$ could be a universal function in any turbulent flow, for sufficiently small values of $r$ (i.e. the Kolmogorov theory). They then reject this possibility, beginning with the statement:
\begin{quote}
`The instantaneous value of $\hat{S}(r,t)$ might in principle be expressed as a universal function of the energy dissipation $\vep$ at the instant considered.'
\end{quote}
Now this is rather an odd statement. Ignoring the fact that the dissipation is not the relevant quantity for inertial-range behaviour, it is really quite meaningless to discuss the universality of a random variable in terms of its relation to a mean variable (i.e. the dissipation). A discussion of universality requires mean quantities. Otherwise it is impossible to test the assertion. The authors have possibly relied on their qualification `at the instant considered'. But how would one establish which instant that was for various different flows?

They then go on:
\begin{quote}
When we average these expressions, however, an important part will be played by the law of variation of $\vep$ over times of the order of the periods of the large eddies (of size $\sim L$), and this law is different for different flows.'
\end{quote}
This seems a rather dogmatic statement but, if $\vep$ is the \emph{mean} dissipation rate, then 
it is clearly wrong for the the broad (and important) class of stationary flows. In such flows, the mean dissipation rate $\vep$ does not vary with time.

Nevertheless, the authors conclude that: `The result of the averaging therefore cannot be universal.'

One has to make allowance for possible uncertainties arising in translation, but nevertheless, the latter argument only makes any sort of sense if the dissipation rate is also instantaneous. Such an assumption appears to have been made by Kraichnan \cite{Kraichnan74}, who provided an interpretation which does not actually depend on the \emph{nature} of the averaging process.
In fact Kraichnan worked with the energy spectrum, rather than the
structure function, and interpreted Landau's criticism of K41 as 
applying to
\beq
E(k) = \alpha\vep^{2/3}k^{-5/3}.
\label{6-K41}
\eeq 
His interpretation of Landau was that the prefactor $\alpha$ may not be a
universal constant because the left-hand side of equation (\ref{6-K41})
is, an average, while the right-hand side is the 2/3 power of an
average. Any average involves the taking of
a limit. Suppose we consider a time average, then we have
\beq
E(k) = \lim_{T\rightarrow\infty}\frac{1}{T}\int^{T}_{0}\widehat{E}(k,t)dt,
\eeq 
where as usual the `hat' denotes an instantaneous value. Clearly the statement
\beq
E(k) = \mbox{a constant};
\eeq
or equally the statement,
\beq
E(k) = f\equiv\langle\hat{f}\rangle, 
\eeq
for some suitable $f$, presents no problem. It is the `2/3' power on the
right-hand side of equation (\ref{6-K41}) which means that we are
apparently equating the operation of taking a limit to the 2/3 power of
taking a limit. However, it has recently been shown \cite{McComb09a}
that this issue is resolved by noting that the pre-factor $\alpha$
itself involves an average over the phases of the system. It turns out that $\alpha$ depends on an ensemble average to the $-2/3$ power and this cancels the dependence on the $2/3$ power on the right hand side of (\ref{6-K41}).

\subsection{The Kolmogorov (1962) theory: a critical view}

As we have already observed, Kolmogorov interpreted Landau's criticism as
referring to the small-scale intermittency of the instantaneous
dissipation rate. His response was to adopt Obukhov's proposal to
introduce a dissipation rate which had been averaged over a sphere of
radius $r$, which may be denoted by $\varepsilon_r$. This procedure runs into an immediate fundamental objection.
In K41A, (or its wavenumber- space equivalent) the relevant
inertial-range quantity for the dimensional analysis is the local (in
wavenumber) energy
transfer. Under certain circumstances, as is well known, this is equal to the mean dissipation rate by the
global conservation of energy\footnote{It is a potent source of
confusion that these theories are almost always discussed in terms of
dissipation $\varepsilon$ when the proper inertial-range quantity is the nonlinear
transfer of energy $\Pi$. The inertial range is defined by the condition $\Pi_{max} = \varepsilon$}. However, as pointed out by Kraichnan \cite{Kraichnan74}, there is no such simple relationship between locally-averaged energy transfer and locally-averaged
dissipation.

Although Kolmogorov presented his 1962 theory as `A refinement of previous hypotheses \dots', following Kraichnan \cite{Kraichnan74}, it is now generally understood that this is incorrect. In fact it is a radical change of approach. The 1941 theory amounted to a general assumption that a cascade of many steps would lead to scales where the mean properties of turbulence were independent of the conditions of formation (i.e. of, essentially, the physical size of the system). Whereas, in 1962, the assumption was, in effect, that the mean properties of turbulence \emph{did} depend on the physical size of the system. We will return to this point presently, but for the moment we concentrate on the preliminary steps.

The 1941 theory relied on a general assumption with an underlying physical plausibility. In contrast, the 1962 theory involved an arbitrary and specific assumption. This was to the effect that the logarithm of $\varepsilon(\mathbf{x},t)$ has a normal distribution for large $L/r$ where $L$ is referred to as an external scale and is related to the physical size of the system. We describe this as `arbitrary' because no physical justification is offered; but in any case it is certainly specific. Then, arguments are developed that lead to a modified expression for the second-order structure function, thus: \begin{equation}S_2(r)=C(\mathbf{x},t)\varepsilon^{2/3}r^{2/3}(L/r)^{-\mu}, \label{62S2}\end{equation} where $C(\mathbf{x},t)$ depends on the macrostructure of the flow.

In addition, Kolmogorov points out that `the theorem of constancy of skewness \dots derived (\emph{sic}) in Kolmogorov (1941b)' is replaced by \begin{equation} S(r) = S_0(L/r)^{3\mu/2},\end{equation} where $S_0$ also depends on the macrostructrure. 

Equation (\ref{62S2}) is rather clumsy in structure, in the way the prefactor $C$ depends on $x$. This is because we have $r=x-x'$, so clearly $C(\mathbf{x},t)$ has an implicit dependence on $r$. A better way of tackling this would be to introduce centroid and relative coordinates, $\mathbf{R}$ and $\mathbf{r}$, such that \begin{equation}\mathbf{R} = (\mathbf{x}+\mathbf{x'})/2; \qquad \mbox{and} \qquad \mathbf{r}= ( \mathbf{x}-\mathbf{x'}).\end{equation} Then we can re-write the prefactor as $C(\mathbf{R}, r; t)$, where the dependence on the macrostructure is represented by the centroid variable, while the dependence on the relative variable holds out the possibility that the prefactor becomes constant for sufficiently small values of $r$. 

Of course, if we restrict our attention to homogeneous fields, then there can be no dependence of mean quantities on the centroid variable. Accordingly, one should make the replacement: \begin{equation}C(\mathbf{R}, r; t)=C(r; t),\end{equation} and the additional restriction to stationarity would eliminate the dependence on time. In fact Kraichnan \cite{Kraichnan74} goes further and replaces the pre-factor with the constant $C$: see his equation (1.9).

For sake of completeness, another point worth mentioning at this stage is that the derivation of
the `4/5' law is completely unaffected by the `refinements' of K62. This
is really rather obvious. The K\'{a}rm\'{a}n-Howarth equation involves only
ensemble-averaged quantities and the derivation of the `4/5' law
requires only the vanishing of the viscous term. This fact was noted by
Kolmogorov \cite{Kolmogorov62}.

Our final point under this heading has not, so far as we know, previously been made in the literature of the subject. That is, the energy spectrum is, in the sense of thermodynamics, an \emph{intensive quantity}. Therefore it should not depend on the system size. This is, as opposed to the total kinetic energy (say) which does depend on the size of the system and is therefore \emph{extensive}. What applies to the energy spectrum also applies to the second-order structure function. If we now consider equation (\ref{62S2}), this contains the factor $L^{-\mu}$. Now, $L$ is only specified as the external scale in K62, but it is necessarily related to the size of the system. Accordingly, taking the limit of infinite system size is related to taking the limit of infinite values of $L$, which is needed in order to have $k=0$ and to be able to carry out Fourier transforms. If we do this, we have three possible outcomes. If $\mu$ is negative, then $S_2 \rightarrow \infty$, as $L \rightarrow \infty$, whereas if $\mu$ is positive, then $S_2$ vanishes in the limit of infinite system size. Hence, in either case, the result is unphysical, both by the standards of continuum mechanics and by those of statistical physics.

However, if $\mu = 0$ then there is no problem. The structure function (and spectrum) exist in the limit of infinite system size. Could this be an argument for K41? At the end of this section, we will discuss a recent investigation which claims to be just that, and which may in some sense be related to the present point.

Lastly, we should mention that McComb and May \cite{McComb18c} have used a plausible method to estimate values of $L$ and, taking a representative value of $\mu=0.1$, have shown that the inclusion of this factor (as in K62) destroys the well-known collapse of spectral data that can be achieved using K41 variables.

\subsection{Anomalous exponents}

The term \emph{anomalous exponents} is used to refer to the case where
the power-law exponents $\zeta_{n}$ of structure functions $S_{n}(r)$
differ from the Kolmogorov values $n/3$. The rise in interest in this
topic has (unsurprisingly) gone hand in hand with the trend in recent
years away from measurements of spectra to measurements of moments and
structure functions in real space. Typically an experimental plot of
exponents $\zeta_{n}$ against $n$ yields a curve in which the difference
between $\zeta_{n}$ and $n/3$ increases with increasing $n$.

The idea of \emph{anomalous exponents} seems to have arisen by analogy
with the concept of \emph{anomalous dimension} in the statistical field
theory of equilibrium critical phenomena. However, such analogies should
be interpreted with caution. In equilibrium, critical exponents
determined by renormalization group methods differ from those obtained
by dimensional analysis. The central role of the dimension of space in
these theories leads to a natural interpretation in terms of anomalous
dimension. But this situation arises because dimensional analysis is a
very weak method in equilibrium problems and requires the introduction
of densities in order to introduce dimensional considerations. In
contrast, non-equilibrium systems are characterised by a
symmetry-breaking current or flux. In the case of turbulence, this is
the inertial transfer flux; and, combined with conservation of energy,
this provides a strong constraint on dimensional analysis. A simple
introduction to these ideas can be found in the book \cite{McComb04}. At the same time, it should be borne in mind that any systematic trend in the dependence of $\zeta_{n}$ on $n$, with increasing $n$,
may be a systematic error due to the increasing importance of rare events with order $n$. This fact has been noted (although taken no further) by Frisch \cite{Frisch95}. More recently, McComb \emph{et al.} employed a standard technique from experimental physics to reduce the effects of systematic error and showed that the exponent of the second-order structure function tended to the Kolmogorov value of $2/3$ as the Reynolds number tended to infinity.

The association of internal intermittency with anomalous exponents has developed strongly
over the last few decades. Originally fractal models were popular (for an introductory
discussion, see \cite{McComb90a}) but later on multifractal models replaced them in 
popularity: for a recent review, see \cite{Boffetta08}. Over the same time, there has
been a growing body of work supporting the obvious explanation for the deviation of
exponents from the Kolmogorov (1941) values: namely that the conditions imposed by the
theory are not fully satisfied at finite Reynolds numbers.

This disagreement is capable of being resolved. As direct numerical simulations increase in size and resolution, an examination of the dependence of $\mu$ on Reynolds number should ultimately settle this question.

However we may conclude this part of our discussion with a salutary
quotation, taken from Kraichnan's 1974 discussion of Kolmogorov's
theories \cite{Kraichnan74}. If $E(k)\sim
k^{-5/3 + \mu}$ is asymptotically valid, then it follows that
\begin{quote} 
'... the value of $\mu$ depends on the details of the
nonlinear interaction embodied in the Navier-Stokes equation and cannot be deduced from
overall symmetries, invariance and dimensionality.'
\end{quote}
In other words, perceived intermittency is an aspect of the solution of the
Navier-Stokes equation.

\subsection{Theoretical and experimental support for Kolmogorov (1941)} 

In addition to the extensive and indeed quite remarkable experimental support for the Kolmogorov spectrum, there are various investigations which, individually and collectively, offer considerable support to K41, rather than K62. Among the earliest in this category is the work by
Effinger and Grossmann on the second-order structure function for the
velocity field \cite{Effinger87}, which was later extended to structure
functions of the temperature field in passive convection
\cite{Effinger89}.

These authors studied the second-order structure function, by
introducing a spatial smoothing operation in which they averaged the
fluid velocity field $\ua\xt$ over a sphere of radius $r$, thus:
\beq
\ur_{\alpha}\xt = \rav{\ua\xt},
\label{raver}
\eeq
where $\ur_{\alpha}$ is referred to as the \emph{super-scale} velocity
field and the superscript on the angle-brackets indicates that the spatial
average is taken over a sphere of radius $r$. The corresponding
\emph{sub-scale} velocity field $\tilur_{\alpha}$ was then obtained by
subtraction, thus:
\beq
\tilur_{\alpha}\xt = \ua\xt - \ur_{\alpha}\xt.
\eeq
The authors drew an analogy between their approach, and that of
Reynolds, in which they operate on the NSE with (\ref{raver}) in order
to derive separate equations of motion for the super-scale and sub-scale
velocity fields. In principle, then, their strategy is to solve the
equation for the sub-scale field and substitute the result into the
equation for the super-scale field. In order to do this, they make a
number of approximations, predominantly of the type used in
renormalization group theory, which their method to some extent
resembles. But, although approximate, their result for $S_2$ agrees well
with experimental results and its asymptotic behaviour supports K41 with
viscous corrections.

As mentioned above, Effinger and Grossmann
\cite{Effinger89} extended this method to the problem of passive scalar
convection. Later, this group presented an analysis of data from DNS
which supported the idea that there are no intermittency corrections to
energy spectra, when their results are extended to very high Reynolds
numbers \cite{Grossmann94}; and more recently they have argued that the
use of nonperturbative renormalization group methods enforces the K41
spectrum for isotropic turbulence \cite{Esser99}.

Chronologically, our next approach is due to Qian, who has published a
series of papers dealing with aspects of the scaling properties of isotropic turbulence;
and, in particular, on deciding whether the second-order exponent
$\zeta_2$ corresponds to normal Kolmogorov scaling ($\zeta_2 = 2/3$) or
anomalous scaling ($\zeta_2 > 2/3$). Here we shall concentrate on just
three of these, that is \cite{Qian00}, and the two papers leading up to
it, \cite{Qian98a,Qian98b}. We may begin by noting that his method is
different from all the other theoretical approaches, in that it is
really a sophisticated form of data analysis, and is based on the use of
exact relationships, combined with well-established data correlations,
in order to extract as much information as possible from experimental
results. Where assumptions are made, careful testing of the effect of
varying these assumptions shows that they are innocuous. 

We will give only a brief impression of this work and concentrate on his
analysis of extended  self-similarity or ESS. Qian shows that a
log-log plot of $S_2$ versus $S_3$ produces the expected straight line.
But when he plots the local gradient of that line against $r/\eta$,
instead of being constant as one would expect, it shows a prominent
peak, only becoming constant at large values of the scale. This in
itself appears to question the validity of ESS. However, extending this
work to higher-order structure functions, Qian demonstrates that the
results from ESS actually support K41 rather than K62. 

We cannot do justice to Qian's analysis here. The interested reader
should consult the original papers which, although hard work, are
rewarding. We now turn to the work of Barenblatt and Chorin, who also
have published extensively on the theory of turbulence, particularly
with reference to scaling and similarity, over a period of years. A good
starting point is their two papers in 1988 \cite{Barenblatt98a,
Barenblatt98b}, which summarise their approach and which cite many
earlier references. Their main emphasis is on the so-called `law of the
wall' in wall-bounded shear flows. However, they also deal with the
inertial-range spectrum and conclude that the classical, unmodified K41
theory gives '.. an adequate description of the local features of
developed turbulent flows'. It is, of course, this latter aspect which
concerns us here.

Essentially, Barenblatt and Chorin discuss the nature of scaling theory
in turbulence. At the point where most people follow K62, and introduce
the external length-scale to fill the gap in the dimensional analysis,
these authors give arguments to support the use of the dissipation
length-scale for this purpose. As a result they conclude that both the
prefactor and the exponent in K41 are subject to corrections which are
dependent on the Reynolds number. Overall, they conclude that ` \dots
there are no intermittency corrections to the Kolmogorov `5/3' spectral
exponent'. 

Next, we consider the first of two asymptotic matching theories: in this
case for the energy spectrum. The work of Gamard and George
\cite{Gamard00} was motivated by the experimental study of Mydlarski and
Warhaft \cite{Mydlarski96}, which reported finite Reynolds number
effects in inertial-range spectra. As with other investigations
discussed here, the paper cited is the outcome of a programme of work
over some years and gives a number of references to previous work by
George and co-workers.

Their starting point is the recognition that the energy spectrum can be
scaled both on K\'{a}rm\'{a}n-Howarth variables (which gives a better collapse
of data at low wavenumbers) and also on Kolmogorov variables (which
gives a better collapse of data at high wavenumbers). Accordingly they
introduced the dimensionless functions $f_{L}$ for low wavenumbers, and
$f_{H}$ for high wavenumbers. Recognising that $f_{L}$ must asymptote to
an inertial-range form for high wavenumbers and $f_{H}$ must asymptote
to an inertial-range form for low wavenumbers, Gamard and George set out
to establish their functional form in a common region such that this
form exists in the limit of infinite Reynolds numbers.  Extension of this to finite Reynolds numbers involved some
approximations, but these were checked by experimental comparisons at
crucial stages. In a convincing analysis, they showed that the
intermittency exponent $\mu$ vanishes as the Reynolds number increases,
in agreement with experiment.

Our second asymptotic analysis is the theory of Lundgren
\cite{Lundgren02}, who adopted a similar strategy to that of Gamard and
George, but who worked in real space with the structure functions. Like
them, he employed both K\'{a}rm\'{a}n-Howarth and Kolmogorov variables to scale
the structure functions, and then matched asymptotic high-Reynolds expansions to
obtain the Kolmogorov '$2/3$' law. However, where Gamard and George
\emph{demonstrate} the point, Lundgren \emph{proves analytically} that
the KH scaling (at large scales) and the Kolmogorov scaling (at small
scales) are both asymptotically valid, as the Reynolds number tends to
infinity. Matching asymptotic expansions, Lundgren concluded that, in the
limit of infinite Reynolds numbers, $S_2 \sim r^{2/3}$ and $S_3 \sim r$,
in accordance with K41. In later work, Lundgren examined the dependence
on Reynolds number in a more general way \cite{Lundgren08}.

Gamard and George on the one hand, and Lundgren on the other, leave very
little room for doubt. Their identical conclusions are to the effect
that the predictions of K41 for $E(k)$ or $S_2(r)$ are subject to
corrections due to finite viscosity and are asymptotically valid in the
limit of infinite Reynolds number. Both can point to experimental
support for their theoretical conclusions.

Returning to the spectral picture, we may make the observation that K41b relies on a
\emph{de facto} closure of the K\'{a}rm\'{a}n-Howarth equation as the viscosity
tends to zero. The same idea can be employed in wavenumber space with
the Lin equation \cite{McComb09a}, but here the effect of the phases can
be taken into account explicitly. It transpires that the spectral
prefactor arises from an integral over the phases and that the presence
of this average resolves the problem of dependence on an average to the
`2/3' power, as noted earlier. This work is related to the resolution of the inertial-range scale-invariance paradox \cite{McComb08} in 2008 (and more recently the introduction of a modified Lin equation which avoids the paradox \cite{McComb20}).

Turning now to approaches based on experiments (including numerical simulations) we have the investigations of Antonia and Burattini \cite{Antonia06} in 2006 and Tchoufag, Sagaut and Cambon \cite{Tchoufag12} in 2012. Both investigations found that the $4/5$ law for $S_3$ was approached asymptotically with increasing Reynolds number and that this approach was much slower in the case of decaying turbulence than for stationary turbulence. Antonia and Burattini cited Lindborg \cite{Lindborg99} for anticipating the difference between decaying and stationary isotropic turbulence; while Tchoufag \emph{et al.} concluded that this difference was due to the non-negligible nature of the term $\partial E(k,t)/\partial t$. Note that McComb and Fairhurst \cite{McComb18a} have used an asymptotic expansion in inverse powers of the Reynolds number to show that this term cannot be neglected by reason of restriction to certain scales nor to large Reynolds numbers. Also relevant, is the lack of a general criterion for the onset of turbulence in free decay, which to some extent hampers comparisons \cite{McComb18b}.

A much fuller treatment of the topics in this subsection may be found in Sections 6.4 and 6.5 of the book \cite{McComb14a}. Here we mention the more recent investigations of Antonia, Djenidi and Tang \cite{Antonia17,Tang19,Djenidi19,Antonia19,Djenidi21}, who have mounted a sustained attack on the problem of clarifying the relevance of the Kolmogorov theory. The last of these is a very remarkable paper which shows that applying H\"{o}lder's inequality to the exponents of the structure functions shows that the assumption $\zeta_3 =1$ leads to $\zeta_2 =2/3$. It is tempting to conjecture that this may be relatable to Lundgren's analysis \cite{Lundgren02} and also to our present use of realizability criteria at the end of subsection 6.4 of the present work.

\section{Conclusion}

It is difficult to avoid concluding that much of the confusion over intermittency corrections arises from the fact that, as we have discussed, K41A is not a completely satisfactory theory.
There is no cascade in real space and the dimensional analysis seems to
rely on some sort of intuitive appeal to what is going on behind the
scenes, as it were, in wavenumber space. There we have the concept of
flux of energy from \emph{all} lower wavenumbers to the wavenumber of
interest. This is due to an exact symmetry of the equations of motion. The necessary
additional concept of scale invariance (which defines the inertial
range) arises inevitably as the Reynolds number increases and the
dissipation is pushed to ever-higher wavenumbers. This also is a
property of the NSE. Perhaps, in principle, it would be better to first carry out the
analysis in wavenumber space and then recover the inertial-range form of
$S_2(r)$ by Fourier transformation? In any case, we have argued here for the combination of the two approaches into the Kolmogorov-Obukhov theory, as indeed was sometimes done many years ago.

In contrast, K41B, as a prediction of the inertial-range form of the
third-order structure function, is incontrovertible. The analyis is
asymptotically exact for stationary turbulence and the Kolmogorov form must apply at infinite
Reynolds numbers. From this, two points arise as a corollary. They are,
as follows: 
\begin{enumerate} 
\item Any theory or procedure which relies on the assumption that the
scaling exponent of $S_3$ is exactly $n=1$ at finite Reynolds numbers is
already subject to error, however small. 
\item The fact that $S_3$ is subject to finite-viscosity corrections
sets a precedent for $S_2$ which is rigorously connected to it by
conservation of energy. 
\end{enumerate}

Overall, our numerical results, in agreement with those from other investigations, support the Obukhov picture of the onset of scale-invariance of the inertial flux of energy to higher wavenumbers and the inevitable consequence that the energy spectrum is given by the `$5/3$' law. The disappearance of the intermittency under ensemble-averaging is a necessary consequence of the isotropy and for those who understand that, this is a reassuring test of the isotropy of our statistical ensemble. For those who do not, it is a demonstration that the internal intermittency has no consequences for the inertial-range flux and hence no consequences for the Kolmogorov-Obukhov theory.

Essentially, we have re-phrased the question about intermittency effects on the energy spectrum. Instead of asking: does intermittency alter the $5/3$ exponent, we have asked: does intermittency affect the onset of scale-invariance of the energy flux? And the answer seems to be that it cannot! However, it should be borne in mind that our present discussion is limited to stationary, homogeneous, isotropic turbulence and our conclusions may have to be modified for any more general situation.
The necessity for this type of examination for other classes of flow
must be borne in mind.

\section*{Acknowledgement}
\emph{Simulations were performed on the Eddie HPC
cluster hosted by the Edinburgh Parallel Computing Centre (EPCC).
Data are available online
[http://dx.doi.org/10.15129/64a4a042-7d0d-48ce-8afa-21f9883d1e84].}

Website for hit3d: http://code.google.com/p/hit3d/


\begin{thebibliography}{10}

\bibitem{Lumley01}
J.~L. Lumley and A.~M. Yaglom.
\newblock {A {C}entury of {T}urbulence}.
\newblock {\em Flow, Turbulence and Combustion}, 66:241, 2001.

\bibitem{Kolmogorov41a}
A.~N. Kolmogorov.
\newblock {The local structure of turbulence in incompressible viscous fluid
  for very large {R}eynolds numbers}.
\newblock {\em C. R. Acad. Sci. URSS}, 30:301, 1941.

\bibitem{Richardson22}
L.~F. Richardson.
\newblock {\em {Weather {P}rediction by {N}umerical {P}rocess}}.
\newblock Cambridge University Press, 1963.

\bibitem{Batchelor47}
G.~K. Batchelor.
\newblock {Kolmogorov's theory of locally isotropic turbulence}.
\newblock {\em Proc. Camb. Philos. Soc.}, 43:533, 1947.

\bibitem{Onsager45}
L.~Onsager.
\newblock {The {D}istribution of {E}nergy in {T}urbulence}.
\newblock {\em Phys. Rev.}, 68:281, 1945.

\bibitem{Obukhov41}
A.~M. Obukhov.
\newblock {On the distribution of energy in the spectrum of turbulent flow}.
\newblock {\em C.R. Acad. Sci. U.R.S.S}, 32:19, 1941.

\bibitem{Anselmet84}
F.~Anselmet, Y.~Gagne, E.~J. Hopfinger, and R.~A. Antonia.
\newblock {High-order velocity structure functions in turbulent shear flows}.
\newblock {\em J. Fluid Mech.}, 140:63, 1984.

\bibitem{Bowman96}
John~C. Bowman.
\newblock {On inertial-range scaling laws}.
\newblock {\em J. Fluid Mech.}, 306:167--181, 1996.

\bibitem{Barenblatt98a}
G.~I. Barenblatt and A.~J. Chorin.
\newblock {New perspectives in turbulence: scaling laws, asymptotics and
  intermittency}.
\newblock {\em SIAM Rev.}, 40:265--291, 1998.

\bibitem{Esser99}
A.~Esser and S.~Grossmann.
\newblock {Nonperturbative renormalisation group approach to turbulence}.
\newblock {\em Eur. Phys. J. B}, 7:467--482, 1999.

\bibitem{Qian00}
J.~Qian.
\newblock {Closure {A}pproach to {H}igh-{O}rder {S}tructure {F}unctions of
  {T}urbulence}.
\newblock {\em Physical Review Letters}, 84(4):646--649, 2000.

\bibitem{Gamard00}
S.~Gamard and W.~K. George.
\newblock {Reynolds number dependence of energy spectra in the overlap region
  of isotropic turbulence}.
\newblock {\em Flow, turbulence and combustion}, 63:443--477, 1999.

\bibitem{Lundgren02}
Thomas~S. Lundgren.
\newblock {Kolmogorov two-thirds law by matched asymptotic expansion}.
\newblock {\em Phys. Fluids}, 14:638, 2002.

\bibitem{McComb09a}
David McComb.
\newblock {Scale-invariance and the inertial-range spectrum in
  three-dimensional stationary, isotropic turbulence}.
\newblock {\em J. Phys. A: Math. Theor.}, 42:125501, 2009.

\bibitem{Kolmogorov41b}
A.~N. Kolmogorov.
\newblock {Dissipation of energy in locally isotropic turbulence}.
\newblock {\em C. R. Acad. Sci. URSS}, 32:16, 1941.

\bibitem{Landau59}
L.~D. Landau and E.~M. Lifshitz.
\newblock {\em {Fluid Mechanics}}.
\newblock Pergamon Press, London, {E}nglish edition, 1959.

\bibitem{Kolmogorov62}
A.~N. Kolmogorov.
\newblock {A refinement of previous hypotheses concerning the local structure
  of turbulence in a viscous incompressible fluid at high {R}eynolds number}.
\newblock {\em J. Fluid Mech.}, 13:82--85, 1962.

\bibitem{McComb14a}
W.~David McComb.
\newblock {\em {Homogeneous, {I}sotropic {T}urbulence: {P}henomenology,
  {R}enormalization and {S}tatistical {C}losures}}.
\newblock Oxford University Press, 2014.

\bibitem{McComb16a}
W.~D. McComb.
\newblock {{Infrared properties of the energy spectrum in freely decaying
  isotropic turbulence}}.
\newblock {\em Phys. Rev. E}, 93:013103, 2016.

\bibitem{McComb10b}
W.~David McComb, Arjun Berera, Matthew Salewski, and Sam~R. Yoffe.
\newblock {Taylor's (1935) dissipation surrogate reinterpreted}.
\newblock {\em Phys. Fluids}, 22:61704, 2010.

\bibitem{McComb18b}
S.~R. Yoffe and W.~D. McComb.
\newblock {Onset criteria for freely decaying turbulence}.
\newblock {\em Phys. Rev. Fluids}, 3:104605, 2018.

\bibitem{McComb14b}
W.~D. McComb, S.~R. Yoffe, M.~F. Linkmann, and A.~Berera.
\newblock {{Spectral analysis of structure functions and their scaling
  exponents in forced isotropic turbulence}}.
\newblock {\em Phys. Rev. E}, 90:053010, 2014.

\bibitem{McComb15a}
W.~D. McComb, A.~Berera, S.~R. Yoffe, and M.~F. Linkmann.
\newblock {{Energy transfer and dissipation in forced isotropic turbulence}}.
\newblock {\em Phys. Rev. E}, 91:043013, 2015.

\bibitem{McComb15b}
W.~D. McComb, M.~F. Linkmann, A.~Berera, S.~R. Yoffe, and B.~Jankauskas.
\newblock {{Self-organization and transition to turbulence in isotropic fluid
  motion driven by negative damping at low wavenumbers}}.
\newblock {\em J. Phys. A Math.Theor.}, 48:25FT01, 2015.

\bibitem{Berera14}
A.~Berera and M.~Linkmann.
\newblock {Magnetic helicity and the evolution of decaying magnetohydrodynamic
  turbulence}.
\newblock {\em Phys. Rev. E}, 90:013007--1--25, 2014.

\bibitem{Linkmann15a}
M.~F. Linkmann, A.~Berera, W.~D. McComb, and M.~E. McKay.
\newblock {Nonuniversality and {F}inite {D}issipation in {D}ecaying
  {M}agnetohydrodynamic {T}urbulence}.
\newblock {\em Phys. Rev. Lett.}, 114:235001, 2015.

\bibitem{Yoffe12}
S.~R. Yoffe.
\newblock {\em {Investigation of the transfer and dissipation of energy in
  isotropic turbulence}}.
\newblock PhD thesis, University of Edinburgh, 2012.

\bibitem{Schumakov07}
S.~G. Chumakov.
\newblock {Scaling properties of subgrid-scale energy dissipation}.
\newblock {\em Phys. Fluids}, 19:058104, 2007.

\bibitem{Schumakov08}
S.~G. Chumakov.
\newblock {A priori study of subgrid-scale flux of a passive scalar in
  turbulence}.
\newblock {\em Phys. Rev. E}, 78:15563, 2008.

\bibitem{Brachet83}
M.~E. Brachet, D.~I. Meiron, S.~A. Orszag, B.~G. Nickel, R.~H. Morf, and
  U.~Frisch.
\newblock {Small-scale structure of the {T}aylor-{G}reen vortex}.
\newblock {\em J. Fluid Mech.}, 130:411--452, 1983.

\bibitem{Wang96}
L.-P. Wang, S.~Chen, J.~G. Brasseur, and J.~C. Wyngaard.
\newblock {Examination of hypotheses in the {K}olmogorov refined turbulence
  theory through high-resolution simulations. Part 1. Velocity field}.
\newblock {\em J. Fluid Mech.}, 309:113, 1996.

\bibitem{Cao99}
N.~Cao, S.~Chen, and G.~D. Doolen.
\newblock {Statistics and structures of pressure in isotropic turbulence}.
\newblock {\em Phys. Fluids}, 11:2235--2250, 1999.

\bibitem{Gotoh02}
T.~Gotoh, D.~Fukayama, and T.~Nakano.
\newblock {Velocity field statistics in homogeneous steady turbulence obtained
  using a high-resolution direct numerical simulation}.
\newblock {\em Phys. Fluids}, 14:1065, 2002.

\bibitem{Kaneda03}
Y.~Kaneda, T.~Ishihara, M.~Yokokawa, K.~Itakura, and A.~Uno.
\newblock {Energy dissipation and energy spectrum in high resolution direct
  numerical simulations of turbulence in a periodic box}.
\newblock {\em Phys. Fluids}, 15:L21, 2003.

\bibitem{Donzis05}
D.~A. Donzis, K.~R. Sreenivasan, and P.~K. Yeung.
\newblock {Scalar dissipation rate and dissipative anomaly in isotropic
  turbulence}.
\newblock {\em J. Fluid Mech.}, 532:199--216, 2005.

\bibitem{Yeung12}
P.~K. Yeung, D.~A. Donzis, and K.~R. Sreenivasan.
\newblock {Dissipation, enstrophy and pressure statistics in turbulence
  simulations at high Reynolds numbers}.
\newblock {\em J. Fluid Mech.}, 700:5--15, 2012.

\bibitem{Orszag71}
S.~A. Orszag.
\newblock {On the Elimination of Aliasing in Finite-Difference Schemes by
  Filtering High-Wavenumber Components}.
\newblock {\em J. Atmos. Sc.}, 28:1074, 1971.

\bibitem{Heun00}
K.~Heun.
\newblock {Neue Methoden zur approximativen Integration der
  Differentialgleichungen einer unabh\"angigen Ver\"anderlichen}.
\newblock {\em Z. Math. Phys.}, 45:23--38, 1900.

\bibitem{Jimenez93}
J.~Jim{\'e}nez, A.~A. Wray, P.~G. Saffman, and R.~S. Rogallo.
\newblock {The structure of intense vorticity in isotropic turbulence}.
\newblock {\em J. Fluid Mech.}, 255:65, 1993.

\bibitem{Machiels97a}
L.~Machiels.
\newblock {Predictability of small-scale motion in isotropic fluid turbulence}.
\newblock {\em Phys. Rev. Lett.}, 79(18):3411--3414, 1997.

\bibitem{Yamazaki02}
Y.~Yamazaki, T.~Ishihara, and Y.~Kaneda.
\newblock {Effects of Wavenumber Truncation on High-Resolution Direct Numerical
  Simulation of Turbulence}.
\newblock {\em J. Phys. Soc. Jap.}, 71:777--781, 2002.

\bibitem{Kaneda06}
Y.~Kaneda and T.~Ishihara.
\newblock {High-resolution direct numerical simulation of turbulence}.
\newblock {\em Journal of Turbulence}, 7:1--17, 2006.

\bibitem{Doering05}
Charles~R. Doering and Nikola~P. Petrov.
\newblock {Low-wavenumber forcing and turbulent energy dissipation}.
\newblock {\em Progress in Turbulence}, 101(1):11--18, 2005.

\bibitem{McComb01a}
W.~D. McComb, A.~Hunter, and C.~Johnston.
\newblock {Conditional mode-elimination and the subgrid-modelling problem for
  isotropic turbulence}.
\newblock {\em Phys. Fluids}, 13:2030, 2001.

\bibitem{Yeung97}
P.~K. Yeung and Y.~Zhou.
\newblock {Universality of the {K}olmogorov constant in numerical simulations
  of turbulence}.
\newblock {\em Phys. Rev. E}, 56:1746, 1997.

\bibitem{Ishihara09}
T.~Ishihara, T.~Gotoh, and Y.~Kaneda.
\newblock {Study of high-{R}eynolds number isotropic turbulence by direct
  numerical simulation}.
\newblock {\em Ann. Rev. Fluid Mech.}, 41:165, 2009.

\bibitem{Gotoh01}
T.~Gotoh and D.~Fukayama.
\newblock {Pressure spectrum in homogeneous turbulence}.
\newblock {\em Phys. Rev. Lett.}, 86:3775, 2001.

\bibitem{Sreenivasan95}
K.~R. Sreenivasan.
\newblock {On the universality of the {K}olmogorov constant}.
\newblock {\em Phys. Fluids}, 7:2778, 1995.

\bibitem{Mydlarski96}
L.~Mydlarski and Z.~Warhaft.
\newblock {On the onset of high-{R}eynolds-number grid-generated wind tunnel
  turbulence}.
\newblock {\em J. Fluid Mech.}, 320:331--368, 1996.

\bibitem{Vincent91}
A.~Vincent and M.~Meneguzzi.
\newblock {The spatial structure and statistical properties of homogeneous
  turbulence}.
\newblock {\em J. Fluid Mech.}, 225:1--20, 1991.

\bibitem{Kerr85}
R.~M. Kerr.
\newblock {Higher-order derivative correlations and the alignment of
  small-scale structures in isotropic numerical turbulence}.
\newblock {\em J. Fluid Mech.}, 153:31--58, 1985.

\bibitem{Sreenivasan97}
K.~R. Sreenivasan and R.~A. Antonia.
\newblock {The phenomenology of small-scale turbulence}.
\newblock {\em Annu. Rev. Fluid Mech.}, 29:435--472, 1997.

\bibitem{Ishihara07}
T.~Ishihara, Y.~Kaneda, M.~Yokokawa, K.~Itakura, and A.~Uno.
\newblock {Small-scale statistics in high-resolution direct numerical
  simulation of turbulence: {R}eynolds number dependence of one-point velocity
  gradient statistics}.
\newblock {\em J. Fluid Mech.}, 592:335--366, 2007.

\bibitem{She93a}
Z.-S. She, S.~Chen, G.~Doolen, R.~H. Kraichnan, and S.~A. Orszag.
\newblock {Reynolds number dependence of isotropic {N}avier-{S}tokes
  turbulence}.
\newblock {\em Phys. Rev. Lett.}, 70:3251, 1993.

\bibitem{Young99}
A.~J. Young.
\newblock {\em {Investigation of renormalization group methods for the
  numerical simulation of isotropic turbulence}}.
\newblock PhD thesis, University of Edinburgh, 1999.

\bibitem{McComb10a}
W.~David McComb, Arjun Berera, Matthew Salewski, and Sam~R. Yoffe.
\newblock {An exact expression for the Reynolds number dependence of the
  dissipation rate in homogeneous isotropic turbulence}.
\newblock {\em arXiv:1002.2131v1[physics.flu-dyn]}, 2010.

\bibitem{Batchelor53}
G.~K. Batchelor.
\newblock {\em {The theory of homogeneous turbulence}}.
\newblock Cambridge University Press, Cambridge, 1st edition, 1953.

\bibitem{Haller05}
G.~Haller.
\newblock {An objective definition of a vortex}.
\newblock {\em J. Fluid Mech.}, 525:1--26, 2005.

\bibitem{Jeong95}
J.~Jeong and F.~Hussain.
\newblock {On the identification of a vortex}.
\newblock {\em J. Fluid Mech.}, 285:69--94, 1995.

\bibitem{Okamoto07}
N.~Okamoto, K.~Yoshimatsu, K.~Schneider, M.~Farge, and Y.Kaneda.
\newblock {Coherent vortices in high resolution direct numerical simulation of
  homogeneous isotropic turbulence}.
\newblock {\em Phys. Fluids}, 19:115109, 2007.

\bibitem{Hunt88}
J.~C.~R. Hunt, A.~A. Wray, and P.~Moin.
\newblock {Eddies, streams and convergence zones in turbulent flows}.
\newblock In {\em Center for {T}urbulence {R}esearch: Proceedings of the
  {S}ummer {P}rogram}, pages 193--208, 1988.

\bibitem{McComb11}
W.~D. McComb.
\newblock {{Kolmogorov's Theory: K41 or K62}}.
\newblock {\em ERCOFTAC Bulletin}, 88, 2011.

\bibitem{Batchelor49}
G.~K. Batchelor and A.~A. Townsend.
\newblock {The nature of turbulent motion at large wavenumbers}.
\newblock {\em Proc. R. Soc. Lond. A}, 199:238, 1949.

\bibitem{McComb90a}
W.~D. McComb.
\newblock {\em {The {P}hysics of {F}luid {T}urbulence}}.
\newblock Oxford University Press, 1990.

\bibitem{Saffman68}
P.~G. Saffman.
\newblock {Lectures on homogeneous turbulence}.
\newblock In N.~Zabusky, editor, {\em Topics in nonlinear physics}, pages
  485--614. Springer-Verlag, 1968.

\bibitem{Kraichnan74}
R.~H. Kraichnan.
\newblock {On {K}olmogorov's inertial-range theories}.
\newblock {\em J. Fluid Mech.}, 62:305, 1974.

\bibitem{Tsinober09}
A.~Tsinober.
\newblock {\em {An {I}nformal {C}onceptual {I}ntroduction to {T}urbulence}}.
\newblock Springer, Dordrecht, 2nd edition, 2009.

\bibitem{Sagaut08}
P.~Sagaut and C.~Cambon.
\newblock {\em {Homogeneous {T}urbulence {D}ynamics}}.
\newblock Cambridge University Press, Cambridge, 2008.

\bibitem{Domaradzki90}
J.~A. Domaradzki and R.~S. Rogallo.
\newblock {Local energy transfer and nonlocal interactions in homogeneous
  isotropic turbulence}.
\newblock {\em Phys, Fluids A}, 2:413, 1990.

\bibitem{Shanmugasundaram92}
V.~Shanmugasundaram.
\newblock {Modal interactions and energy transfers in isotropic turbulence as
  predicted by local energy transfer theory}.
\newblock {\em Fluid. Dyn. Res}, 10:499, 1992.

\bibitem{Yeung95}
P.~K. Yeung, J.~G. Brasseur, and Qunzhen Wang.
\newblock {Dynamics of direct large-scale couplings in coherently forced
  turbulence: concurrent physical- and Fourier-space views}.
\newblock {\em J. Fluid Mech.}, 283:43--95, 1995.

\bibitem{Batchelor71}
G.~K. Batchelor.
\newblock {\em {The theory of homogeneous turbulence}}.
\newblock Cambridge University Press, Cambridge, 2nd edition, 1971.

\bibitem{McComb18c}
W.~D. McComb and M.~Q. May.
\newblock {The effect of {Kolmogorov} (1962) scaling on the universality of
  turbulence energy spectra}.
\newblock {\em arXiv:1812.09174[physics.flu-dyn]}, 2018.

\bibitem{McComb04}
W.~D. McComb.
\newblock {\em {Renormalization Methods: {A} {G}uide for {B}eginners}}.
\newblock Oxford University Press, 2004.

\bibitem{Frisch95}
U.~Frisch.
\newblock {\em {{T}urbulence: the legacy of {A}. {N}. {K}olmogorov}}.
\newblock Cambridge University Press, 1995.

\bibitem{Boffetta08}
G.~Boffetta, A.~Mazzino, and A.~Vulpiani.
\newblock {Twenty-five years of multifractals in fully developed turbulence: a
  tribute to {G}iovanni {P}aladin}.
\newblock {\em J. Phys. A: Math. Theor.}, 41:363001, 2008.

\bibitem{Effinger87}
H.~Effinger and S.~Grossmann.
\newblock {Static {S}tructure {F}unction of {T}urbulent {F}low from the
  {N}avier-{S}tokes {E}quations}.
\newblock {\em Z. Phys. B}, 66:289--304, 1987.

\bibitem{Effinger89}
H.~Effinger and S.~Grossmann.
\newblock {Prandtl number dependence of turbulent temperature structure
  functions: {A} unified theory}.
\newblock {\em Phys. Fluids A}, 1:1021--1026, 1989.

\bibitem{Grossmann94}
S.~Grossmann and D.~Lohse.
\newblock {Universality in fully developed turbulence}.
\newblock {\em Phys. Rev. E}, 50:2784, 1994.

\bibitem{Qian98a}
J.~Qian.
\newblock {Scaling exponents ofthe second-order structure function of
  turbulence}.
\newblock {\em J. Phys. A: Math. Gen.}, 31:3193--3204, 1998.

\bibitem{Qian98b}
J.~Qian.
\newblock {Normal and anomalous scaling of turbulence}.
\newblock {\em Physical Review E}, 58(6):7325--7329, 1998.

\bibitem{Barenblatt98b}
G.~I. Barenblatt and A.~J. Chorin.
\newblock {Turbulence: an old challenge and new perspectives}.
\newblock {\em Meccanica}, 33:445--468, 1998.

\bibitem{Lundgren08}
Thomas~S. Lundgren.
\newblock {Turbulent scaling}.
\newblock {\em Phys. Fluids}, 20:31301, 2008.

\bibitem{McComb08}
David McComb.
\newblock {Scale-invariance in three-dimensional turbulence: a paradox and its
  resolution}.
\newblock {\em J. Phys. A: Math. Theor.}, 41:75501, 2008.

\bibitem{McComb20}
W.~D. McComb.
\newblock {A modified {L}in equation for the energy balance in isotropic
  turbulence}.
\newblock {\em Theoretical \& Applied Mechanics Letters}, 10:377--381, 2020.

\bibitem{Antonia06}
R.~A. Antonia and P.~Burattini.
\newblock {Approach to the 4/5 law in homogeneous isotropic turbulence}.
\newblock {\em J. Fluid Mech.}, 550:175, 2006.

\bibitem{Tchoufag12}
J.~Tchoufag, P.~Sagaut, and C.~Cambon.
\newblock {Spectral approach to finite {R}eynolds number effects on
  {K}olmogorov's 4/5 law in isotropic turbulence}.
\newblock {\em Phys. Fluids}, 24:015107, 2012.

\bibitem{Lindborg99}
Erik Lindborg.
\newblock {Correction to the four-fifths law due to variations of the
  dissipation}.
\newblock {\em Phys. Fluids}, 11:510, 1999.

\bibitem{McComb18a}
W.~D. McComb and R.~B. Fairhurst.
\newblock {The dimensionless dissipation rate and the Kolmogorov (1941)
  hypothesis of local stationarity in freely decaying isotropic turbulence}.
\newblock {\em J. Math. Phys.}, 59:073103, 2018.

\bibitem{Antonia17}
R.~A. Antonia, L.~Djenidi, L.~Danilla, and S.L. Tang.
\newblock {Small scale turbulence and the finite {R}eynolds number effect}.
\newblock {\em Phys. Fluids}, 29:020715, 2017.

\bibitem{Tang19}
Shunlin Tang, Robert~A. Antonia, Lyazid Djenidi, and Yu~Zhou.
\newblock {Can small-scale turbulence approach a quasi-universal state?}
\newblock {\em Physical Review Fluids}, 4:024607, 2019.

\bibitem{Djenidi19}
L.~Djenidi, R.~A. Antonia, and S.~L. Tang.
\newblock {Scale invariance in finite Reynolds number homogeneous, isotropic
  turbulence}.
\newblock {\em J. Fluid Mech.}, 864:244, 2019.

\bibitem{Antonia19}
R.~A. Antonia, S.L. Tang, L.~Djenidi, and Y.~Zhou.
\newblock {Finite {R}eynolds number effect and the 4/5 law}.
\newblock {\em Physical Review Fluids}, 4:084602, 2019.

\bibitem{Djenidi21}
L.~Djenidi, R.~A. Antonia, and S.~L. Tang.
\newblock {Mathematical constraints on the scaling exponents in the inertial
  range of fluid turbulence}.
\newblock {\em Phys. Fluids}, 33:031703, 2021.

\end{thebibliography}

\end{document}